\documentclass{pasa}%

\usepackage{hyperref}
\usepackage{breakurl}
\usepackage{framed}

\usepackage{float}
\usepackage{caption}
\usepackage[subrefformat=parens,labelformat=parens]{subcaption}

\let\stdcaption\caption
\usepackage{float}
\let\caption\stdcaption

\usepackage[authoryear]{natbib}

\usepackage{supertabular}

\usepackage{hyperref} 
\usepackage{color}
\usepackage{color,hyperref}
\definecolor{darkblue}{rgb}{0.0,0.0,0.5}
\definecolor{yell}{rgb}{.8,.9,.0}
\hypersetup{colorlinks,breaklinks,
            linkcolor=darkblue,urlcolor=darkblue,
            anchorcolor=darkblue,citecolor=darkblue}

\usepackage[figuresleft]{rotating}
\usepackage{animate}

\newcommand{\arcdeg}{\ensuremath{^{\circ}}}

\def\fig{Figure}

\title[MWA ionosphere calibration]{Ionospheric modelling using GPS to calibrate the MWA. 1: Comparison of first order ionospheric effects between GPS models and MWA observations}
\def\Curtin{$^{1}$}
\def\USydney{$^{2}$}
\def\SKASA{$^{3}$}
\def\DPERU{$^{4}$}
\def\CfA{$^{5}$}
\def\ANU{$^{6}$}
\def\CASS{$^{7}$}
\def\CAASTRO{$^{8}$}
\def\RRI{$^{9}$}
\def\MIT{$^{10}$}
\def\UWA{$^{11}$}
\def\UW{$^{12}$}
\def\Victoria{$^{13}$}
\def\ASU{$^{14}$}
\def\UMelbourne{$^{15}$}
\def\Tata{$^{16}$}
\def\Haystack{$^{17}$}
\def\UWisc{$^{18}$}
\def\UMichigan{$^{19}$}
\def\NRAO{$^{20}$}
\def\SKAO{$^{21}$}

\author[Arora et al.]{B.~S.~Arora\Curtin, 
J.~Morgan\Curtin, 
S.~M.~Ord\Curtin,
S.~J.~Tingay\Curtin,
N.~Hurley-Walker\Curtin,
M.~Bell\USydney,
G.~Bernardi\SKASA$^,$\DPERU$^,$\CfA,
R.~Bhat\Curtin,
F.~Briggs\ANU,
J.~R.~Callingham\CASS$^,$\USydney$^,$\CAASTRO, 
A.~A.~Deshpande\RRI, 
K.~S.~Dwarakanath\RRI,
A.~Ewall-Wice\MIT, 
L.~Feng\MIT, 
B.-Q.~For\UWA,
P.~Hancock\Curtin$^,$\CAASTRO, 
B.~J.~Hazelton\UW, 
L.~Hindson\Victoria,
D.~Jacobs\ASU,
M.~Johnston-Hollitt\Victoria, 
A.~D.~Kapi\'{n}ska\UWA$^,$\CAASTRO,
N.~Kudryavtseva\Curtin,
E.~Lenc\USydney$^,$\CAASTRO, 
B.~McKinley\ANU$^,$\CAASTRO, 
D.~Mitchell\UMelbourne$^,$\CASS$^,$\CAASTRO, 
D.~Oberoi\Tata,
A.~R.~Offringa\ANU,
B.~Pindor\UMelbourne,
P.~Procopio\UMelbourne$^,$\CAASTRO, 
J.~Riding\UMelbourne,
L.~Staveley-Smith\UWA$^,$\CAASTRO, 
R.~B.~Wayth\Curtin$^,$\CAASTRO, 
C.~Wu\UWA,
Q.~Zheng\Victoria,
J.~D.~Bowman\ASU, 
R.~J.~Cappallo\Haystack, 
B.~E.~Corey\Haystack, 
D.~Emrich\Curtin,
R.~Goeke\MIT,
L.~J.~Greenhill\CfA,
D.~L.~Kaplan\UWisc, 
J.~C.~Kasper\UMichigan$^,$\CfA, 
E.~Kratzenberg\Haystack, 
C.~J.~Lonsdale\Haystack, 
M.~J.~Lynch\Curtin, 
S.~R.~McWhirter\Haystack,
M.~F.~Morales\UW, 
E.~Morgan\MIT, 
T.~Prabu\RRI, 
A.~E.~E.~Rogers\Haystack, 
A.~Roshi\NRAO, 
N.~Udaya~Shankar\RRI, 
K.~S.~Srivani\RRI, 
R.~Subrahmanyan\RRI$^,$\CAASTRO, 
M.~Waterson\SKAO,
R.~L.~Webster\UMelbourne$^,$\CAASTRO,  
A.~R.~Whitney\Haystack, 
A.~Williams\Curtin
and C.~L.~Williams\MIT \\
\\
\affil{$^{1}$International Centre for Radio Astronomy Research, Curtin University, Bentley, WA 6102, Australia}%
\affil{$^{2}$Sydney Institute for Astronomy, School of Physics, The University of Sydney, NSW 2006, Australia}%
\affil{$^{3}$Square Kilometre Array South Africa (SKA SA), 3rd Floor, The Park, Park Road, Pinelands, 7405, South Africa}%
\affil{$^{4}$Department of Physics and Electronics, Rhodes University, PO Box 94, Grahamstown, 6140, South Africa}%
\affil{$^{5}$Harvard-Smithsonian Center for Astrophysics, Cambridge, MA 02138, USA}%
\affil{$^{6}$Research School of Astronomy and Astrophysics, Australian National University, Canberra, ACT 2611, Australia}%
\affil{$^{7}$CSIRO Astronomy and Space Science (CASS), PO Box 76, Epping, NSW 1710, Australia}%
\affil{$^{8}$ARC Centre of Excellence for All-Sky Astrophysics (CAASTRO)}%
\affil{$^{9}$Raman Research Institute, Bangalore 560080, India}%
\affil{$^{10}$Kavli Institute for Astrophysics and Space Research, Massachusetts Institute of Technology, Cambridge, MA 02139, USA}%
\affil{$^{11}$International Centre for Radio Astronomy Research (ICRAR), University of Western Australia, Crawley, WA 6009, Australia}
\affil{$^{12}$Department of Physics, University of Washington, Seattle, WA 98195, USA}%
\affil{$^{13}$School of Chemical \& Physical Sciences, Victoria University of Wellington, Wellington 6140, New Zealand}%
\affil{$^{14}$School of Earth and Space Exploration, Arizona State University, Tempe, AZ 85287, USA}%
\affil{$^{15}$School of Physics, The University of Melbourne, Parkville, VIC 3010, Australia}%
\affil{$^{16}$National Centre for Radio Astrophysics, Tata Institute for Fundamental Research, Pune 411007, India}%
\affil{$^{17}$MIT Haystack Observatory, Westford, MA 01886, USA}%
\affil{$^{18}$Department of Physics, University of Wisconsin--Milwaukee, Milwaukee, WI 53201, USA}%
\affil{$^{19}$Department of Atmospheric, Oceanic and Space Sciences, University of Michigan, Ann Arbor, MI 48109, USA}%
\affil{$^{20}$National Radio Astronomy Observatory, Charlottesville and Greenbank, USA}%
\affil{$^{21}$SKA Organisation Headquarters, Jodrell Bank Observatory, Manchester, UK}%
}%
\jid{PASA}
\doi{10.1017/pas.\the\year.xxx}
\jyear{\the\year}

\begin{document}%
\begin{abstract}
We compare first order (refractive) ionospheric effects seen by the Murchison Widefield Array (MWA) with the ionosphere as inferred from Global Positioning System (GPS) data. The first order ionosphere manifests itself as a bulk position shift of the observed sources across an MWA field of view. These effects can be computed from global ionosphere maps provided by GPS analysis centres, namely the Center for Orbit Determination in Europe (CODE), using data from globally distributed GPS receivers. However, for the more accurate local ionosphere estimates required for precision radio astronomy applications, data from local GPS networks needs to be incorporated into ionospheric modelling. For GPS observations, the ionospheric parameters are biased by GPS receiver instrument delays, among other effects, also known as receiver Differential Code Biases (DCBs). The receiver DCBs need to be estimated for any non-CODE GPS station used for ionosphere modelling, a requirement for establishing dense GPS networks in arbitrary locations in the vicinity of the MWA. In this work, single GPS station-based ionospheric modelling is performed at a time resolution of 10 minutes. Also the receiver DCBs are estimated for selected Geoscience Australia (GA) GPS receivers, located at Murchison Radio Observatory (MRO1), Yarragadee (YAR3), Mount Magnet (MTMA) and Wiluna (WILU). The ionospheric gradients estimated from GPS are compared with the ionospheric gradients inferred from radio source position shifts observed with the MWA. The ionospheric gradients at all the GPS stations show a correlation with the gradients observed with the MWA. The ionosphere estimates obtained using GPS measurements show promise in terms of providing calibration information for the MWA. 

\end{abstract}
\maketitle%
\section{INTRODUCTION }
\label{sec:intro}
The Earth's ionosphere, being a dispersive medium at radio wavelengths, causes a change in the propagation velocity of radio waves, among other effects. The ionosphere is directly dependent on the solar activity through the high energy far ultraviolet and X-rays. The ionosphere further varies depending on various transportation and depletion processes, namely, due to the influence of tides and atmospheric (gravity) waves, solar winds, and vertical transport through eddy's diffusion and the geomagnetic disturbances \citep{Zol14}. Considering the ionosphere to be highly variable and uncertain, it becomes important to monitor the ionosphere on a regular basis for a number of applications. \\

Early ground-based ionospheric sensors like ionosondes were effective for understanding the bottomside ionosphere. Ground based Coherent and Incoherent radars operating at High Frequency (HF, 3-30 MHz), Very High Frequency (VHF, 30-300 MHz) and Ultra High Frequency (UHF, 300-3000 MHz), are able to probe the middle and upper ionosphere. Example of coherent radars include the Super Dual Auroral Radar Network (SuperDARN)  consisting of around 34 HF radars located at mid-latitudes and extending into the polar regions. The radars are pointed towards the North and the South poles to study ionospheric convection \citep{Zol14}. Incoherent radars include the Jicamarca Radio Observatory located along the geomagnetic equator in Lima, Peru, (49.92 MHz), the Arecibo dish in Puerto Rico (430 MHz), and EISCAT (European Incoherent SCATter), northern Scandinavia, operating at UHF (931 and 500 MHz) and VHF (224 MHz) \citep{Zol14}. To access information about the topside ionosphere, satellite based topside sounders have also been used \citep{Hun13}. With the advent of radio communication satellites, the signals from such satellite systems were used to obtain spatial and temporal information on the ionosphere \citep{Lei84}.\\

The Global Positioning System (GPS) was designed by the Department of Defence, U.S., in the early 1970s to fulfil U.S. military requirements \citep{FRP}. It has since been used in various civilian applications, from navigation to precise geodetic positioning. The promise of GPS to operate in all weather conditions, 24$\times$7, in addition to its multi-frequency transmission, has made it a useful tool to monitor ionospheric parameters \citep{El06}. In comparison to ground-based ionospheric sensors like radars and ionosondes, the satellite based system like GPS can provide continuous near real-time global coverage of the ionosphere.\\

For precise positioning applications using GPS, new control points can be established by applying constraints using positions from the established control points. Global GPS data processing centres like the International GNSS (Global Navigation Satellite System) Service (IGS), with international multi-agency members, provide support for such global geodetic activities. In addition to the positions of the global network of GPS/GNSS stations, a number of products such as precise satellite ephemerides, satellite clock parameters, Earth rotation parameters, global ionosphere maps, and zenith tropospheric path delays are routinely generated \citep{Beu99}. Global ionospheric maps are also generated by various IGS analysis centres, namely, the Center for Orbit Determination in Europe (CODE), the Astronomical Institute at the University of Bern (AIUB), Switzerland \citep{Rot97}, and the Jet Propulsion Laboratory (JPL), California Institute of Technology, U.S \citep{Man98,Kom05}, among others. The temporal and spatial resolution of global maps is generally of 2 hours and 5$^{\circ}$/2.5$^{\circ}$ in longitude/latitude, respectively.\\

Real-time GPS positioning accuracy can be improved by providing ionospheric and other corrections directly to the user. This is the role of Space Based Augmentation Systems (SBAS) such as the Wide Area Augmentation System (WAAS) in the US \citep{Par96}. Vertical ionosphere corrections are generated at temporal and spatial resolution of three minutes and 5$^{\circ}$/5$^{\circ}$ in latitude/longitude, respectively\footnote{\url{http://www.nstb.tc.faa.gov/RT_WaasSIGPStatus.htm}}. \\

The Global GPS data are also processed at the MIT Haystack Observatory by the MAPGPS software package in order to generate global TEC maps \citep[see][]{Rid06}. The TEC maps are generated with a greater temporal and spatial resolution of 10 minutes and 1$^{\circ}$/1$^{\circ}$ in longitude/latitude, respectively and are distributed through an open source, web based system\footnote{\url{http://madrigal.haystack.mit.edu/}}. To achieve even finer spatial resolution, regional ionosphere modelling must be performed. Regional ionosphere data centres like the Royal Observatory of Belgium (ROB) are able to model the ionosphere over small spatial areas with higher sampling of ground GPS stations with a temporal and spatial resolution of 15 minutes and 0.5$^{\circ}$/0.5$^{\circ}$ in longitude/latitude, respectively  \citep{Che13}. \\

Since the early days of radio astronomy, the ionosphere has been found to have a profound effect on astrometric observations. In a study of discrete radio sources, fluctuations in the signal from Cygnus A at observing frequency of 68 MHz were reported by \citet{Hey46}. Further investigation confirmed this to be a result of ionospheric structures. \citet{Hew51} and \citet{Boo58}, among others, used this information to quantify ionospheric fluctuations and present some insights into the nature and behaviour of the ionosphere. Radio astronomy can be used to extract information on the ionosphere, however an external source for deriving information on the ionosphere is needed to calibrate astrometric observations.\\

\cite{Ros00} evaluated the quality of GPS-based ionosphere corrections for Very Long Baseline Interferometry (VLBI) at 8.4 and 2.3 GHz. In his study, \citeauthor{Ros00} corrected the ionosphere for continental ($\sim$200 to $\sim$700 km) and inter-continental ($\sim$7800 to $\sim$8300 km) baseline lengths and concluded that GPS maps of Total Electron Content (TEC) can usefully contribute to VLBI astrometric analysis. \cite{Eri01} made use of the ionospheric corrections generated from an experimental set-up of four GPS receivers at the Very Large Array (VLA) site to correct for the ionospheric effect. \citeauthor{Eri01} found that large scale structures ($>$1000 km) could be resolved observing at frequencies of 322 and 333 MHz, whereas small scale fluctuations ($<$100 km) could not be seen using a global model; it was noted that global ionosphere models perform averaging over the ionosphere and hence lose their capacity to monitor small scale ionospheric changes. \citeauthor{Eri01} argues that a dense GPS network is required to correct for small scale fluctuations in the ionosphere.\\

In work by \citet{Sot13}, CODE and ROB maps were used to compute the Rotation Measure due to the ionosphere for polarised sources that were observed by LOFAR (the Low-Frequency Array for radio astronomy). However, the CODE maps presented in \citeauthor[see][(Figure 2)]{Sot13} show an error in the location of the equatorial anomaly. This error is further investigated and discussed in Appendix \ref{app} of this paper. This error is also discussed by \citet{Her14}.\\

GPS-based estimation of the ionosphere has also been carried out by \citet{Her13}. High-fidelity GPS systems were specially deployed by the US Air Force for measurement of ionospheric TEC and scintillation indices at the location of the Murchison Radio Observatory (MRO) and Australian Space Academy campus (Meckering, Western Australia). Two periods of time corresponding to low (2008-2009) and high (2012-2013) ionospheric activity were studied. During 2008-2009, the F10.7 Index varied between 65 to 82 ($10^{-22}Wm^{-2}Hz^{-1}$), the monthly variation was found to be insignificant for most of the period. The Ap index reached a maximum of 35 during this period. However, during high solar activity (2012-2013), F10.7 led between 85 to 190, the monthly variation was significant throughout the period. Ap index reached a maximum of 90. The ionosphere at the MRO was expected to exhibit very low levels of scintillation \citep{Ausion05}. For Kp=9, scintillation at MRO reached about 0.2. In \citeauthor{Her13}, during 2012-2013, scintillation was found to reached a maximum of about 0.3.\\

The Murchison Widefield Array (MWA) is a recently operational low-frequency radio telescope located within the MRO in Western Australia. The instrument exemplifies exceptional wide-field imaging capability at low frequencies (see \citet{Lon09} and \citet{2013PASA...30....7T} for a technical overview of the instrument). These capabilities make it ideal for a range of science investigations with an emphasis on: detection of the 21\,cm line from the Epoch of Reionoisation; large-scale surveys; searches for radio transients; continuum surveys; and Solar, Heliosphere and Ionospheric studies \citep{2013PASA...30...31B}. In the recent survey by \citet{mwacs}, over 14,000 radio sources were detected in just four nights of observing.\\

Since a fundamental observable of an interferometer is the phase difference between elements, such instruments are unable to measure the Total Electron Content (TEC) towards a particular source directly (though it may be inferred indirectly from polarisation measurements). However, for a single source located at infinity, the radiation reaching each element of an interferometer will have passed through a different part of the ionosphere providing information on the differential ionosphere or ionospheric gradients.\\

Excellent summaries of how the ionosphere affects radio interferometers are given in \citet{lonsdale:2004} and \citet{2010ISPM...27...30W}. In the specific case of the MWA, the instrument has a very wide field of view, but relatively short baselines. Thus, while the effects of the ionosphere may change across the instruments' field of view, the complex structure of the ionosphere is on larger spatial scales than the dimensions of the array, and the wavefront arriving at the array from a source will be largely coherent. However more complicated effects may be seen. Unless explicitly corrected in the calibration process, a TEC gradient across the array will manifest itself as a shift in source position; the incoming radiation gets refracted by the ionosphere. \\

The MWA has recently commenced operations and ionospheric effects are routinely detected. For example, observations have revealed variations in the Rotation Measure of polarised point sources and the diffuse galactic background (Lenc, private communication). \citet{Loi2015a} demonstrate the utility of the MWA as a powerful imager for studying high-altitude irregularities, revealing a population of field-aligned density ducts that appear regularly over the observatory. The spatial distributions of celestial source refractive offsets over the field of view, combined with a novel parallax technique for altitude measurement, enable the 3D characterisation of ionospheric structures at high temporal cadence. Travelling ionospheric disturbances (TIDs), sinusoidal perturbations with wavelengths of 100-1000 km and periods of several tens of minutes to an hour, are also often observed in MWA data. A technique for the spatio-temporal power spectrum analysis of ionospheric gradients, presented by \citet{Loi2015b}, enables characteristic wavelengths and periods of fluctuation to be measured. A quantitative study of the statistical properties of ionosphere-induced position and amplitude variations of sources in MWA images \citep{Loi2015c} has yielded information about characteristic density gradients and the diffractive scale in the ionosphere. This illustrates the capability of the MWA to probe ionospheric structure in great detail.\\

All previous attempts at GPS-based ionosphere calibration for radio astronomy made use of ionosphere information from TEC maps which have a temporal resolution of 2 hours. Ionospheric structures such as TIDs have a period of several tens of minutes, hence it is important to generate ionospheric information with higher resolution. We present here a method where publicly available data from GPS stations are used to generate location-specific ionosphere models at 10 minutes intervals. This requires the calibration of GPS receiver/transmitter instrumental biases in order to accurately model the ionosphere, also known as Differential Code Bias (DCB).\\

We perform a simultaneous fit for both GPS DCBs and ionospheric parameters. While the $\textsc{Bernese}$ software provides the GPS DCBs and ionospheric parameters \citep{Beu07}, our work allows much higher time resolution of the ionospheric parameters than is commonly obtained, while simultaneously solving for the DCBs.\\

This paper is organised as follows. In section \ref{sec:met}, the methodology concerning the GPS-based ionospheric modelling is fully described. In this section, the basic GPS observation equations, estimating the Vertical Total Electron Content ($VTEC$) are formulated. In particular, we show how the $VTEC$ and DCBs are determined simultaneously through Kalman filtering. Section \ref{sec:codemaps} briefly describes the methodology adopted by CODE to generate ionosphere maps. Section \ref{sec:mwaiono} presents the procedure to model the ionosphere seen by the MWA in the form of position shifts. The results are presented and discussed in section \ref{sec:results}. Results from the IGS analysis centre, CODE, serve as reference to validate our $VTEC$ results. In particular, we compare our estimated receiver Differential Code Biases (DCBs), the inter-frequency biases on code observables, with those determined by the $\textsc{Bernese}$ processing software employed by CODE. The ionosphere observed by the MWA and the GPS are analysed and the agreement discussed. Finally, we make some concluding remarks in section \ref{sec:conc}.

\section{ESTIMATION OF THE IONOSPHERE USING GPS OBSERVATIONS}
\label{sec:met}

A GPS constellation consists of up to 32 operational satellites placed in 6 orbital planes at an altitude of 20,200 km above the Earth's surface. With this constellation geometry and an orbital period of about 12 hours, 4 to 10 GPS satellites are visible anywhere in the world at any given time. The earlier GPS satellites (Block IIA (2nd generation, ``Advanced'') and Block IIR (``Replenishment'') transmitted on two carrier frequencies ($L1$ and $L2$), each encoded with one or two digital codes, and navigation messages \citep{El06}. The BLOCK IIR(M) (``Modernized'') satellites have an additional civil signal on $L2$ ($L2C$) \citep{Hof93}. The BLOCK IIF (``Follow-on'') satellites transmit an additional third frequency, $L5$, which has higher transmitted power and greater bandwidth, to support high-performance applications \citep{El06}.\\

The signals transmitted by the GPS satellites form the observables, namely the phase (of the carrier frequency) and code (digital code) measurements. A phase measurement is the number of cycles at the corresponding carrier frequency between the satellite and the receiver. The phase delay between the receiver and the satellite is obtained by multiplying the number of phase cycles with the wavelength of the corresponding carrier. The phase measurements are biased by an unknown number of phase cycles, in addition to other errors. When a GPS receiver is switched on or tracks a newly risen satellite, it cannot determine the total number of complete cycles between the receiver and the satellite. Hence the initial number of complete cycles remains ambiguous, known as phase ambiguity bias.\\

The GPS code measurements have two types of code observables, namely the C/A-code (Coarse/Acquisition code, modulated only on the $L1$ carrier, denoted as $C1$) and P-code (Precise code, modulated on both $L1$ and $L2$ carriers, denoted as $P1$ and $P2$, respectively). The code modulation is different for each GPS satellite, with code signals sometimes also referred to as PRN (Pseudo Random Noise). 
The C/A-code measurement is less precise than the P-code, since the bit rate of C/A-code is 10 times lower than P-code \citep{Lan93}. The GPS receiver generates replicas of the transmitted code signals. By comparing the code signal to its replica, the signal travel time is obtained \citep{Hof93}.  
For a more detailed description, one can refer to \citet{Hof93}, \citet{Lan93} and \citet{El06}. This work uses two GPS carrier frequencies, namely $f_{1}$ = 1575.42 MHz (for carrier $L1$) and $f_{2}$ = 1227.60 MHz (for $L2$).

\subsection{The GPS observation equation}
\label{sec:GPSmodel}
The terms contributing to the GPS phase and code observables can be given as follows \citep{Teu98}:

\begin{eqnarray}\label{eq:gnssobs1}
\displaystyle E(\Phi_{r,j}^{s}) &=& \displaystyle \overline{\rho}_{r}^{s} -  \iota_{r,j}^{s} + c \cdot ( \delta_{r,j} -\delta^{s}_{,j}) + \lambda_{j} M_{r,j}^{s}, \\ \nonumber
& & \\ \label{eq:gnssobs2}
\displaystyle E(P_{r,j}^{s}) &=& \displaystyle \overline{\rho}_{r}^{s} +  \iota_{r,j}^{s}  + c \cdot(d_{r,j} - d^{s}_{,j}).
\end{eqnarray}

\begin{table*}
\begin{center}
\caption{GPS model parameters and definitions}
\tablefirsthead{%
\hline \hline
Parameter & Definition & Units\\
\hline}
\tablelasttail{%
\hline
\hline}
\begin{supertabular}{p{3cm} p{9cm} p{2cm}}
$E(\cdot)$ & Expectation operator &\\
$\Phi_{r,j}^{s} $ & Phase observables &(metres)\\
$P_{r,j}^{s}$ & Code observables &(metres)\\
$\overline{\rho}_{r}^{s}$ & `Geometry' parameters, $\rho_{r}^{s}$ + $\tau_{r}^{s}$ + $c \cdot dt_{r}^{s}$ &(metres)\\
$\rho_{r}^{s}$ & Receiver-Satellite range &(metres)\\
$\tau_{r}^{s}$ & Tropospheric path length &(metres)\\
$c \cdot dt_{r}^{s}$ & Receiver and satellite clock errors &(metres)\\
$\iota_{r,j}^{s}$ & Receiver-Satellite ionospheric delay &(metres)\\
$\mu_{j}=1/f_{j}^{2}$ & Inverse of frequency square &\\
$f_{j}$ & GPS frequencies, $j=1,2$ &\\
$M_{r,j}^{s}$ & Non-integer phase ambiguity & (cycles)\\
$N_{r,j}^{s} $ & Integer phase ambiguity & (cycles)\\
$\phi_{r,j} (t_{0})$ & Initial receiver phase offset &(cycles)\\
$\phi_{,j}^{s} (t_{0})$ & Initial satellite phase offset & (cycles)\\
$\lambda_{j}$ & wavelength at frequency $j$ & (metres) \\
$c \cdot d^{s}_{,j}$ & Code satellite instrument delay &(metres)\\
$c \cdot d_{r,j}$ & Code receiver instrument delay &(metres)\\
$c \cdot \delta^{s}_{,j}$ & Phase satellite instrument delay &(metres)\\
$c \cdot \delta_{r,j}$ & Phase receiver instrument delay &(metres)\\
$\Phi_{r,21}^{s}$ & Frequency-difference of phase observables & (metres)\\
$P_{r,21}^{s}$ & Frequency-difference of code observables & (metres)\\
$\iota_{r,21}^{s}$ & Frequency-difference of ionospheric delay  & (metres)\\
$\mathrm{C}_{r}^{s}$ & Constant term over each satellite arc & (metres)\\
$c \cdot d_{r,21} $ & Receiver DCB & (metres)\\
$c \cdot d^{s}_{,21}$ & Satellite DCB & (metres)\\
$STEC$ & Slant Total Electron Content & (TECU)\\
$z$ & Zenith angle of satellite & (radians)\\
$z'$ & Zenith angle at IPP & (radians)\\
$R_{e}$ & Radius of Earth & (metres)\\
$H_{ion}$ & Height of the ionospheric layer & (metres)\\
$F^{s}$ & Mapping function & \\
$\mu_{21}=\mu_{1} - \mu_{2}$ & Frequency-difference of $\mu_{j}$ & \\
$VTEC$ & Vertical Total Electron Content & (TECU)\\
$\varphi_{m}$ & Geomagnetic latitude at IPP & (radians)\\
$s$ & Sun-fixed longitude at IPP & (radians)\\
$\lambda_{m}$ & Geomagnetic longitude at IPP & (radians)\\
$VTEC_{0}$ & $VTEC$ at receiver location & (TECU)\\
$\varphi_{m_{0}}$ & Latitude at receiver location & (radians)\\
$s_{0}$ & Sun-fixed longitude at receiver location & (radians)\\
$f'$ & First order derivative & \\
$f''$ & Second order derivative & \\
$\widetilde{DCB^{s}} $ & Receiver DCBs lumped with satellite DCBs($= c \cdot(d_{r,21} - d^{s}_{,21})$)& (metres)\\
\end{supertabular}
\label{tab:paradef}
\end{center}
\end{table*}

\noindent The parameters used in sections \ref{sec:GPSmodel} and \ref{sec:ionomodel} are listed in Table \ref{tab:paradef}. Here subscripts $s,r$, and $j$ indicate satellite, receiver and GPS frequency number, respectively. \\

$M_{r,j}^{s}$ are the non-integer ambiguities on the phase observables which contain the unknown integer ambiguities, $N_{r,j}^{s}$, and the non-integer initial phase offsets for the receiver ($\phi_{r,j} (t_{0})$) and satellite ($\phi_{,j}^{s} (t_{0})$), i.e.,

\begin{equation}
M_{r,j}^{s} = N_{r,j}^{s} + \phi_{r,j} (t_{0}) - \phi_{,j}^{s} (t_{0}).
\end{equation}

\noindent The phase ambiguities remain constant for any given receiver, frequency, and continuous satellite arc unless there is a loss of signal lock.\\

Since the effect of the ionospheric delay is a function of frequency, a frequency difference of phase and code observables can be formed which retains the ionospheric delay while the geometry-related parameters are eliminated. Along with the ionospheric delay ($\iota_{r}^{s}$), the phase and code instrumental delays ($\delta_{r,j}, \delta^{s}_{,j}, \;\;  d_{r,j},d^{s}_{,j}$) and phase ambiguities ($\lambda_{j} M_{r,j}^{s}$) remain. The frequency-difference phase and code observation equations for dual frequency GPS observables, formed using equations \eqref{eq:gnssobs1} and \eqref{eq:gnssobs2}, are given as follows:

\begin{eqnarray} \label{eq:sdphase1}
\displaystyle E(\Phi_{r,21}^{s}) = \displaystyle \Phi_{r,1}^{s}-\Phi_{r,2}^{s} &=&  -  \iota_{r,21}^{s} + \mathrm{C}_{r}^{s}, \\ \nonumber 
&&\\   \label{eq:sdphase2}
\displaystyle E(P_{r,21}^{s}) = \displaystyle P_{r,1}^{s}-P_{r,2}^{s} &=&  \iota_{r,21}^{s} + c \cdot(d_{r,21} - d^{s}_{,21}). 
\end{eqnarray}

Here, $d^{s}_{,21}$ and  $\ d_{r,21}$ are the inter-frequency code delays for the satellite and the receiver and the constant phase term $\mathrm{C}_{r}^{s}$ is given as follows: \\

\begin{equation}
\mathrm{C}_{r}^{s} =\left[ c \cdot ( \delta_{r,21} -\delta^{s}_{,21} ) + \left(\lambda_{1} M_{r,1}^{s} -\lambda_{2} M_{r,2}^{s}\right) \right].
\end{equation}

The slant ionospheric delay, $\iota_{r,j}^{s}$, is related to the Slant Total Electron Content ($STEC$) as follows \citep{Hof93}:

\begin{eqnarray}\label{eq:stecvtec}
\displaystyle \iota_{r,j}^{s} &=& \displaystyle \frac{40.3 \;\; STEC}{f^{2}_{j}}.
\end{eqnarray}

\begin{figure}[h]
 \centering
\includegraphics[scale=0.32]{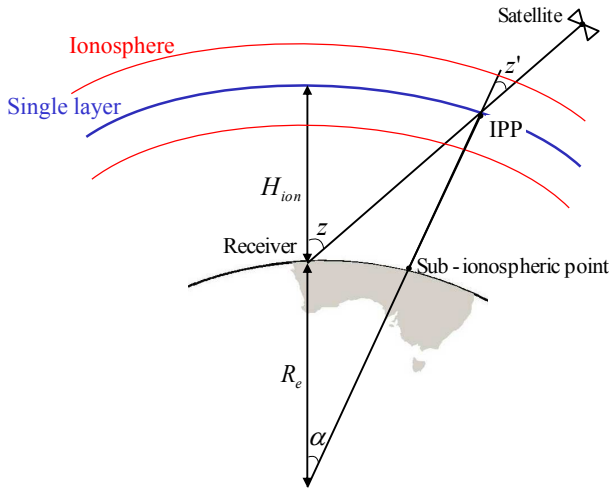}
     \caption{Ionosphere single layer model representation.
        \label{fig:slm}}
\end{figure}

As shown in Figure \ref{fig:slm}, the $STEC$ can be mapped to the Vertical Total Electron Content, $VTEC$, using the obliquity factor $F^{s}$, also known as the ionosphere mapping function. $F^{s}$ is a function of zenith angle at the Ionosphere Pierce Point (IPP), $z'$, given as follows:\\

\begin{eqnarray}\label{eq:Fs}
\left. \begin{array}{lll}
\displaystyle STEC  &=& \displaystyle VTEC \cdot F^{s} \\
\\
\displaystyle F^{s} &=& \displaystyle \frac{1}{\cos(z')} =\frac{1} {\sqrt{1-\sin^{2} z'}} \\
\\
 \displaystyle \sin z' &=& \displaystyle \frac{R_{e}}{R_{e}+H_{ion}} \sin(z)
\end{array}\right\}
\end{eqnarray}

\noindent where $z'$ is the zenith angle at the IPP, $R_{e}$ is the mean radius of the Earth, considered to be 6371 km and assuming a spherical Earth, $H_{ion}$ is the height at the sub-ionospheric  point, assumed to be 450 km, and $z$ is the zenith angle of the satellite as seen by the receiver. The geometry of the model is illustrated in Figure \ref{fig:slm}. This study aims to compare and analyse ionosphere gradients, the height, $H_{ion}$, of 450 km was chosen in order to compare the $VTEC$ with CODE analysis center published values.\\

We can now map the GPS observables to the $VTEC$ as follows

\begin{eqnarray}\label{eq:sdphase3}
\left. \begin{array}{lll}
\displaystyle E\left(\frac{\Phi_{r,21}^{s}}{40.3 \ \mu_{21}}\right)  &=&  \displaystyle -F^{s} \ VTEC +  \displaystyle \frac{\mathrm{C}_{r}^{s}}{40.3 \ \mu_{21}}  \\
&&\\
\displaystyle E\left(\frac{P_{r,21}^{s}}{40.3 \ \mu_{21}}\right)  &=&  \displaystyle F^{s} \ VTEC + \displaystyle \frac{c \cdot(d_{r,21} - d^{s}_{,21})}{40.3 \ \mu_{21}} 
\end{array}\right\}
\end{eqnarray}

\noindent Here $c \cdot(d_{r,21})$ and $ c \cdot(d^{s}_{,21})$ constitute terms known as Differential Code Bias (DCB) for the receiver and the satellite, respectively, where

\begin{eqnarray}
\left. \begin{array}{lll}
\mu_{21} &=& \mu_{1} - \mu_{2}\\
&&\\
& \textrm{and}&\\
&&\\
\mu_{1} =  \displaystyle \frac{1}{f_{1}^{2}} & & \mu_{2} =  \displaystyle \frac{1}{f_{2}^{2}}
\end{array}\right\}
\end{eqnarray}

\subsection{GPS ionospheric modelling}
\label{sec:ionomodel}
The $VTEC$ can be modelled by assuming that the ionosphere is concentrated at a single layer at height $H_{ion}$ as illustrated in Figure \ref{fig:slm} \citep{Sch99, Wan14}. The intersection of the GPS receiver-satellite line-of-sight with the ionospheric layer is called as IPP. The slant TEC is mapped to the vertical TEC by an obliquity factor (equation \eqref{eq:Fs}), which is a function of the zenith angle at the IPP, $z'$. $VTEC$ is modelled as a function of geomagnetic latitude, $\varphi_{m}$, and Sun fixed longitude, $s$, using the following polynomial function \citep{Wan14}

\begin{eqnarray}\label{eq:polmod2} \nonumber
VTEC (\varphi_{m},s) &=& VTEC_{0} +  (\varphi_{m} - \varphi_{m_{0}})f' \varphi + (s - s_{0})f' s +\\ \nonumber
\\ \nonumber
& & + (\varphi_{m} - \varphi_{m_{0}})^{2}f'' \varphi_{m} \varphi_{m}  + (s - s_{0})^{2}f''  s s \\\nonumber
\\
&& + (\varphi_{m} - \varphi_{m_{0}}) (s - s_{0}) f'' \varphi_{m} s .
\end{eqnarray}

The Sun fixed longitude, $s$, is related to the local solar time ($LT$) as $s = \lambda_{m} + LT - \pi$, where $\lambda_{m}$ is the geomagnetic longitude at IPP, $LT$ is in radians, $VTEC_{0}$ is the $VTEC$ at the receiver location and $ f' s, \ f' \varphi_{m}, \ f'' s s, \  f'' \varphi_{m} \varphi_{m}, \  f'' \varphi_{m} s $ are the first and second order derivatives of $VTEC$ along the Sun fixed longitude and latitude, respectively.\\

The single layer ionospheric model is a computation efficient model to estimate local ionospheric gradients at the zenith. However, the ionospheric features, for example those along the magnetic field lines, cannot be resolved using this model.\\

\subsection{VTEC determination through Kalman filtering}
\label{sec:kf}
The GPS model to estimate unknowns can be formed from equation \eqref{eq:sdphase2}. In our method, the DCBs for receiver and satellite are estimated as a single parameter, $\widetilde{DCB^{s}} = c \cdot(d_{r,21} - d^{s}_{,21})$, since the model presented in equation \eqref{eq:sdphase3} is rank deficient. The parameters, namely the receiver and the satellite DCBs, cannot be separated from each other, hence cannot be independently estimated. A minimum set of parameters, known as the $\mathcal{S}$-basis, are chosen which can be lumped with the remaining parameters in order to overcome the rank deficiency in the underlying model \citep{Teu84}. For $m$ satellites seen by receiver $r$, a weighted least square model is formed using equations \eqref{eq:sdphase2} and \eqref{eq:polmod2}, given as follows

\begin{eqnarray}\label{eq:lsq}
\begin{array}{lll}
\displaystyle E({y_{i}}) = {A_{i}} \ {x_{i}} \\
\\
\displaystyle D(y_{i}) = Q_{yi}\\
\\
\displaystyle x_{i} = \scriptstyle \left[ \mathrm{C}_{r}^{s} \ \widetilde{DCB^{s}} \ VTEC_{0} \ f' \varphi_{m}  \ f' s \ f'' \varphi_{m} \varphi_{m} \ f'' s s \  f'' \varphi_{m} s   \right]^{T}
\end{array}
\end{eqnarray}

\noindent where $E(\cdot)$ and $D(\cdot)$ denote the expectation and dispersion operators, $y_{i}$ denotes a vector of observables of size $[2m \times 1]$ at time stamp $i$, $A_{i}$ is the design matrix of size $[2m \times (2m+6)]$, the unknowns given by $x_{i}$ are of size $[(2m+6) \times 1]$, and $Q_{yi}$ is the stochastic model for observables in $y_{i}$ given as 

\begin{equation}\label{eq:qy}
Q_{yi} = \left(\frac{1}{40.3 \ \mu_{21}}\right)^{2} \frac{\sigma_{0}^{2}} {\sin^{2} (el)},
\end{equation}

\noindent $el$ is the elevation angle of the satellite, and $\sigma_{0}$ is the measurement noise of the observables.\\

\begin{figure}
 \centering
\includegraphics[scale=0.5]{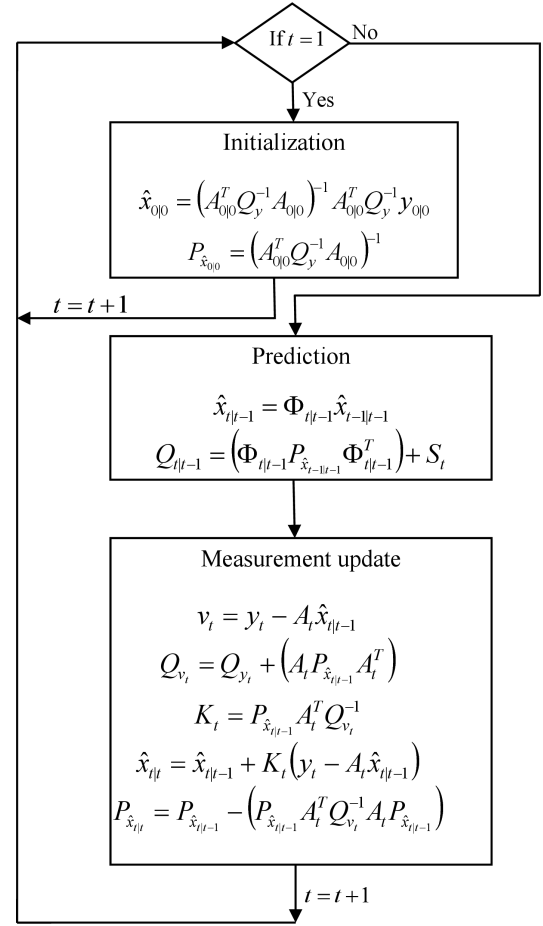}
     \caption{Flow chart describing the implementation of a Kalman Filter.
        \label{fig:kf}}
\end{figure} 

To solve the above model, a cut-off for the satellite zenith angle of $70^{\circ}$ was chosen. The model given in equation \eqref{eq:lsq} cannot be solved in a single epoch. An ionosphere refreshing interval of 20 epochs (1 epoch = 30 seconds) is chosen over which the ionosphere is assumed to remain constant. The phase bias term $\mathrm{C}_{r}^{s}$ constitutes an integer phase ambiguity and non-integer receiver and satellite phase bias terms. The integer phase ambiguities remain constant for a single continuous satellite arc unless there is a loss of signal lock, hence $\mathrm{C}_{r}^{s}$ is assumed to remain constant for a continuous satellite arc. The GPS observables were free of multipath and cycle slips. The code bias terms, $\widetilde{DCB^{s}}$, are assumed to remain constant over a period of 24 hours for any satellite. \\

With the above considerations, the unknowns given in equation \eqref{eq:lsq} are estimated using the Kalman Filter approach \citep{Kal60,Kai81,Gre11}. In this work, Kalman filter is used in data assimilation mode. A Kalman filter can be described as a three step procedure, with initialisation, prediction, and measurement update executed at different time steps $t$, $t=1,\cdots, t_{max}$. The implementation of the Kalman filter is shown in Figure \ref{fig:kf}, where, $\hat{x}_{0|0}, \mathrm{A}_{0|0}^{T}, Q_{y}$, and $y_{0|0}$ indicate the unknowns, the design matrix, the Variance Covariance (VC) matrix of the measurements, and the measurements, respectively. $\Phi_{t|t-1}$ is the transition matrix which relates the unknowns at the current $\hat{x}_{t|t-1}$ and previous $\hat{x}_{t-1|t-1}$ time step with a system noise given by the VC matrix. $ S_{t}$, $v_{t}$ are the predicted residuals and $Q_{v_{t}}$ its VC matrix. $K_{t}$ is the Kalman gain which is used to compute the  measurement update given by $\hat{x}_{t|t}$ and its VC matrix $P_{\hat{x}_{t|t}}$. The parameters used in Kalman filter are presented in Table \ref{tab:kfpara}.

\begin{table*}
\begin{center}
\caption{Kalman Filter parameters and definition}
\begin{tabular*}{\textwidth}{p{3cm} p{9cm} p{2cm}}
\hline \hline
Parameter & Definition & Units\\
\hline
$el$ & Elevation angle of satellite & (radians)\\
$t$ & Time step & \\
$t_{max}$ & Maximum number of time steps & \\
$\hat{x}_{0|0}$ & Initial estimate of unknowns & \\
$P_{\hat{x}_{0|0}}$ & Variance matrix of $\hat{x}_{0|0}$ & \\
$\mathrm{A}_{0|0}^{T}$ & Design matrix & \\
$Q_{y}$ & Stochastic model of observations & \\
$y_{0|0}$ & Observation vector & \\
$\Phi_{t|t-1}$ & Transition matrix & \\
$\hat{x}_{t|t-1}$ & Predicted unknowns & \\
$\hat{x}_{t-1|t-1}$ & Unknowns from previous time step & \\
$S_{t}$ & Variance matrix of the system noise & \\
$v_{t}$ & Predicted residuals & \\
$Q_{v_{t}}$ & Variance matrix of predicted residual& \\
$K_{t}$ & Kalman gain matrix & \\
$\hat{x}_{t|t}$ & Updated unknowns & \\
$P_{\hat{x}_{t|t}}$ & Variance matrix of $\hat{x}_{t|t}$& \\
\hline\hline
\end{tabular*}
\label{tab:kfpara}
\end{center}
\end{table*}

\subsection{GPS DATA PREPARATION}
\begin{figure}[h]
 \centering
\includegraphics[scale=0.8]{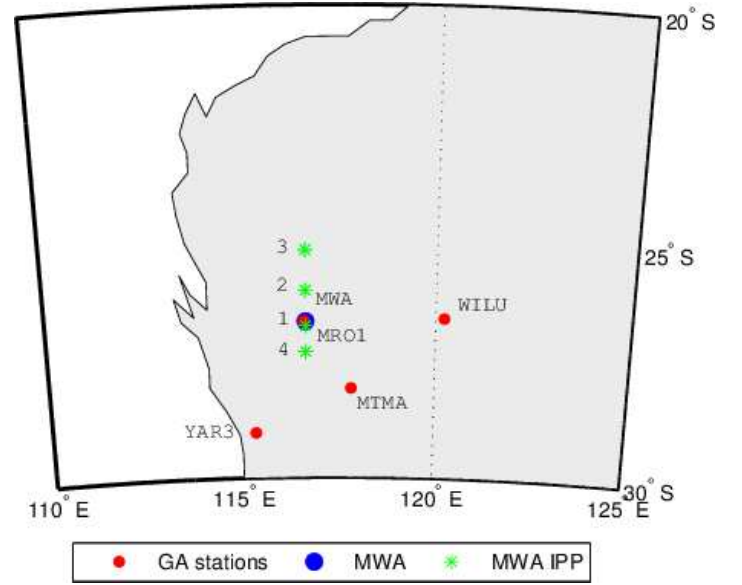}
     \caption{Selected GPS station locations from Geoscience Australia's network (red), MWA location (blue) and MWA IPP (green) for the four MWA observation nights (DOY 062, 063, 065 and 075 marked by 1 to 4).
        \label{fig:WAGPSnet}}
\end{figure}

The data from the three GPS/GNSS stations nearest to the Murchison Radioastronomy Observatory (MRO) were used for this analysis, namely the Geoscience Australia (GA) stations MRO1 (Murchison), MTMA (Mount Magnet), YAR3 (Yarragadee), and WILU (Wiluna) were chosen (Figure \ref{fig:WAGPSnet}). A description of the selected GPS/GNSS stations is given in Table \ref{tab:descGPSnet}. The data for the selected GA GNSS network were downloaded from the GA archive\footnote{\url{ftp://ftp.ga.gov.au/geodesy-outgoing/gnss/data/daily/}$yyyy/yyddd/xxxxddd0.yyd.Z$ The abbreviations $yyyy$ and $yy$ are the four and two digit year, $ddd$ is the DOY, $xxxx$ represents the four character GPS station id, $d$ stands for Hatanaka compressed \citep{Hat08} Receiver INdependent EXchange (RINEX) format \citep{Gur07}, and $Z$ indicates compressed/zipped file. The Hatanaka compressed files can be decompressed by the software available at \url{ftp://terras.gsi.go.jp/software}} for 3$^{rd}$, 4$^{th}$, 6$^{th}$ and 16$^{th}$ March 2014 corresponding to Day of Year (DOY) 062, 063, 065 and 075, respectively. The four days chosen for this analysis were the first four nights of GLEAM observations for which suitable MWA data was available. Also, by choosing data from the year 2014, data from the recently active GPS receiver MRO1 could be included. Only YAR3 is included in CODE analysis, illustrating why our analysis is required to establish dense GPS networks near the MWA. Table \ref{tab:indices} presents the summary of ionospheric weather on the selected four days, the parameters, F10.7 solar flux, and Planetary Kp indices are presented. The data presented in Table \ref{tab:indices} indicate quiet ionospheric conditions, which are ideal for testing the methods described in this paper.\\

\begin{table*}
\caption{Description of the selected GA network GPS/GNSS stations and the MWA} 
\begin{center}
\begin{tabular*}{\textwidth}{@{}c\x c\x c\x c\x c\x c\x c@{}}
\hline \hline
 Station & Receiver type & Antenna type & Observables & Location & Observing sessions\\
         &               &              & used        & (degrees)&  (Year,Day of Year) \\
\hline
 MRO1 & TRIMBLE NETR9 & TRM59800.00 & $L1$, $L2$, $C1$, $P2$ & 26.70$^{\circ}$ S 116.37$^{\circ}$ E & 2014, 062, 063, 065, 075\\
 MTMA & LEICA GRX1200+GNSS & LEIAR25.R3 & $L1$, $L2$, $C1$, $P2$ & 28.11$^{\circ}$ S 117.84$^{\circ}$ E & 2014, 062, 063, 065, 075\\
 YAR3 & LEICA GRX1200GGPRO & LEIAR25   & $L1$, $L2$, $C1$, $P2$ &29.04$^{\circ}$ S 115.34$^{\circ}$ E & 2014, 062$^a$, 063, 065, 075\\
 WILU & LEICA GRX1200+GNSS & LEIAR25.R3 & $L1$, $L2$, $C1$, $P2$ &26.62$^{\circ}$ S 120.21$^{\circ}$ E & 2014, 062, 063, 065, 075\\
 MWA & - & - & - & 26.70$^{\circ}$ S 116.67$^{\circ}$ E & 2014, 062, 063, 065, 075 \\
\hline \hline
\end{tabular*}\label{tab:descGPSnet}
\end{center}
\tabnote{$^a$ Partial data available, from 00:00:00 UTC to 18:07:00 UTC}
\end{table*}

\begin{table*}
\caption{Daily solar and geomagnetic indices for the selected MWA observation days} 
\begin{center}
\begin{tabular*}{16.5cm}{@{} c c | p{0.8cm} p{0.8cm} p{0.8cm} p{0.8cm} p{0.8cm} p{0.8cm} p{0.8cm} p{0.8cm} c@{}}
\hline \hline
 Year, DOY & Solar Flux at 10.7 cm (F10.7$^a$) & \multicolumn{8}{c}{Planetary K index (Kp$^b$) } \\
           &    $10^{-22} Wm^{-2}Hz^{-1}$         &  \multicolumn{8}{c} {3 hourly, from 00 to 24 UTC, ranging from 0-9 (low-high) } \\
\hline
 2014, 062 & 161  & 3 & 2 & 2 & 2 & 2 & 1 & 1 & 1  \\
 2014, 063 & 158  & 2 & 1 & 0 & 2 & 3 & 2 & 2 & 3  \\
 2014, 065 & 149  & 1 & 1 & 1 & 2 & 2 & 2 & 3 & 1  \\
 2014, 075 & 136  & 1 & 1 & 1 & 0 & 0 & 0 & 1 & 0  \\
\hline \hline
\end{tabular*}\label{tab:indices}
\end{center}
\tabnote{$^a$ National Geophysical Data Center (NGDC), National Oceanic and Atmospheric Administration (NOAA) {\url{ftp://ftp.ngdc.noaa.gov/STP/GEOMAGNETIC_DATA/INDICES/KP_AP/2014}}}
\tabnote{$^b$ Space Weather Prediction Center (SWPC), NOAA, BOULDER, USA {\url{ftp://ftp.swpc.noaa.gov/pub/warehouse/2014}}}
\end{table*}

\section{CODE IONEX maps}
\label{sec:codemaps}
The daily ionosphere maps from the CODE are based on a global network of $\sim$200 GPS/GNSS stations. The line of sight GPS ionospheric delay is mapped to $VTEC$ using a Modified Single Layer Model (MSLM) mapping function approximating the Jet Propulsion Laboratory (JPL) Extended Slab Model (ESM) \citep[for ESM mapping function see,][]{Cos92}. The MSLM\footnote{\url{www.aiub.unibe.ch/download/users/schaer/igsiono/doc/mslm.pdf}} is given as follows:

\begin{equation}\label{eq:mslm}
\displaystyle F_{MSLM} = \frac{1}{\cos z'} ; \;\;\; \displaystyle \sin z' =  \left(\frac{R} {R + H} \cdot \sin (\alpha z) \right).
\end{equation}

\noindent where $H$ = 506.7 km and $\alpha$ = 0.9782. \\

The $VTEC$ is modelled using spherical harmonic coefficients of order and degree 15 in the solar magnetic reference frame as snapshots with an ionosphere refreshing interval of 2 hours \citep{Sch96}. The spatial resolution of CODE maps is $5^{\circ}$/$2.5^{\circ}$ in longitude/latitude, respectively. The spherical harmonic model used by CODE to interpret the global ionosphere is described in \citet{Sch96}. \\

 \begin{figure*}
 \centering
{\includegraphics[scale=0.9]{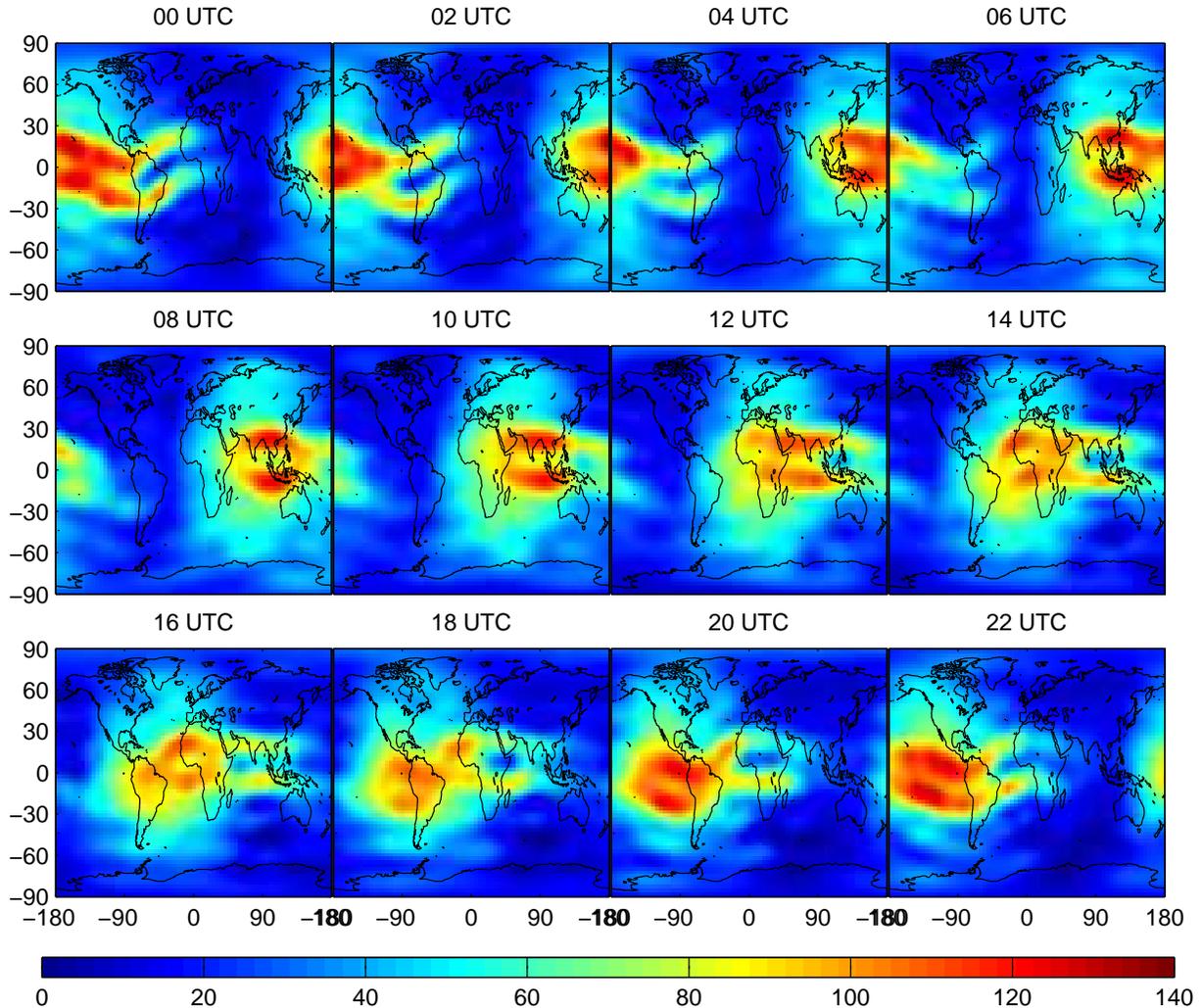}}
     \caption{Global TEC (TECU) from CODE IONEX maps for DOY 062, year 2014.
        \label{fig:ionex}}
\end{figure*}

\begin{figure}[h]
 \centering
{\includegraphics[scale=0.5]{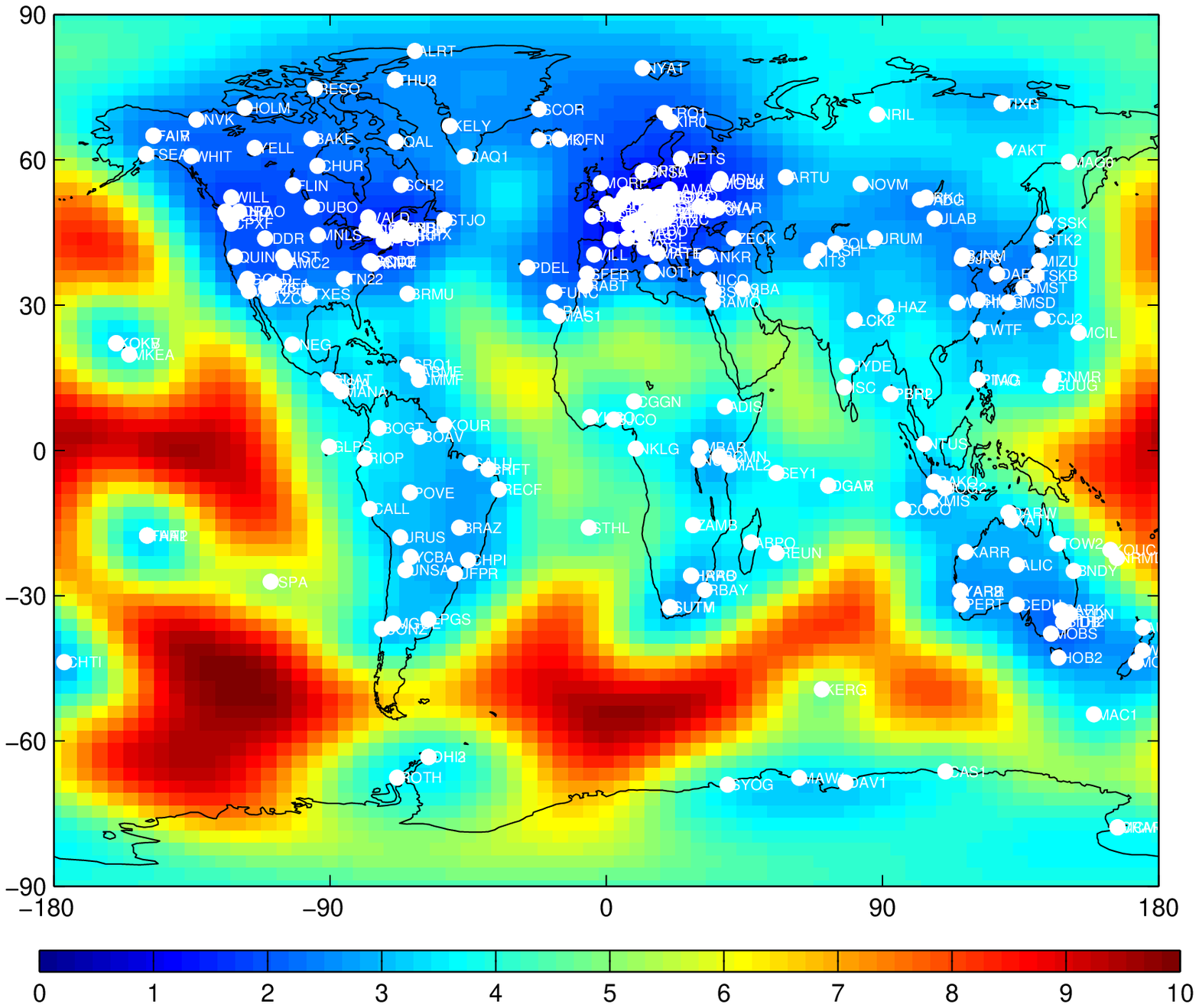}}
     \caption{Average RMS (1-$\sigma$ uncertainties) in TECU, of CODE IONEX maps for DOY 062, year 2014, marked in white are the GPS/GNSS stations considered for the solution.
        \label{fig:avgionexrms}}
\end{figure} 

CODE maps are available as daily final solutions in CODE's online archive\footnote{\url{ftp://ftp.unibe.ch/aiub/CODE/}$yyyy/CODGddd.yyI$, where $yyyy$ and $yy$ are the four and two digit year, $ddd$ is the DOY. The ionosphere maps are  exchanged in IONosphere Map EXchange Format (IONEX), see \cite{Sch98} for detailed description of IONEX data format and its interpolation.}. Figure \ref{fig:ionex} presents the 2 hour snapshots for the day 03-03-2014 (DOY 062), which is one day of interest for the MWA observations. The average CODE $VTEC$ RMS is shown in Figure \ref{fig:avgionexrms}. The RMS of the $VTEC$ fit is higher over oceans and regions with sparse GPS/GNSS receivers coverage, due to limited data points available for the fit. \\

\section{MEASUREMENT OF THE IONOSPHERE USING MWA OBSERVATIONS }
\label{sec:mwaiono}
\subsection{Observations}
\begin{figure}[h]
 \centering
\includegraphics[scale=0.7]{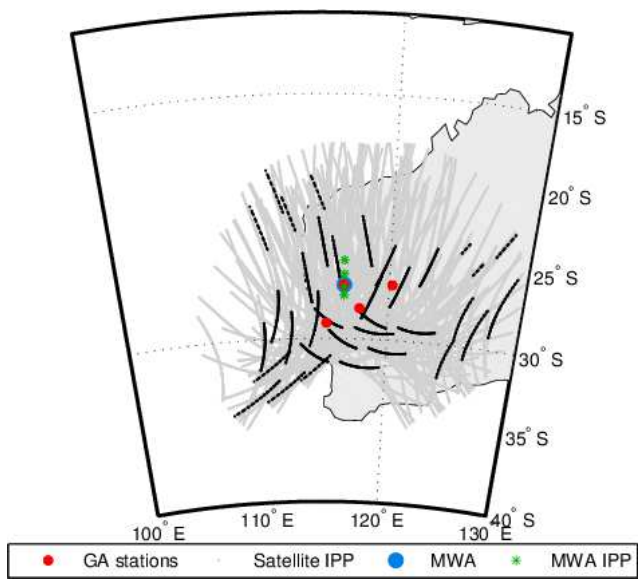}
     \caption{GA station locations (red) with satellite IPPs in earth fixed reference frame over a period of 24 hours (grey) and for 1 hour during MWA observations (black). MWA is marked in blue, whereas the IPP of MWA is shown in green. 
        \label{fig:mwaipp}}
\end{figure}

For comparison with our GPS modelling of the ionosphere, we used observations from the GLEAM survey (\citet{Way15}; Hurley-Walker et al. in prep.).
In this survey, the MWA observes in meridian drift-scan mode, where the telescope remains pointing at a single point on the meridian throughout the night.
Four nights from March 2014 were chosen, when the telescope was pointed close to the zenith.
The ionosphere pierce point corresponding to the pointing centre of the telescope is shown for each night in (\fig~\ref{fig:mwaipp}).

During GLEAM observations, the instrument cycles five frequency bands, centred on approximately (88\,MHz, 118\,MHz, 154\,MHz 185\,MHz and 215\,MHz), with a dwell time of 2 minutes on each band, each band having an instantaneous bandwith of 30.72\,MHz.
Each two-minute observation is then imaged with separate images being generated for 4 subbands (each having a bandwidth of 7.68\,MHz).
The position of a prepared list of bright sources was then determined for each of these images.

\subsection{Ionospheric Modelling}

By comparing the position of each source in all four subbands, it is possible to quantify the effect of the ionosphere, separating it from instrumental and calibration effects. 
By making an least-squares fit to all four points, the contribution of the ionosphere (the gradient) can be determined for each source.
The offset in position at each frequency is shown in \fig~\ref{fig:example_source_fit} for a single strong source.
It can be seen that the change in apparent position of the source depends precisely on  $\lambda^2$, exactly as would be expected from ionospheric refraction.

\begin{figure*}
\centering
	\includegraphics[width=\textwidth]{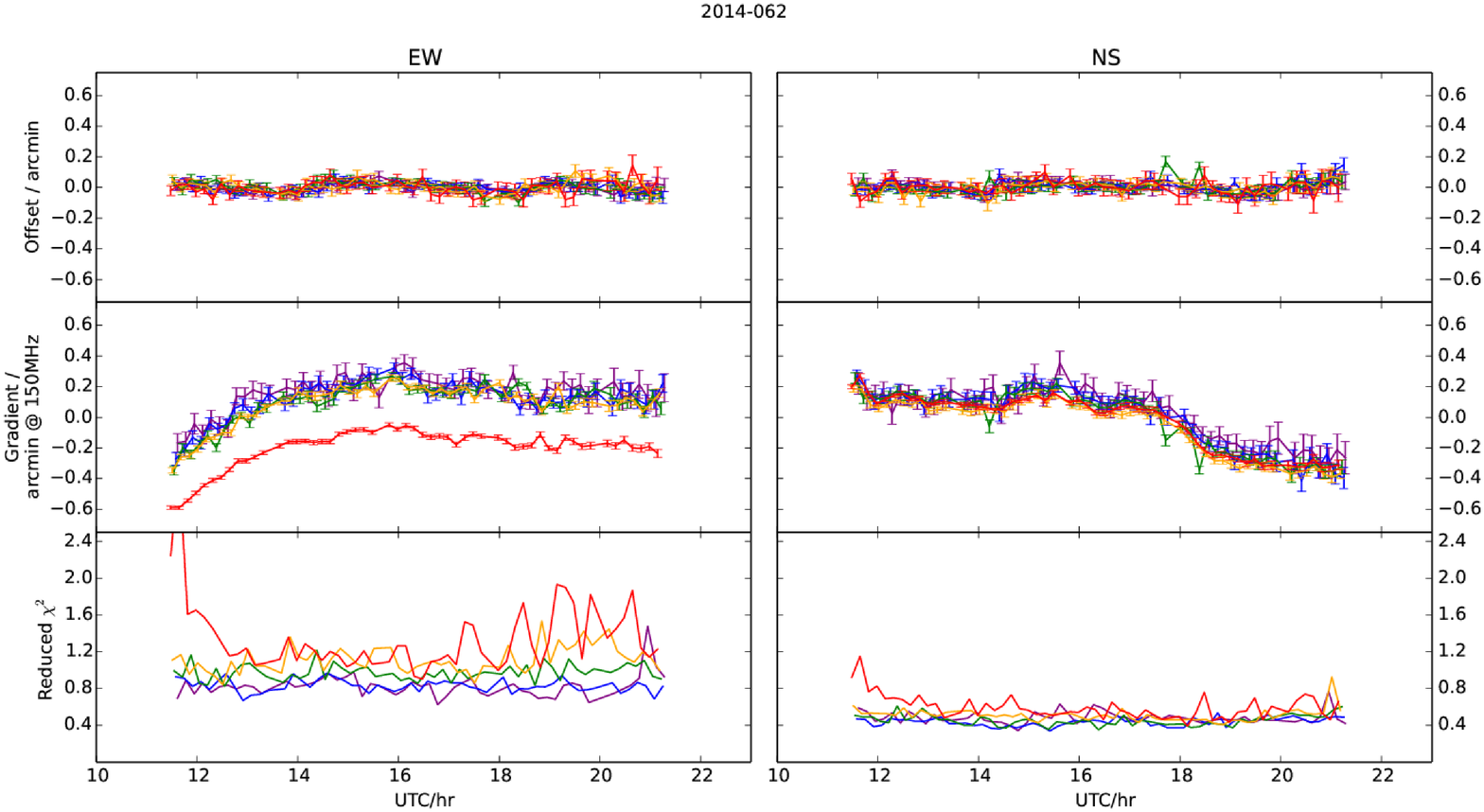}
	\caption{Offset (extrapolated to $\lambda=0$), gradient, and reduced $\chi^2$ of a fit of source position offset as a function of $\lambda^2$. Each point is for a single observation, all quantities are averaged over all (~100) sources detected in that observation. Left panels are for the East-West position offset (Right Ascension) right panels are for the North-South position offset (Declination). Red, Yellow, Green, Blue and Purple are for the 88\,MHz, 118\,MHz, 154\,MHz 185\,MHz and 215\,MHz bands respectively. Note that the gradients (in arcmin m$^1$) have been multiplied by 4, representing an offset at a wavelength of 2\,m (=150MHz).}
    \label{fig:mwa_ionosphere}
\end{figure*}

\begin{figure*}
\centering
        \includegraphics[width=\textwidth]{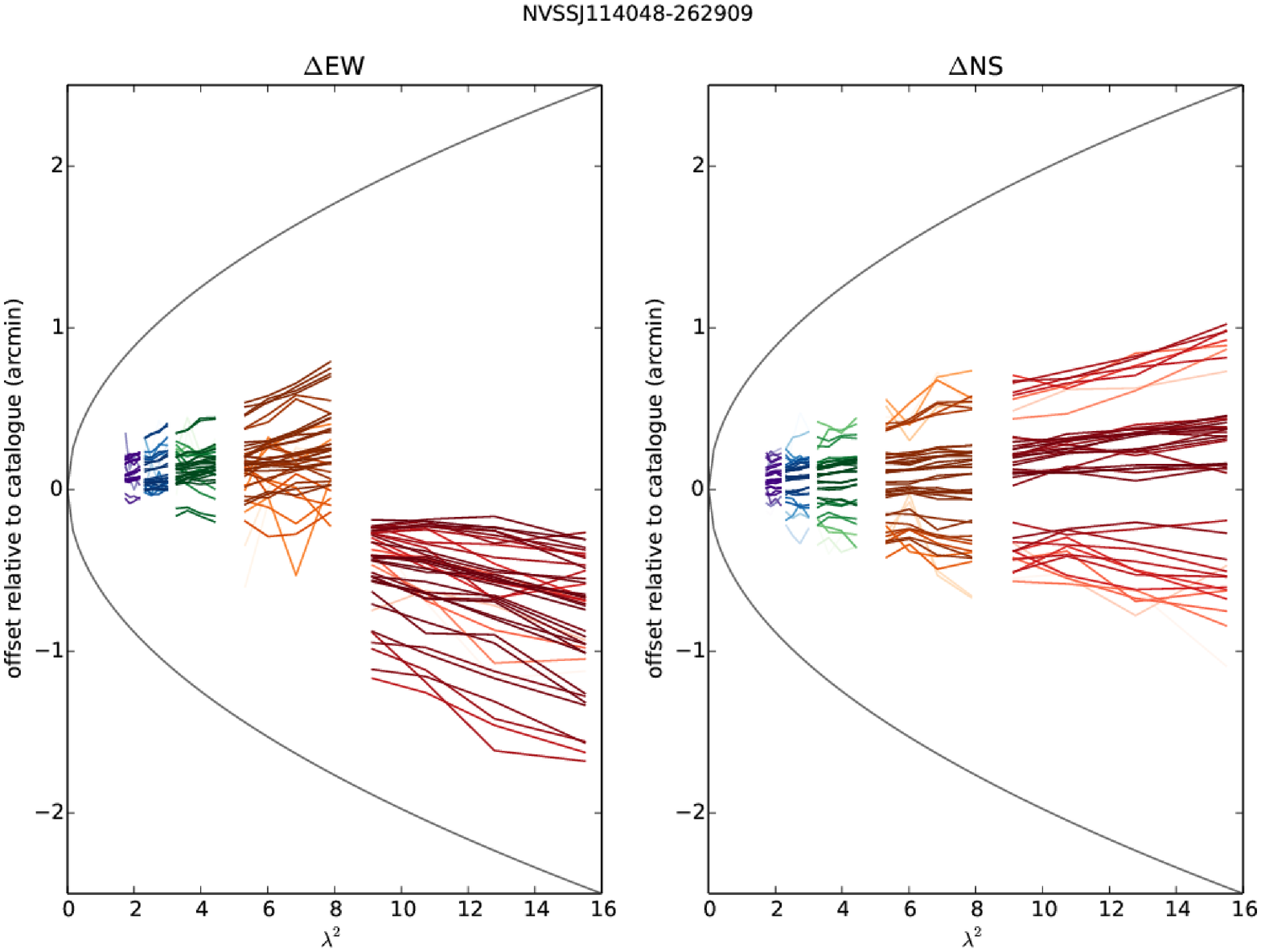}
		\caption{Source position offset against $\lambda^2$ (m$^2$) for a strong source. Each line represents a measurement of the source in each of four subbands in a single 2-minute observation. There are many lines since the source in multiple observations as it passes through the field of view.  Left panels are for the East-West position offset (Right Ascension) right panels are for the North-South position offset (Declination). Red, Yellow, Green, Blue and Purple are for the 88\,MHz, 118\,MHz, 154\,MHz 185\,MHz and 215\,MHz bands respectively. More significant detections are given a darker colour.}
    \label{fig:example_source_fit}
\end{figure*}

A comprehensive analysis of this dataset using MWA observations is underway (Morgan et al. in preparation). For this analysis, ionospheric gradient were estimated over all sources detected in each snapshot.
This is shown in \fig~\ref{fig:mwa_ionosphere} shows the gradient of this fit for each source, scaled by $\lambda^2$ for the highest frequency offset shown in the left panel.
The fact that the average reduced $\chi^2$ for each observation is $\sim$ 1, and the fact that $\lambda=0$ position of the sources remains at zero throughout the night, both serve to reinforce the hypothesis that the shift in sources is largely due to the ionosphere.

For simplicity, only the lowest-frequency data were used, since these observations are the most sensitive to ionospheric effects, and still yield sufficient time resolution.
This yields two time series (on in the NS direction and one in the EW direction) representing the average shift due to the ionosphere of $\sim$100 sources within a 35\arcdeg radius of the pointing centre.
In order that these shifts could be compared directly with GPS measurements, both measurements were scaled to a common reference frequency of 150\,MHz and the (angular) offset was converted to radians.

\section{RESULTS AND DISCUSSION}
\label{sec:results}
\subsection{Comparison of VTEC with CODE IONEX}
\begin{figure*}
 \centering
 \subcaptionbox{$VTEC$ at station MRO1\label{fig:vtecmro1062}}{\includegraphics[scale=0.38]{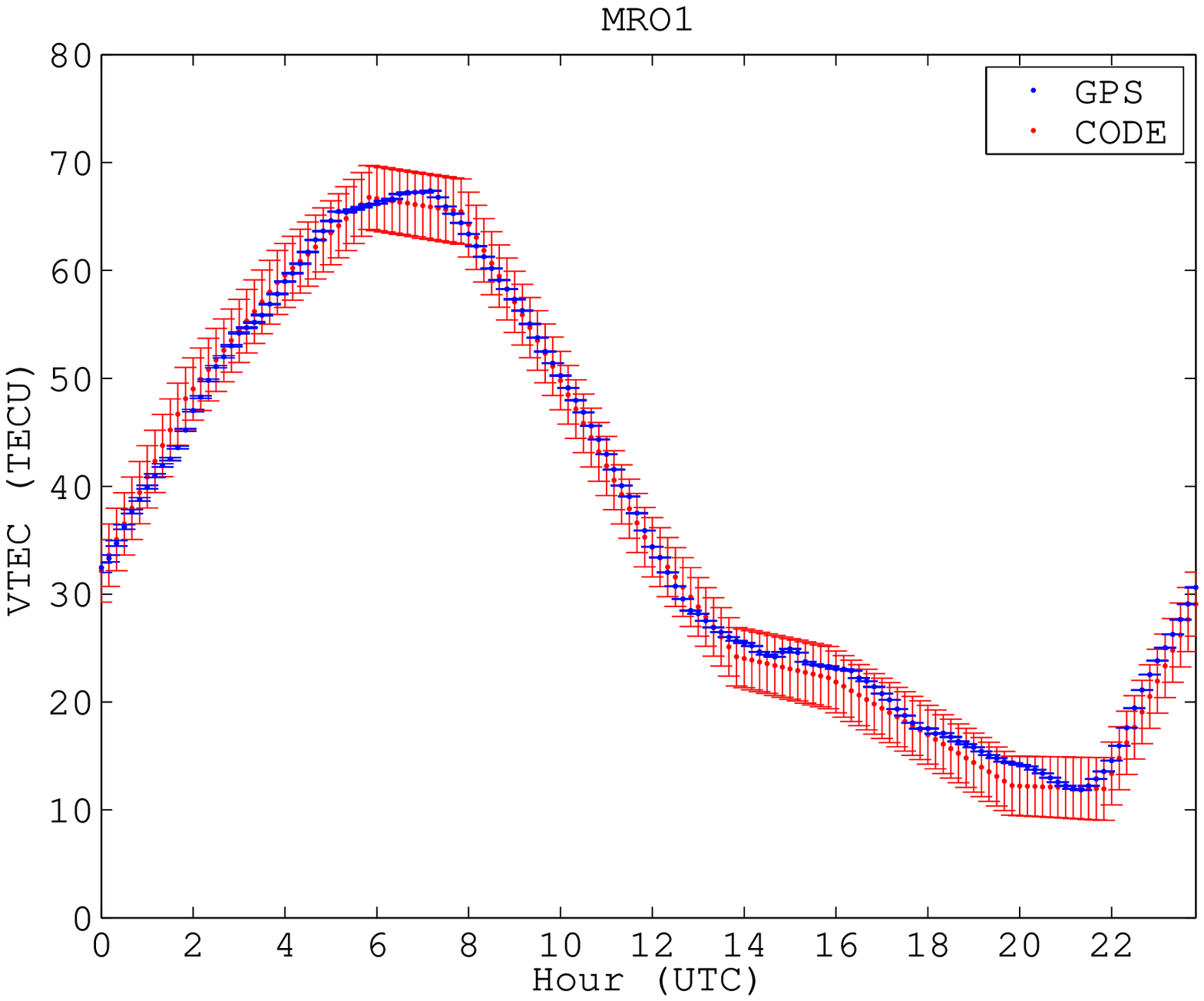}}
  \subcaptionbox{$VTEC$ at station MTMA\label{fig:vtecmtma062}}{\includegraphics[scale=0.38]{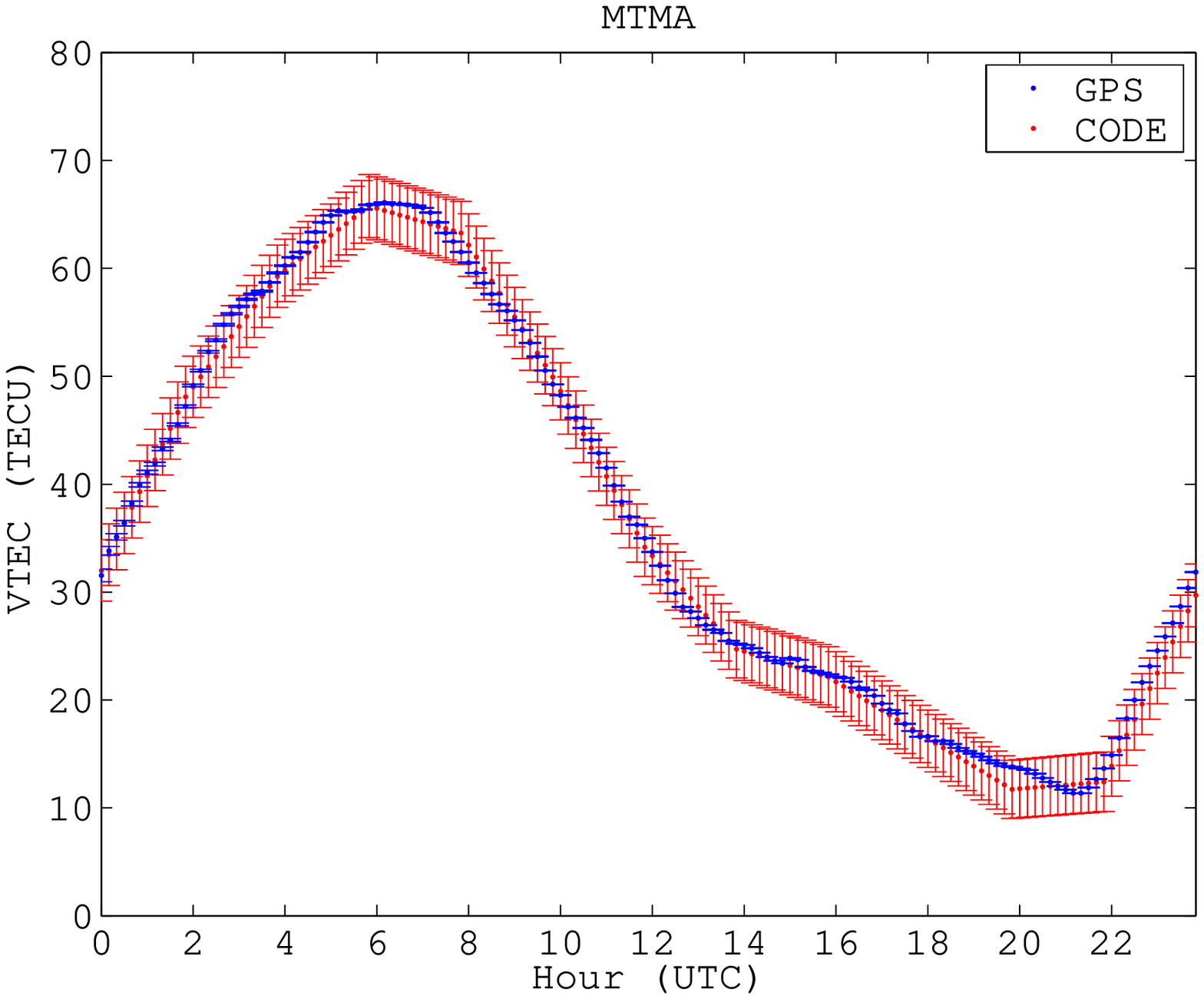}}
  \subcaptionbox{$VTEC$ at station YAR3\label{fig:vtecYAR3062}}{\includegraphics[scale=0.38]{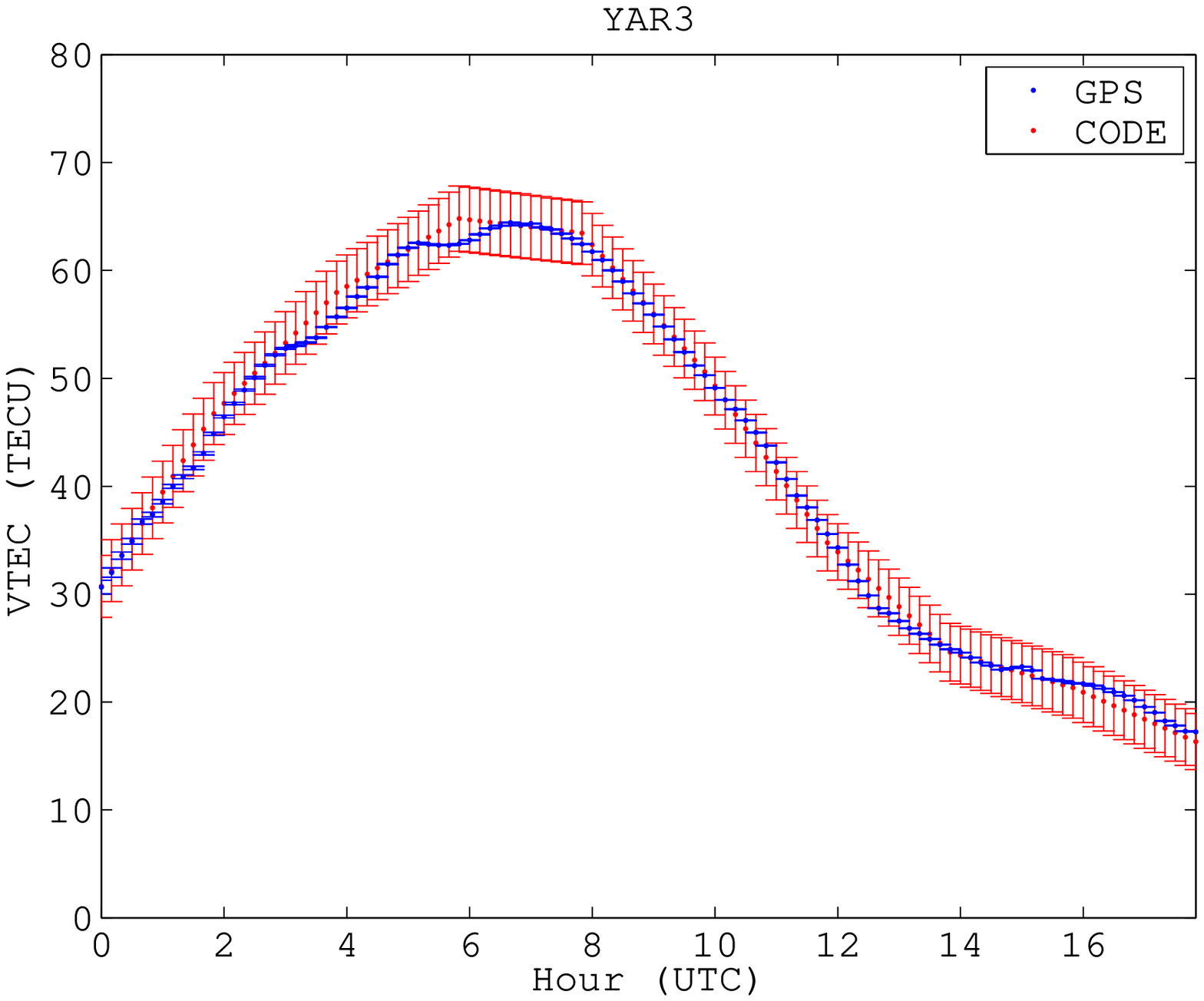}}
   \subcaptionbox{$VTEC$ at station WILU\label{fig:vtecwilu062}}{\includegraphics[scale=0.38]{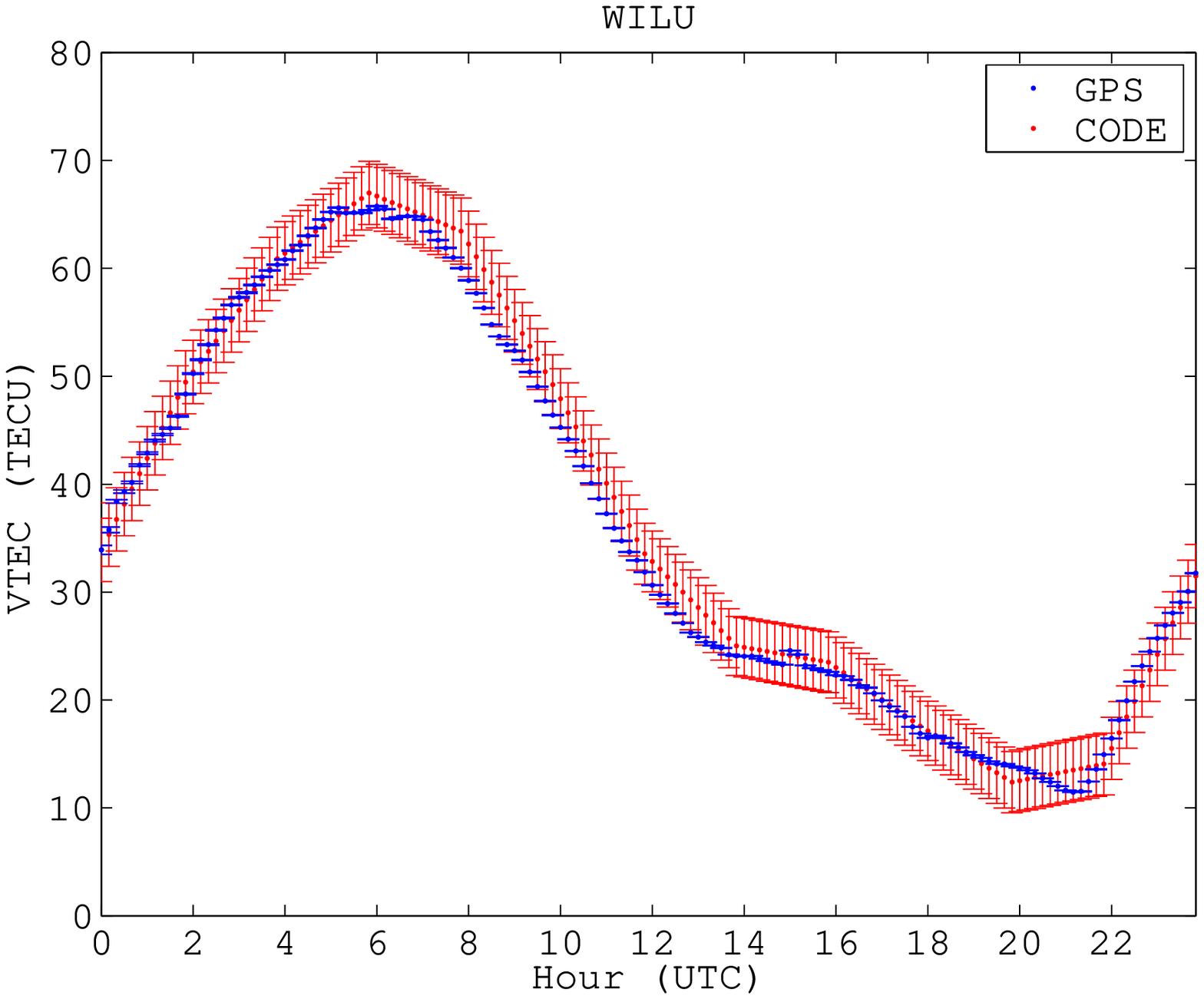}}
     \caption{$VTEC$ at stations MRO1, MTMA, YAR3, and WILU estimated using the method described in the text (blue curve) and CODE IONEX (red curve) on DOY 062, year 2014.
        \label{fig:vtec062}}
\end{figure*}

\begin{figure*}
 \centering
   \subcaptionbox{Differences in $VTEC$ on DOY 062\label{fig:dvtec062}}{\includegraphics[scale=0.38]{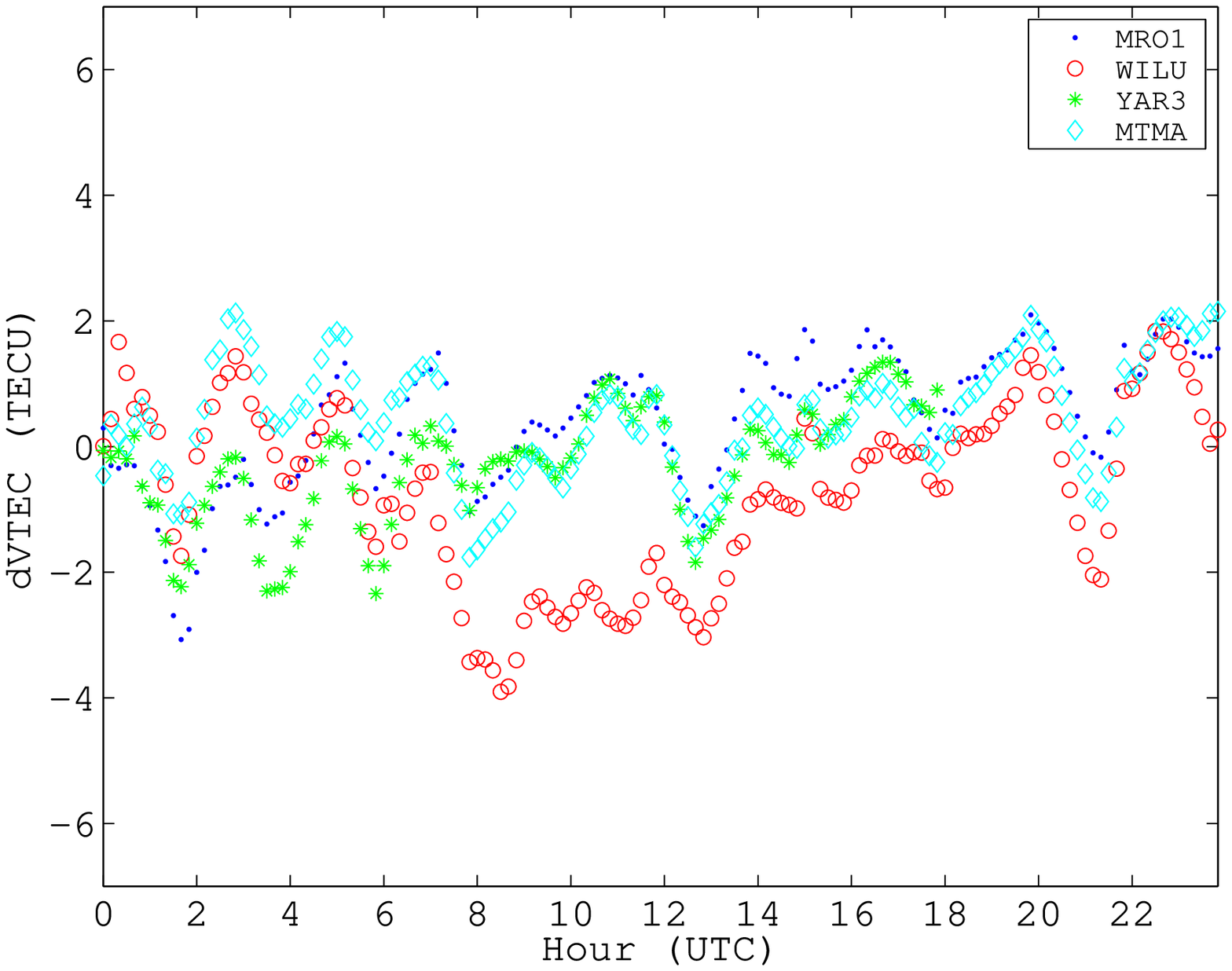}}
   \subcaptionbox{Differences in $VTEC$ on DOY 063\label{fig:dvtec063}}{\includegraphics[scale=0.38]{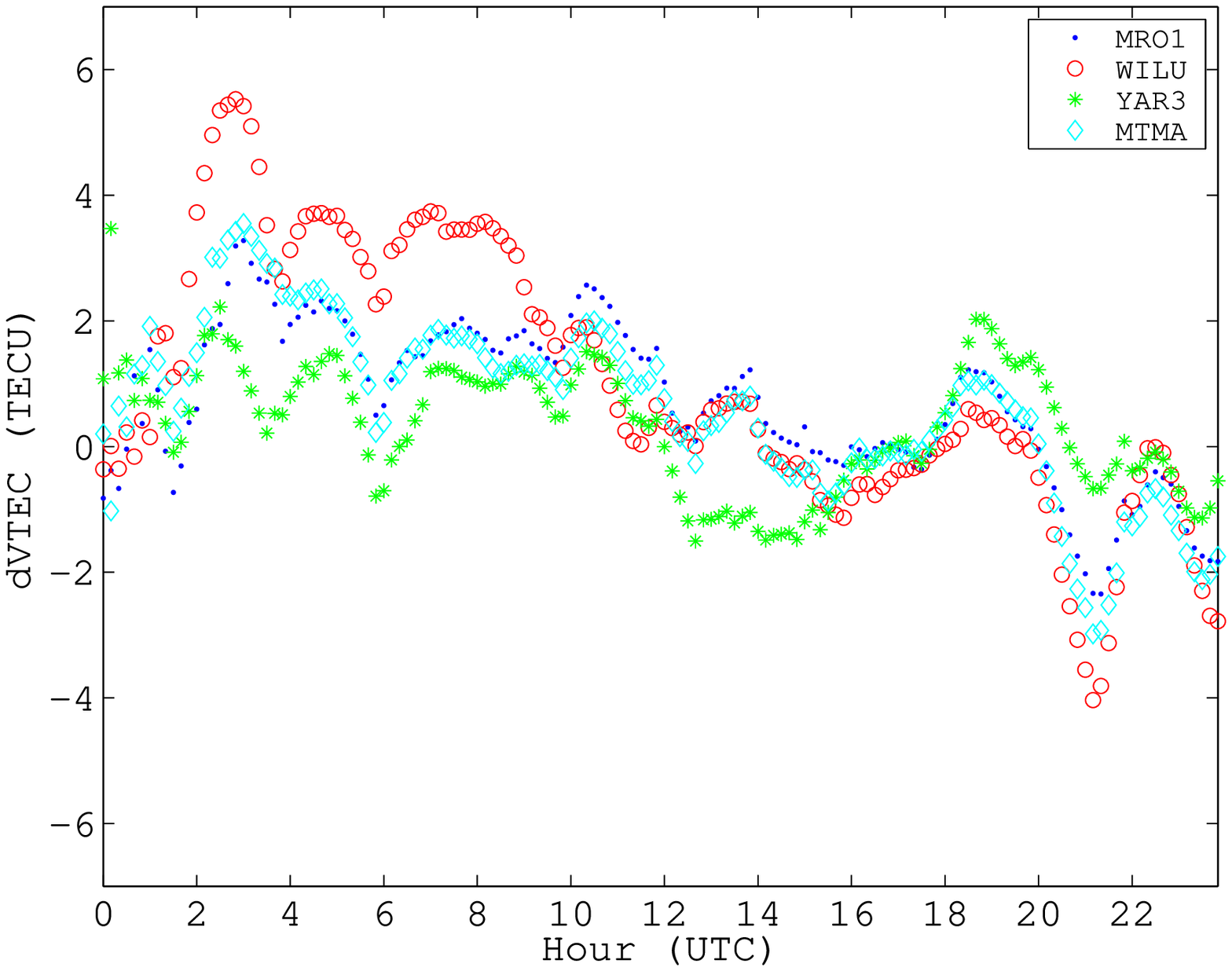}}
    \subcaptionbox{Differences in $VTEC$ on DOY 065\label{fig:dvtec065}}{\includegraphics[scale=0.38]{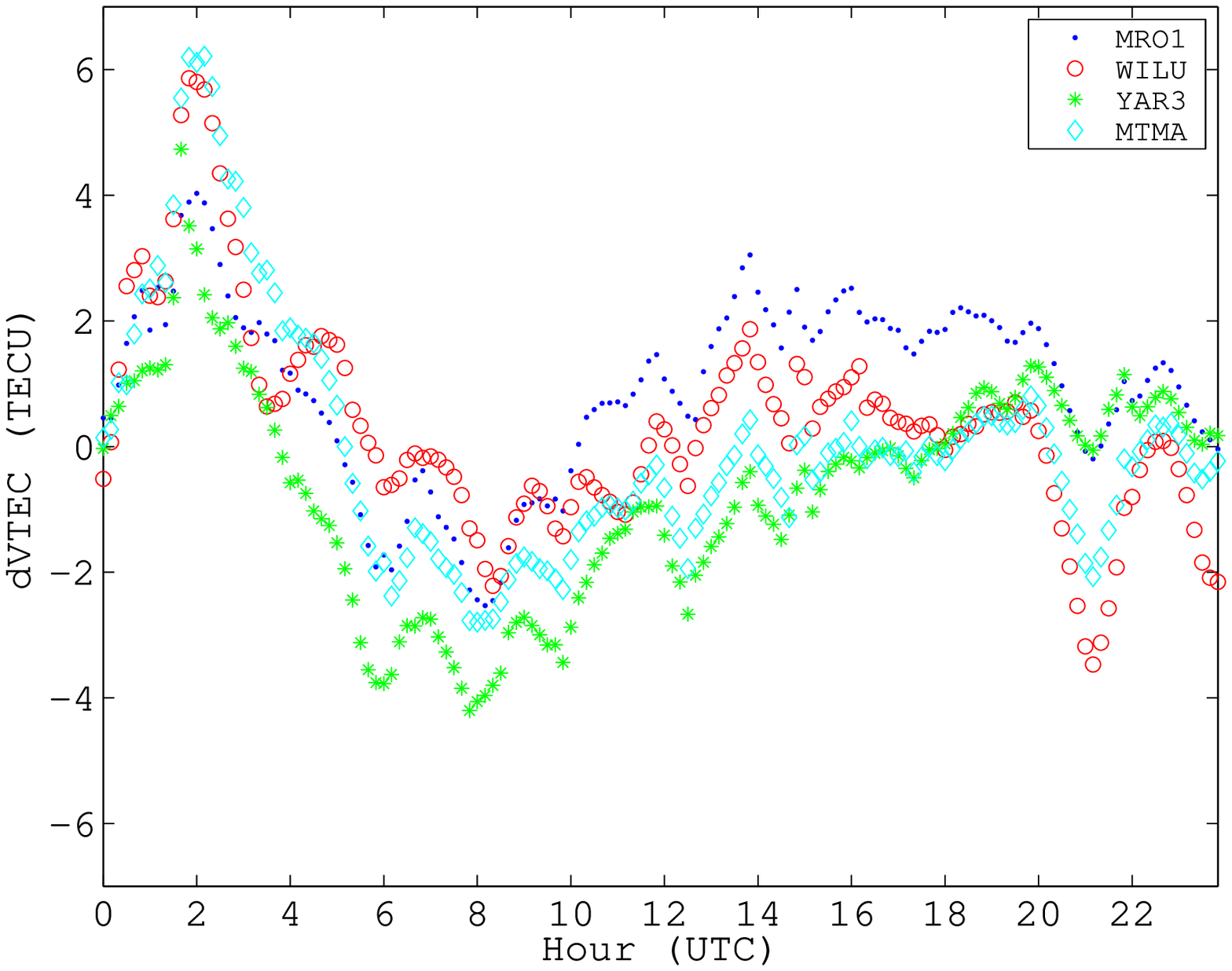}}
     \subcaptionbox{Differences in $VTEC$ on DOY 075\label{fig:dvtec075}}{\includegraphics[scale=0.38]{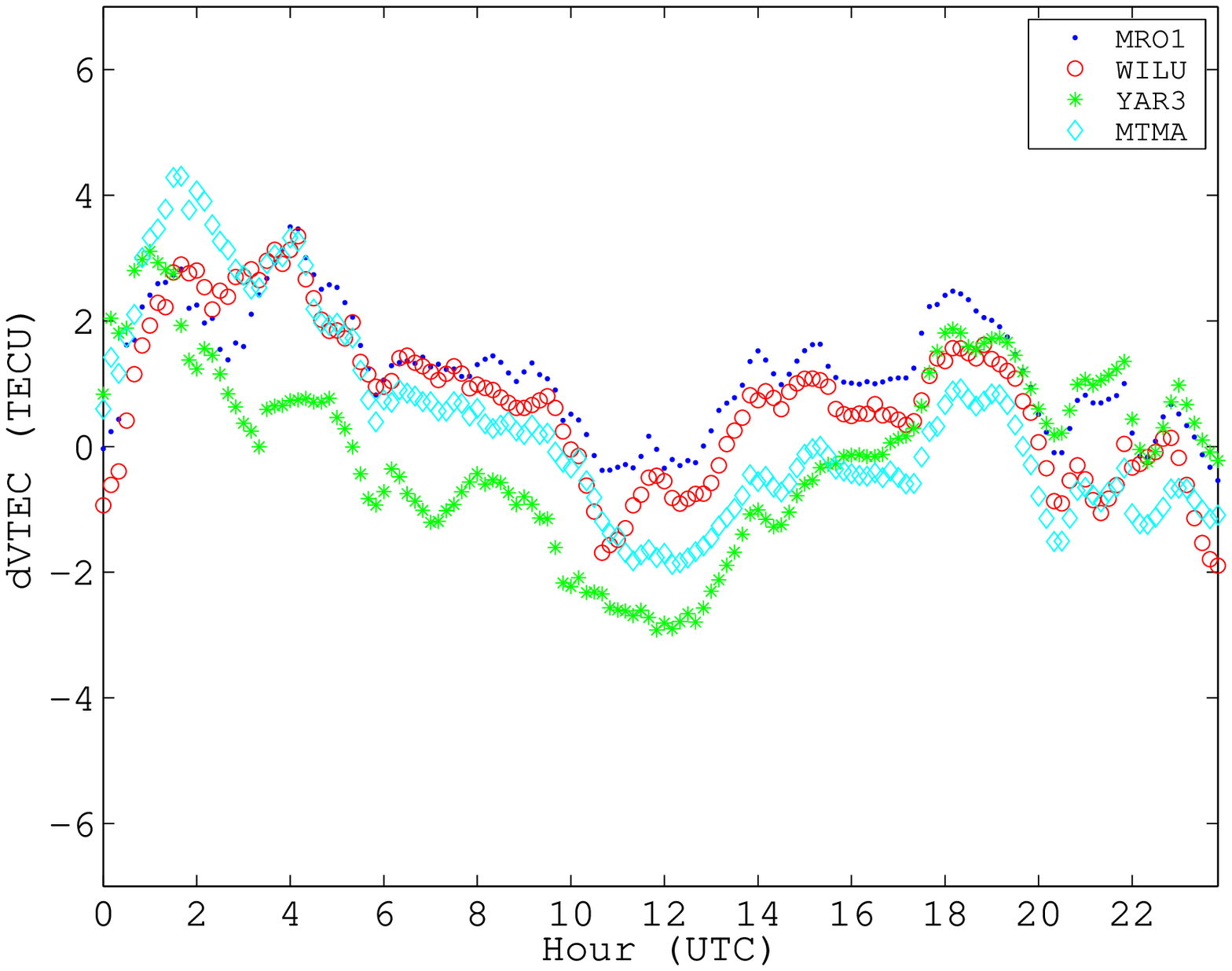}}    
     \caption{Differences in $VTEC$ w.r.t CODE for stations MRO1, MTMA, YAR3, and WILU for DOY 062, 063, 065 and 075 year 2014.
        \label{fig:dvtec}}
\end{figure*}

The $VTEC$ was estimated by our software at a time resolution of 10 minutes as described in section \ref{sec:met} with software developed in Matlab for the location of the four GPS stations, MRO1, MTMA, YAR3, and WILU. The CODE values of $VTEC$ are available at intervals of 2 hours. Values of the $VTEC$ corresponding to our time resolution were interpolated from CODE $VTEC$ maps \citep[see][]{Sch98} for each of the four GPS locations. Our estimated values of $VTEC$ along with the CODE $VTEC$ values are presented in Figure \ref{fig:vtec062} for DOY 062, year 2014. Figures \ref{fig:vtec062}\subref{fig:vtecmro1062}, \ref{fig:vtec062}\subref{fig:vtecmtma062}, \ref{fig:vtec062}\subref{fig:vtecYAR3062}, \ref{fig:vtec062}\subref{fig:vtecwilu062}, present VTEC for MRO1, MTMA, YAR3, and WILU respectively for DOY 062. Figure \ref{fig:dvtec} shows the difference between our $VTEC$ estimates with respect to CODE.  
The F10.7 solar flux, presented in Table \ref{tab:indices}, is highest on DOY 062 and lowest on 075. This is reflected in the $VTEC$ values, they reach a maximum of $\sim$68 TECU on DOY 062 and $\sim$57 TECU on DOY 075 for MRO1, refer Figures \ref{fig:vtec062}\subref{fig:vtecmro1062} and \ref{fig:vtec075}\subref{fig:vtecmro1075}. The 1$\sigma$ uncertainties in CODE maps reach a maximum of 8 TECU (Figure \ref{fig:avgionexrms}). The differences between CODE and our $VTEC$ are found to lie within the errors, with the differences ranging between -6 to 6 TECU for four different days of observations, (Figure \ref{fig:dvtec}). 

\subsection{Comparison of Receiver DCBs}
The receiver DCBs given by $c \cdot d_{r,21}$ in equation \eqref{eq:sdphase2} are the inter-frequency biases on code GPS data. Estimation of receiver DCBs is important for correct estimation of ionospheric parameters from GPS/GNSS observables \citep{Gap92,Sar94,Teu98}. The receiver DCBs were estimated as described in section \ref{sec:met}. In addition, the $\textsc{Bernese}$ GNSS data processing software \citep{Beu07} was used to estimate receiver DCBs for the selected GA GPS/GNSS stations given in Table \ref{tab:descGPSnet}. $\textsc{Bernese}$ 5.0 Precise Point Positioning (PPP) processing estimates one set of station specific ionosphere parameters for the entire session as well as receiver DCBs using the script $\texttt{PPP\underline{ }ION}$ \citep{Beu07}. With $\textsc{Bernese}$ PPP it is possible to obtain centimetre level station positions using precise satellite orbits, satellite clock and Earth Orientation Parameters (EOP) along with the receiver DCB.\\

The station specific ionosphere parameters are estimated by $\textsc{Bernese}$ PPP by forming frequency-differenced geometry-free observables. In this work, frequency-differenced observables were used to estimate station specific parameters. Unlike $\textsc{Bernese}$ PPP where quotidian ionosphere parameters are estimated for a given session, we make use of a Kalman filter to estimate the ionosphere every 10 minutes and a single value of receiver-satellite DCB for an entire session.\\

Since there exists a rank deficiency between the receiver and satellite DCBs. To overcome this rank deficiency, CODE\footnote{\url{http://igscb.jpl.nasa.gov/igscb/center/analysis/archive/code_20080528.acn}} assumes a zero-mean condition over all the satellite DCBs, $\sum \limits_{s=1}^{m} DCB^{s} =0 $. By assuming a zero-mean condition it implies that the DCB results may be shifted by a common offset value \citep[see][Chapter 13]{Beu07}, which is a function of total number of satellites $m$ considered for the solution. Hence an independent $\mathcal{S}$-basis was formed for estimation of DCBs.\\

Our estimated receiver DCBs and $\textsc{Bernese}$ PPP estimates are compared in this section. CODE also provides estimates of receiver DCBs which are estimated while performing a global fit for ionosphere parameters. However, of the selected stations used in this study, only receiver DCBs for station YAR3 are available from CODE. \\

\begin{figure*}
 \centering
  \subcaptionbox{Receiver DCBs for DOY 062, year 2014\label{fig:rxdcb062}}
  {\includegraphics[scale=0.44]{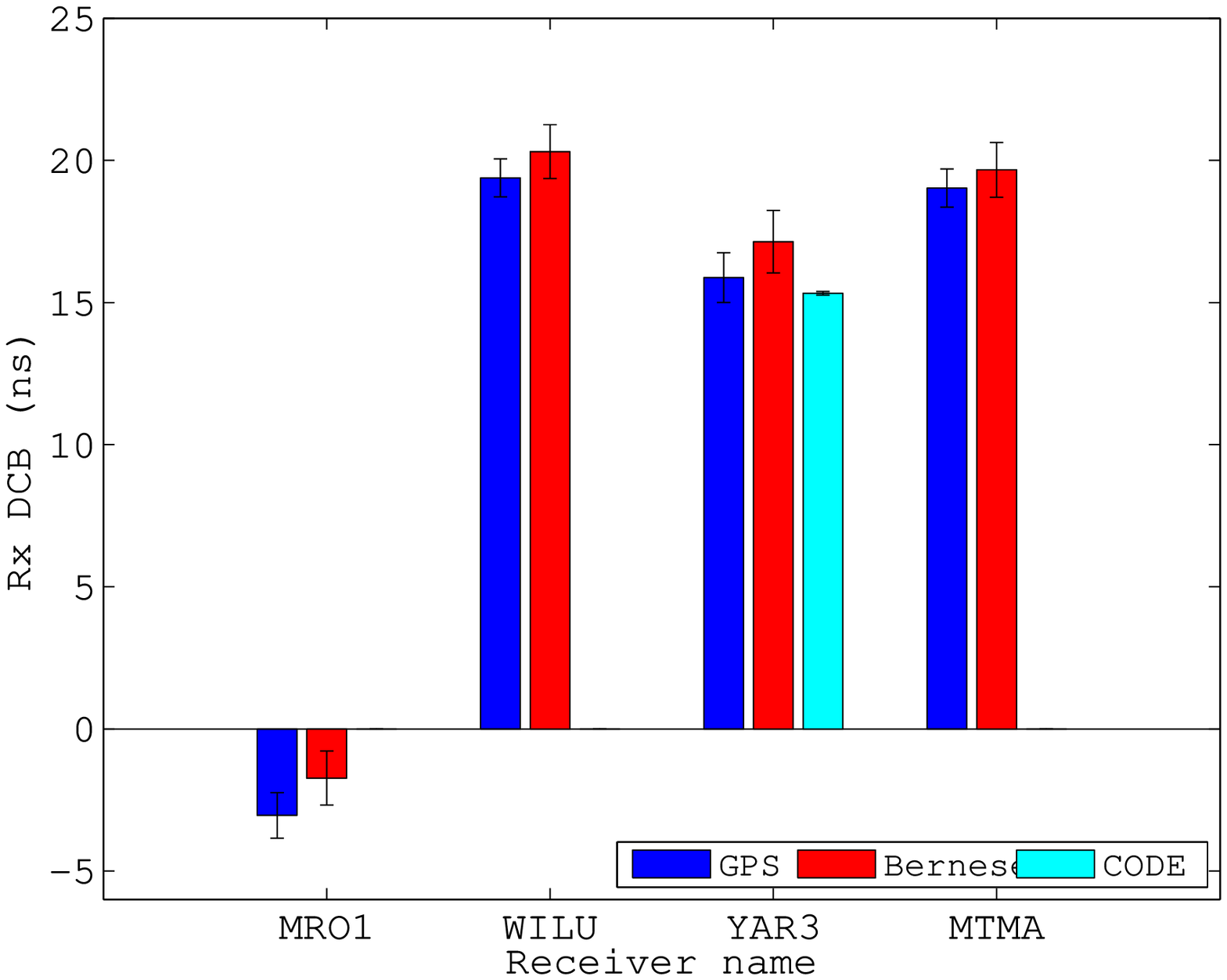}}
  \subcaptionbox{Receiver DCBs for DOY 063, year 2014\label{fig:rxdcb063}}
  {\includegraphics[scale=0.44]{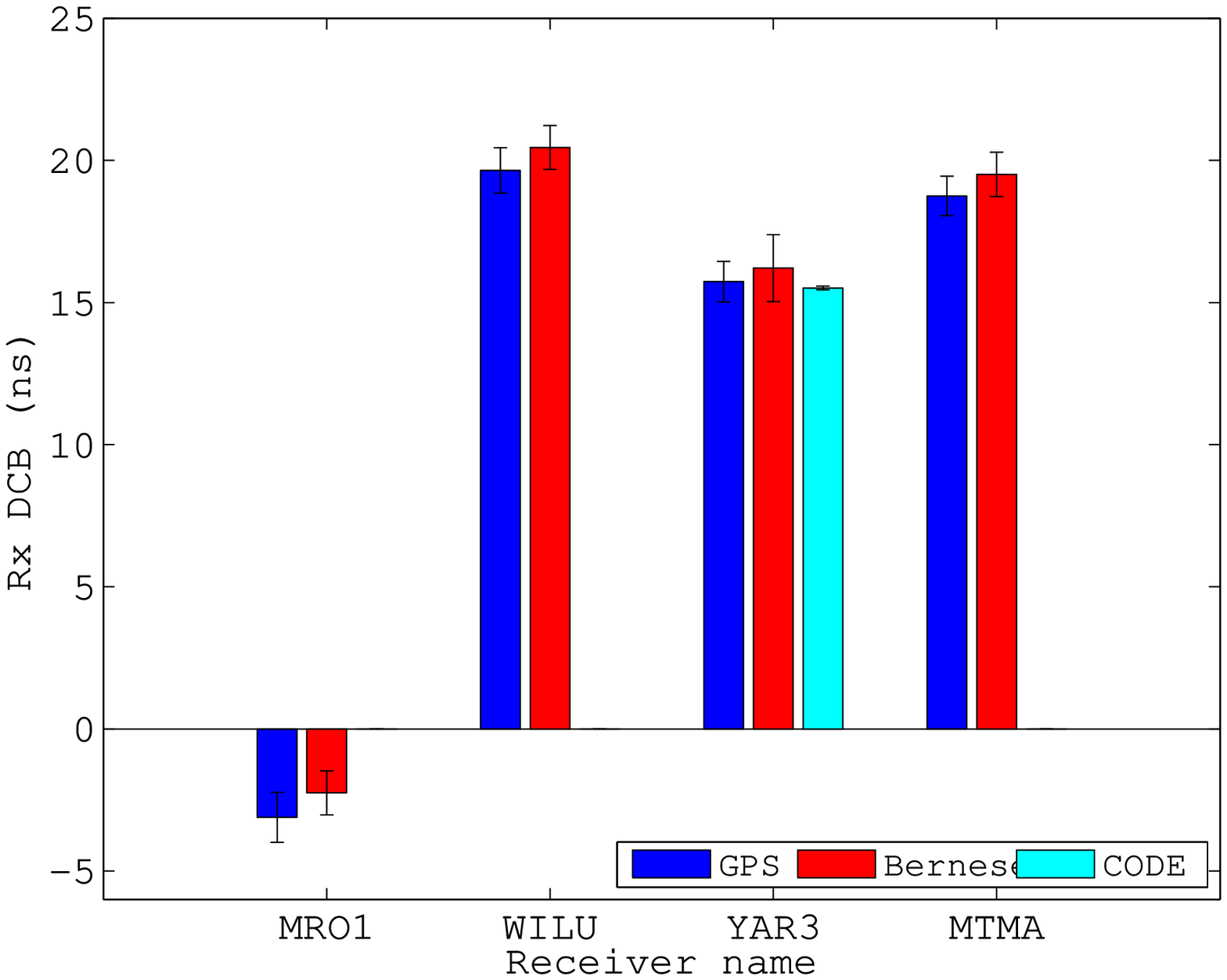}}
  \subcaptionbox{Receiver DCBs for DOY 065, year 2014\label{fig:rxdcb065}}
  {\includegraphics[scale=0.44]{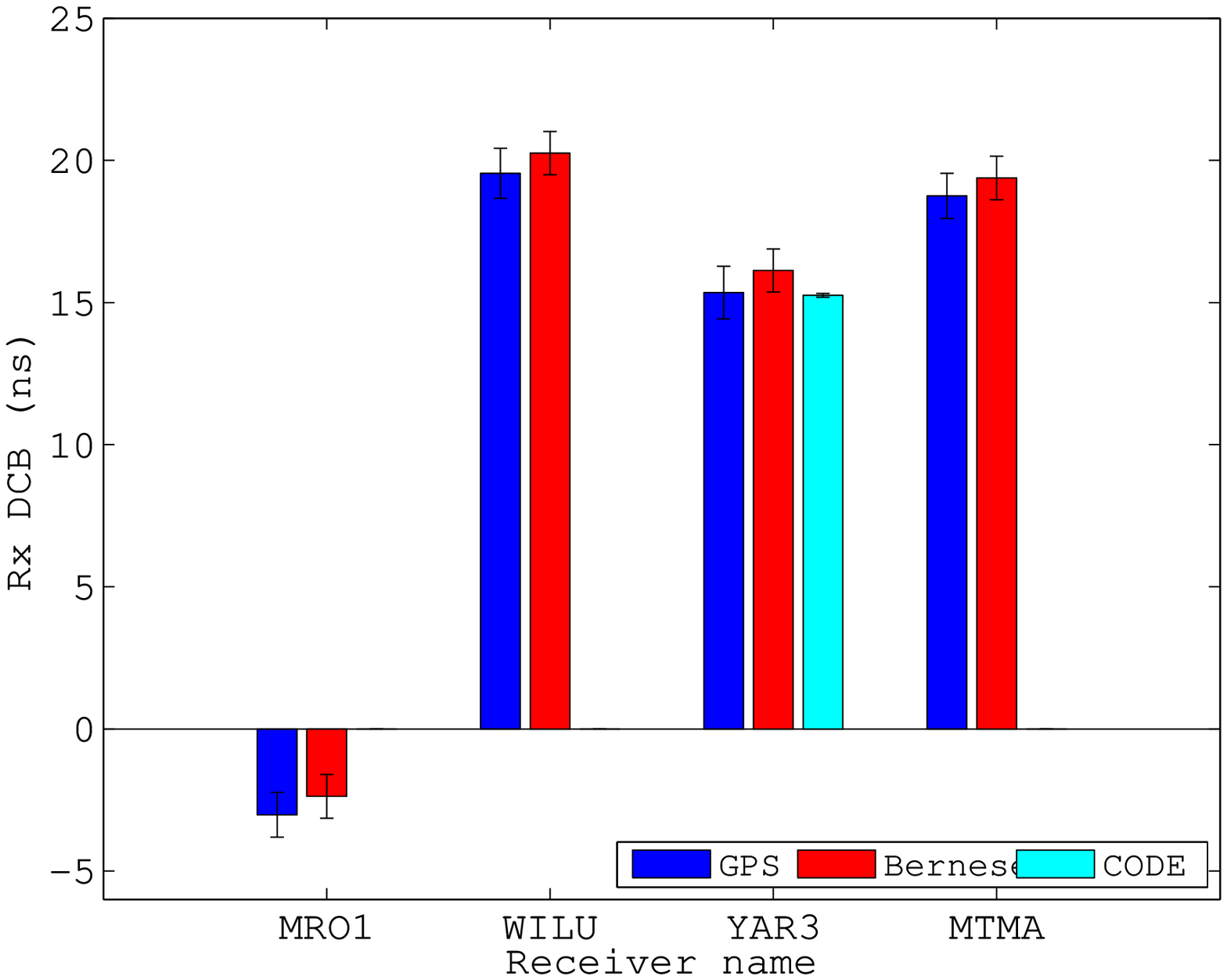}}
  \subcaptionbox{Receiver DCBs for DOY 075, year 2014\label{fig:rxdcb075}}
  {\includegraphics[scale=0.44]{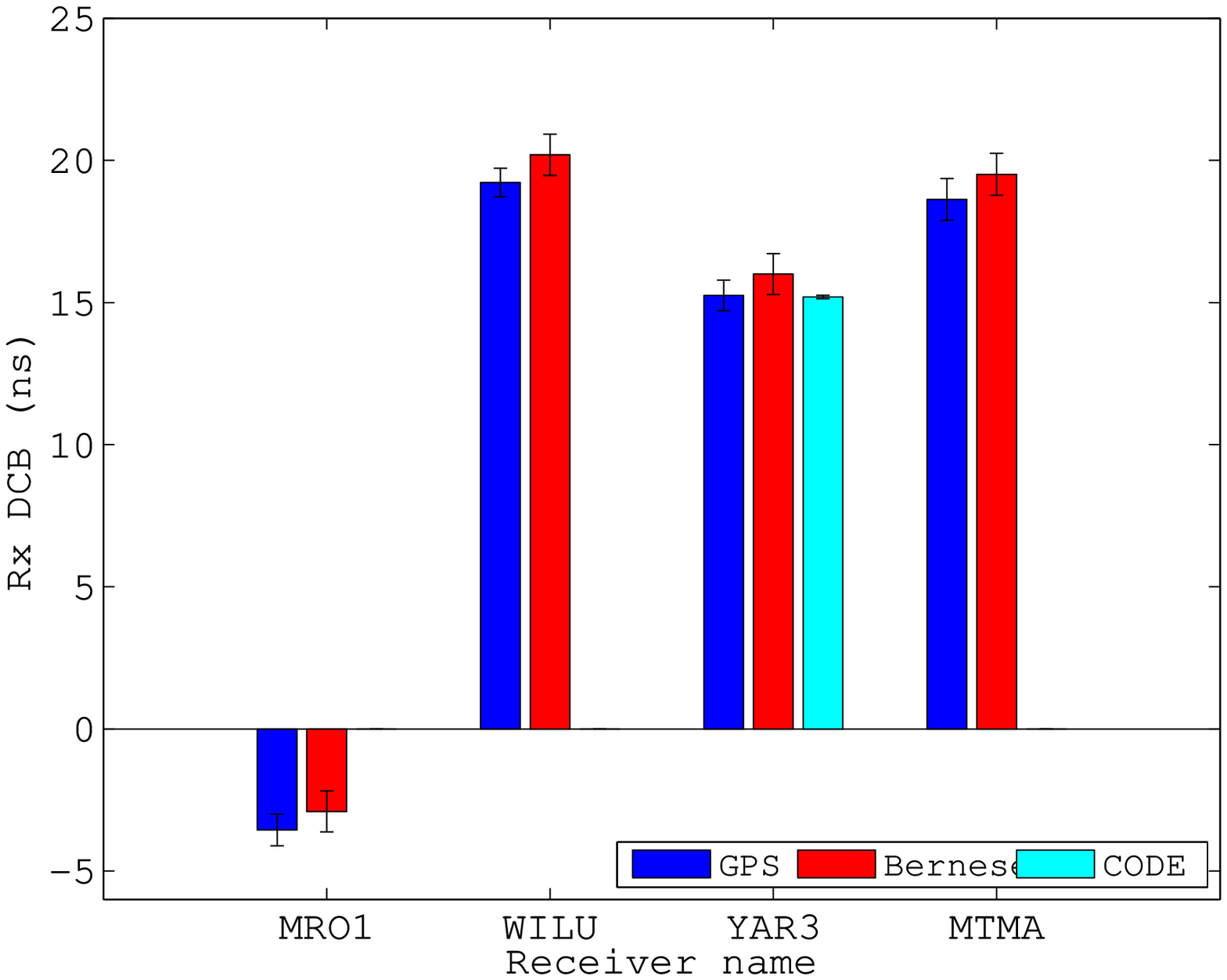}}
     \caption{Comparison of our estimated receiver DCBs (blue) with $\textsc{Bernese}$ estimates (red) and CODE DCBs (cyan) for DOY 062, 063, 065 and 075, year 2014.
        \label{fig:rxdcb}}
\end{figure*}

Figure \ref{fig:rxdcb} shows our estimated receiver DCBs (blue), with $\textsc{Bernese}$ (red) and from CODE (cyan) for DOY 062. Table \ref{tab:dcbdiff} presents the differences between our estimated receiver DCBs and $\textsc{Bernese}$ and CODE values. In comparison with $\textsc{Bernese}$ estimated DCBs, differences are between -0.475 to -1.311 ns which corresponds to -1.356 to -3.743 TECU. Whereas comparing to CODE DCBs, the DCB for YAR3 differed by 0.055 ns to 0.553 ns, 0.157 to 1.579 TECU. \\

\cite{Hon08} estimated the receiver DCBs by initially estimating single differenced DCBs. Further, by finding the time $t_{0}$ for which the single difference geometric range is zero, absolute receiver DCBs were computed. \citeauthor{Hon08} compared the estimated receiver DCBs with $\textsc{Bernese}$ values and found the maximum difference to lie around 15cm or 0.5ns. \cite{Ari08} estimated the receiver DCBs using IONOLAB-BIAS method and compared them with CODE estimates. \citeauthor{Ari08} found the differences in receiver DCBs to lie between -0.552ns and 0.110ns for different receivers.

\begin{table*}
\caption{Differences between our estimated receiver DCBs with $\textsc{Bernese}$ and CODE for DOY 062, 063, 065 and 075, year 2014.} 
\begin{center}
\begin{tabular}{@{}c c c c c | c c c c c@{}}
\hline \hline
 Station & \multicolumn{4}{c}{Difference (ns)} & \multicolumn{4}{c}{Difference (ns)} \\
         & \multicolumn{4}{c}{w.r.t $\textsc{Bernese}$ }& \multicolumn{4}{c}{w.r.t CODE} \\ \cline{2-9}%
         & \multicolumn{4}{c}{Year 2014, DOY} & \multicolumn{4}{c}{Year 2014, DOY}\\
      & 062 & 063 & 065  & 075 & 062 & 063  & 065 & 075 \\
\hline%
 MRO1 & -1.311 & -0.861 & -0.651 & -0.648 & - & - & - & - \\
 MTMA & -0.640 & -0.760 & -0.629 & -0.882 & - & - & - & - \\
 YAR3 & -1.262 & -0.475 & -0.778 & -0.756 & 0.553 & 0.230 & 0.095 & 0.055\\
 WILU & -0.924 & -0.806 & -0.711 & -0.980 & - & - & - & - \\
\hline\hline
\end{tabular}
\end{center}
\label{tab:dcbdiff}
\end{table*}

\subsection{Comparison of GPS ionosphere gradients with MWA observations}

\begin{figure*}
 \centering
\subcaptionbox{EW gradient for DOY 062\label{fig:EWcorr062}}{\includegraphics[scale=0.38]{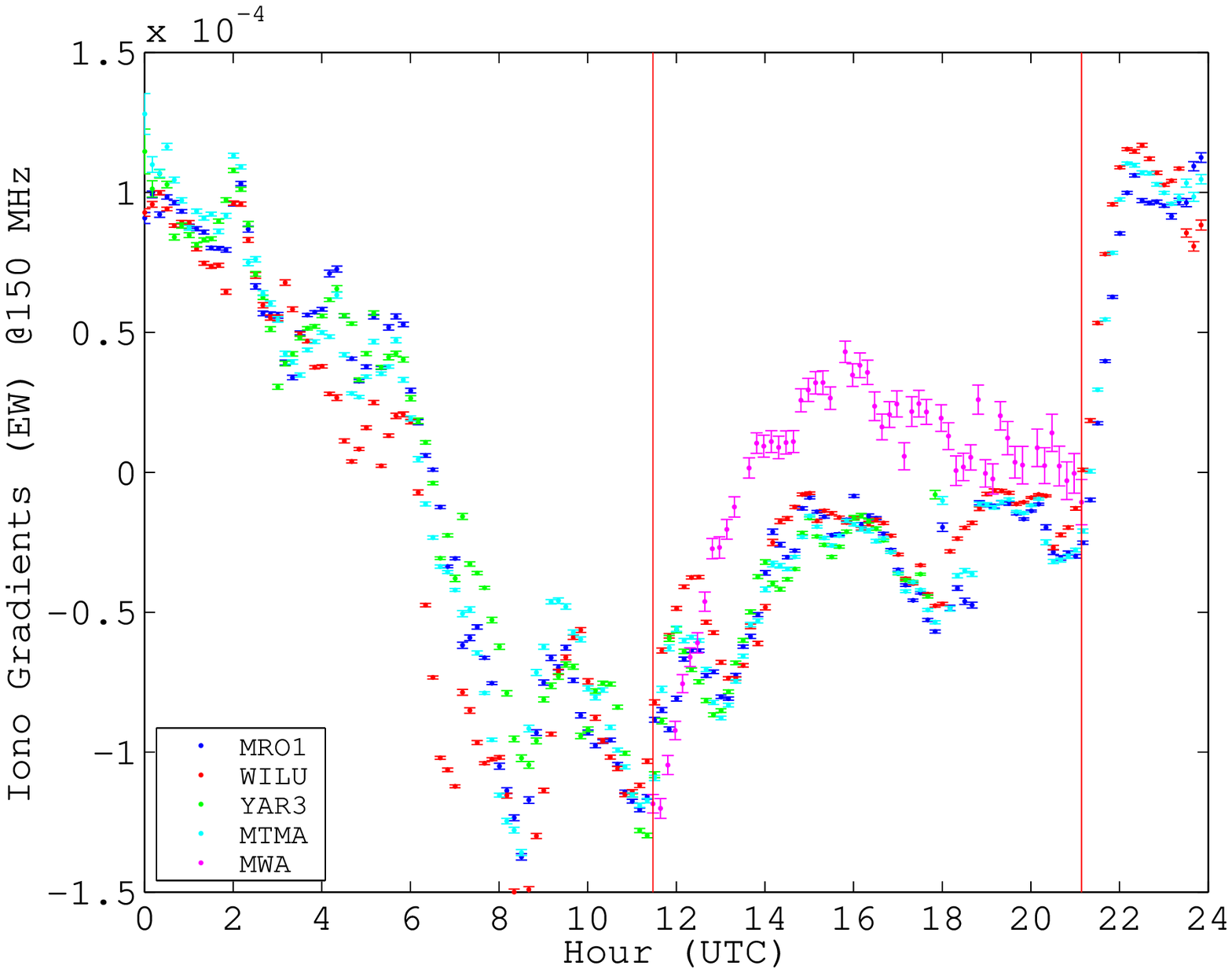}}
\subcaptionbox{NS gradient for DOY 062\label{fig:NScorr062}}{\includegraphics[scale=0.38]{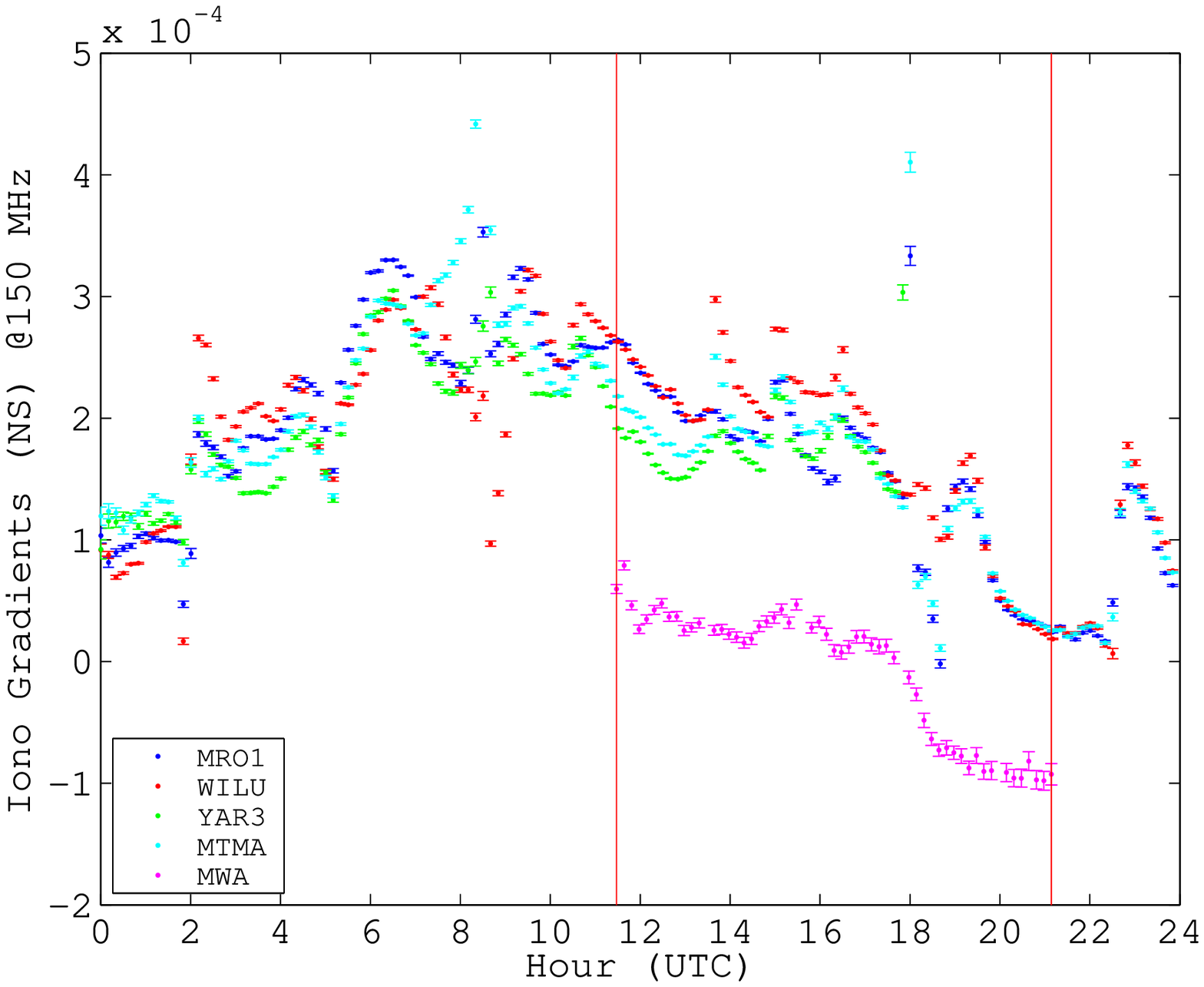}}
\subcaptionbox{EW gradient for DOY 063\label{fig:EWcorr063}}{\includegraphics[scale=0.38]{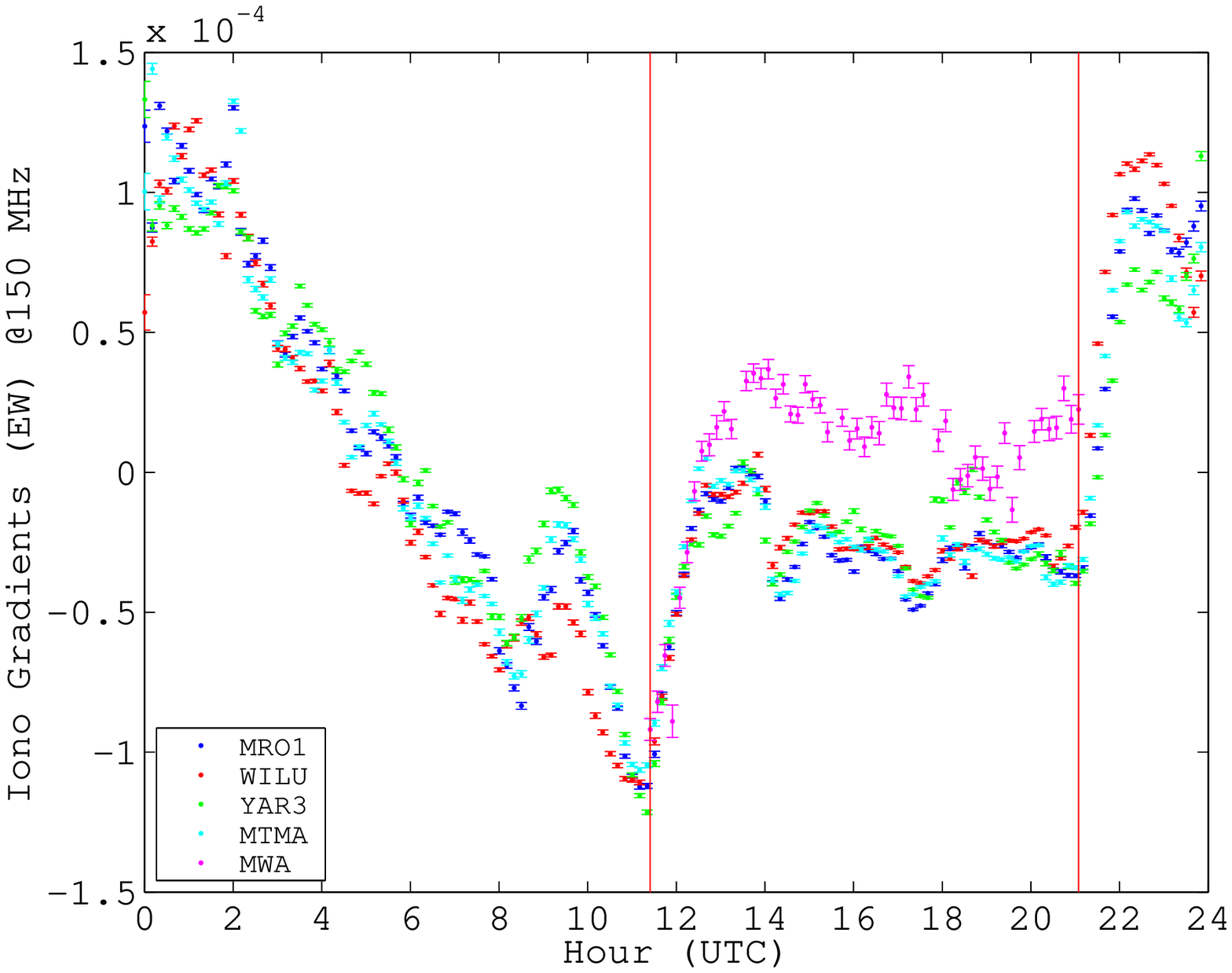}}
\subcaptionbox{NS gradient for DOY 063\label{fig:NScorr063}}{\includegraphics[scale=0.38]{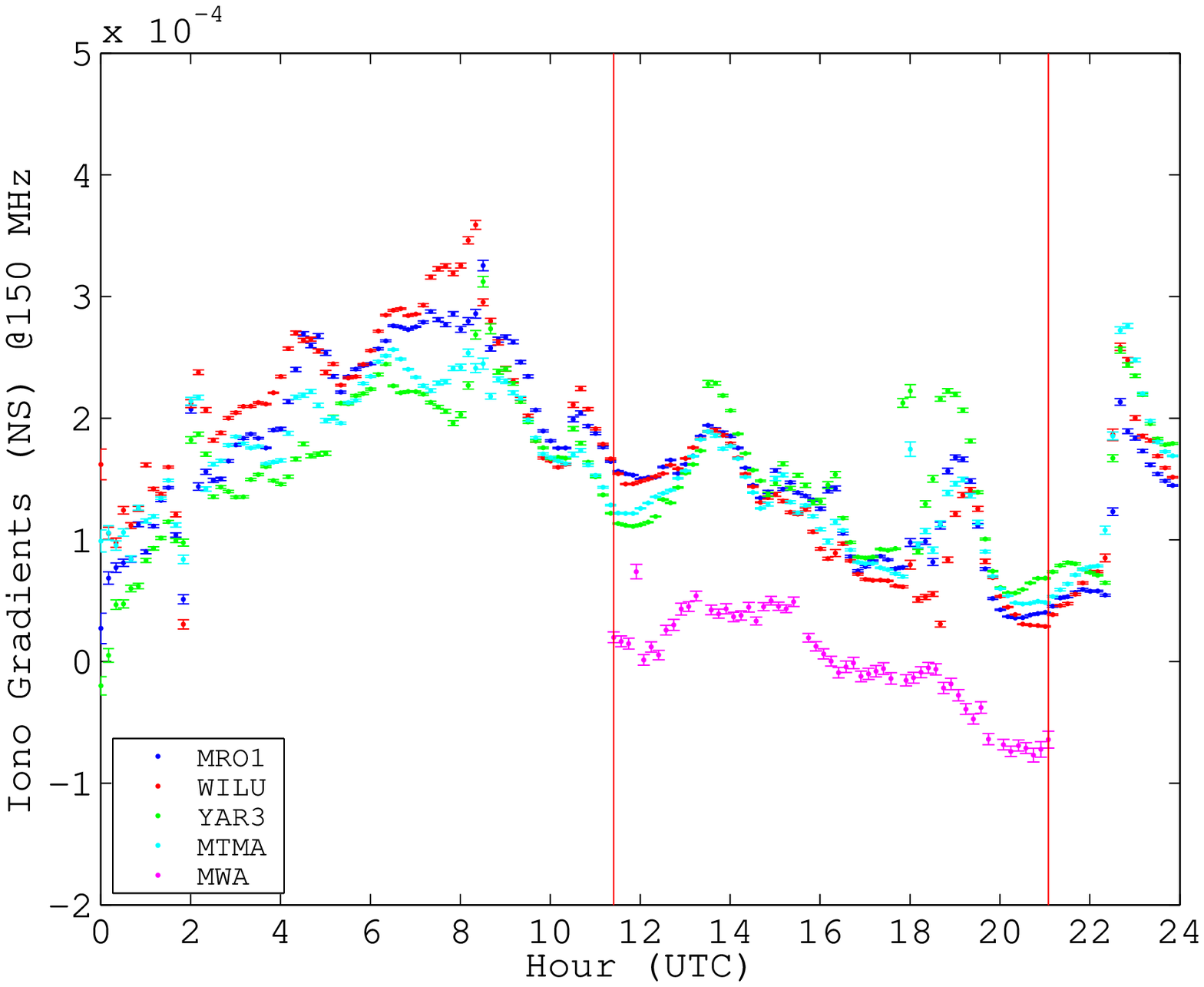}}
\subcaptionbox{EW gradient for DOY 065\label{fig:EWcorr065}}{\includegraphics[scale=0.38]{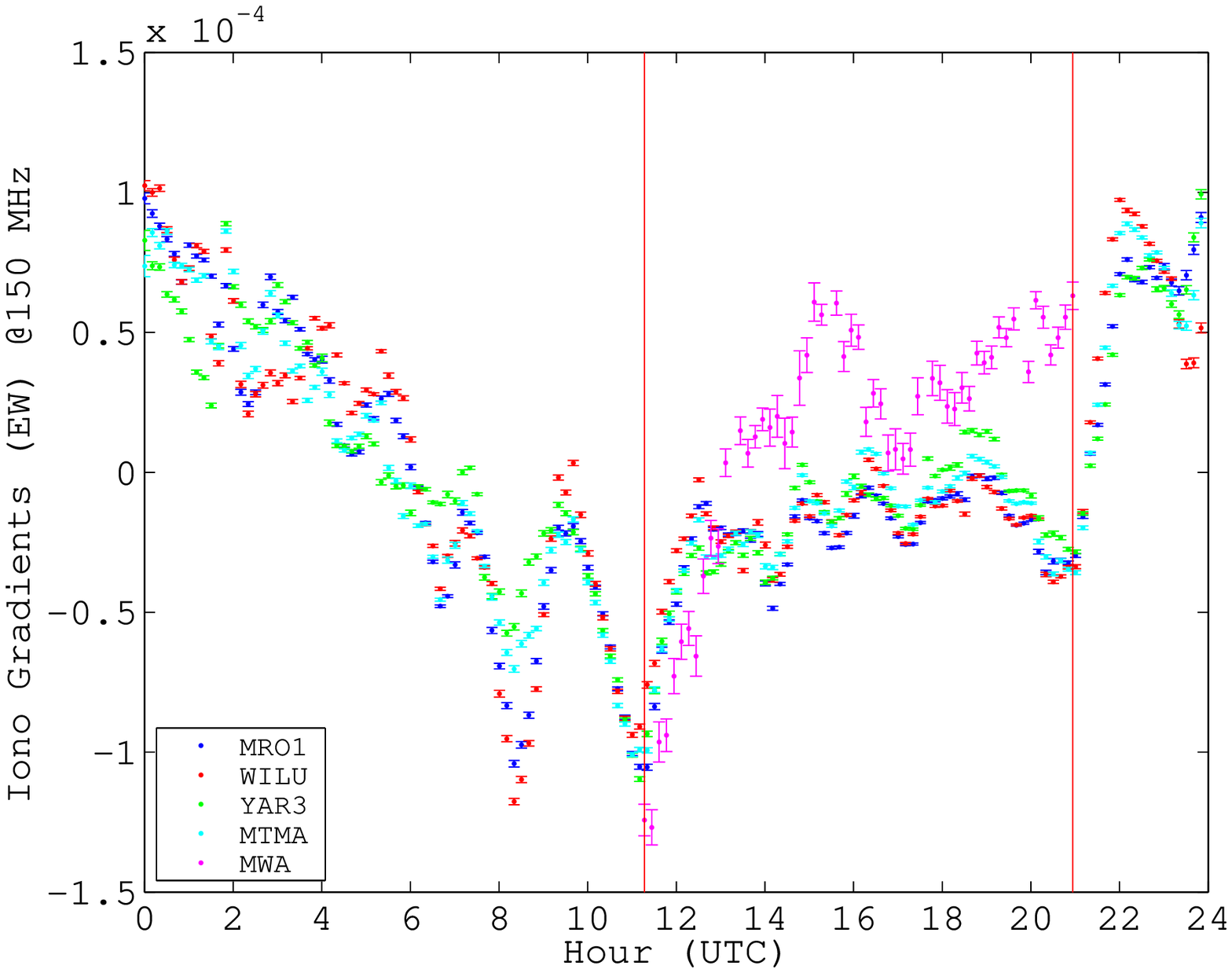}}
\subcaptionbox{NS gradient for DOY 065\label{fig:NScorr065}}{\includegraphics[scale=0.38]{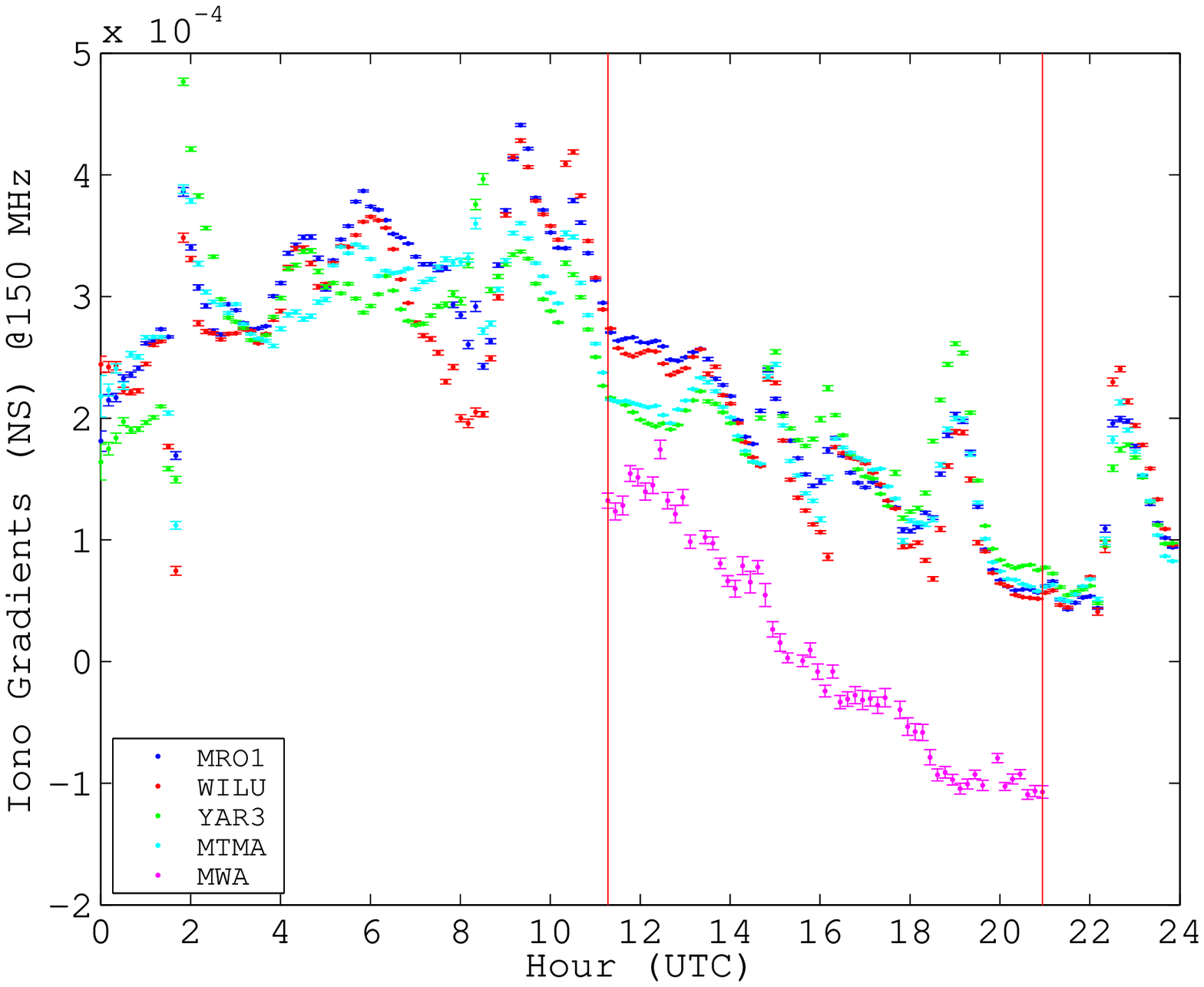}}
\subcaptionbox{EW gradient for DOY 075\label{fig:EWcorr075}}{\includegraphics[scale=0.38]{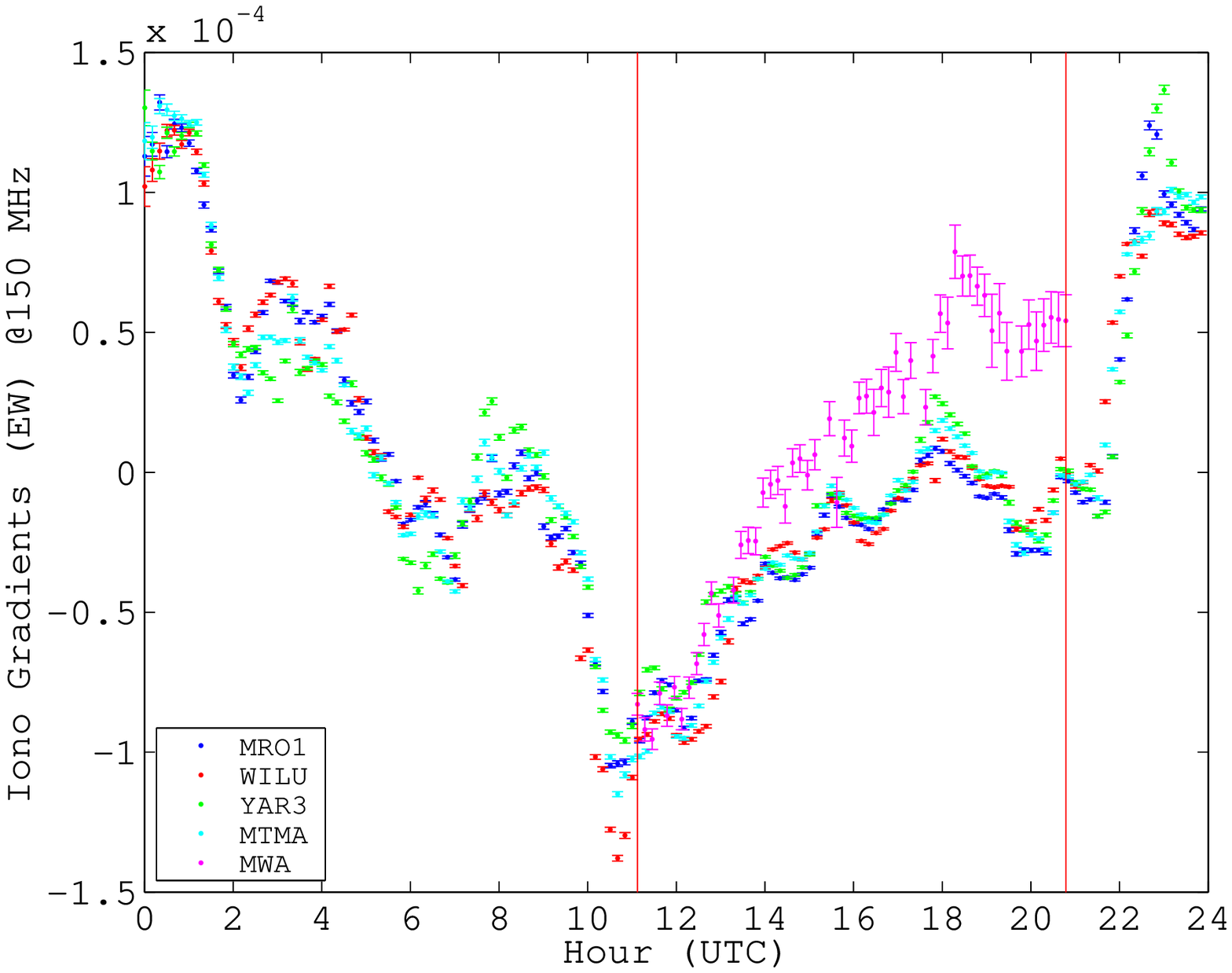}}
\subcaptionbox{NS gradient for DOY 075\label{fig:NScorr075}}{\includegraphics[scale=0.38]{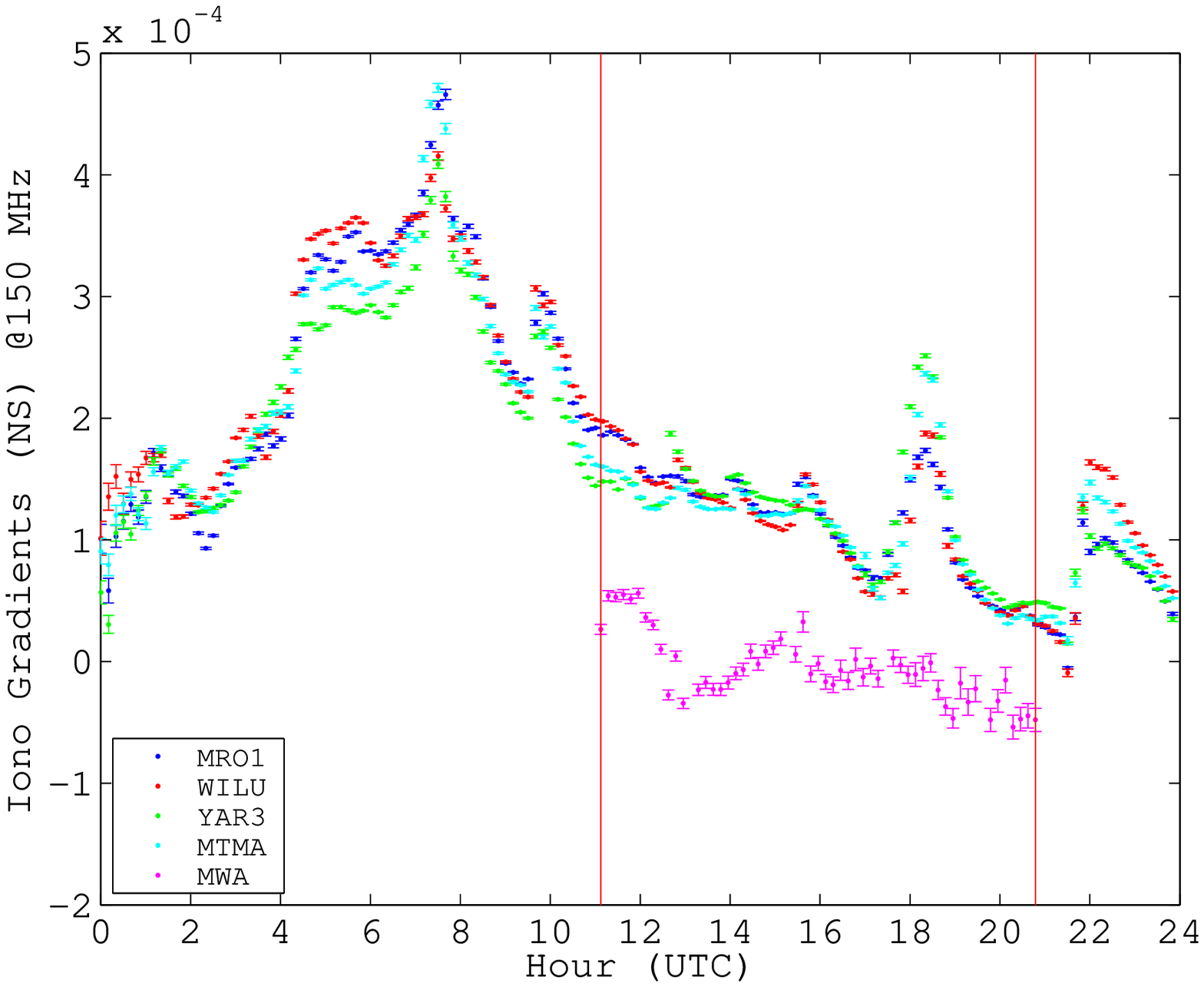}}
     \caption{EW and NS ionosphere gradients for selected GA stations and MWA for DOY 062, 063, 065 and 072 for year 2014. The time window related to MWA observations is shown by red line.
        \label{fig:corr-offset}}
\end{figure*}

The gradients in the East-West (EW) and North-South (NS) directions were computed from the ionosphere first order coefficients given in equation \eqref{eq:polmod2} for each of the selected stations and compared with MWA observed gradients. Figure \ref{fig:corr-offset}  presents each of the EW and NS gradients for all the GA GPS stations and MWA. Figures \ref{fig:corr-offset}\subref{fig:EWcorr062}, \ref{fig:corr-offset}\subref{fig:EWcorr063}, 
\ref{fig:corr-offset}\subref{fig:EWcorr065},and \ref{fig:corr-offset}\subref{fig:EWcorr075}, show the EW gradients for DOY 062, 063, 065 and 075 of year 2014, respectively. The NS gradients are presented in Figures \ref{fig:corr-offset}\subref{fig:NScorr062}, \ref{fig:corr-offset}\subref{fig:NScorr063}, 
\ref{fig:corr-offset}\subref{fig:NScorr065},and \ref{fig:corr-offset}\subref{fig:NScorr075} for DOY 062, 063, 065 and 075 of year 2014, respectively. In each of the subplots, along with the GPS ionospheric gradients, the MWA observed gradients are shown. 

Table \ref{tab:gradcorr} presents the correlation between GPS and MWA gradients in the EW and NS directions for the four days of observations. The IPP separations in longitude ($|\Delta \lambda_{IPP}|$) and latitude ($|\Delta \phi_{IPP}|$) for each of the GPS stations and MWA are presented in Table \ref{tab:gradcorr}. The correlation between the GPS and MWA ionosphere gradients is computed using Pearson's coefficient of correlation, $r$, for an assumed mean method along with the standard error, $\sigma_{r}$, refer \citet{Fis36}.

There is a high correlation in the EW and NS gradient between GPS and MWA for most of the GPS stations for all the days (Table \ref{tab:gradcorr}). The NS gradients had a weak correlation with YAR3 for most days. The correlation was highest on DOY 075 while the IPP was closest among the four days of observation (Table \ref{tab:gradcorr}). The EW gradient showed consistent good correlation with MRO1 GPS stations, whereas correlation with WILU seemed to be most inconsistent (Table \ref{tab:gradcorr}). A general trend seemed to show that the EW gradient was proportional to the longitudinal difference and the NS gradient to the latitude difference. For DOY 075 while the solar activity was the lowest among the selected days (Table \ref{tab:indices}), the EW gradient was found to be high for all the stations. NS gradient did not have a similar behaviour. \\

Table \ref{tab:gradcorr2} and Figure \ref{fig:corr-offset} present the comparison of zenith EW and NS gradients between GPS stations. Correlation for the EW and NS gradients, presented in Table \ref{tab:gradcorr2}, were computed between each of the selected GPS stations for the time window of MWA observations (marked by the red line in Figure \ref{fig:corr-offset}). Table \ref{tab:gradcorr2} summarises the inter-station EW and NS gradient correlations. Inter-station correlation is strong for the EW and NS gradient for almost all the days between all the stations. The EW gradient is consistently strong between all the stations, however the NS gradient is found to be weakest between WILU and YAR3 which have a distinguished latitudinal and longitudinal separation (Figure \ref{fig:WAGPSnet}). \\

\begin{table*}
\caption{Correlation between the GPS and MWA observed gradients in EW ($r_{EW}$) and NS ($r_{NS}$) components, its standard error ($\sigma_{r}$), and IPP separations in longitude ($|\Delta \lambda_{IPP}|$) and latitude ($|\Delta \phi_{IPP}|$) of GPS stations and MWA ($\Delta$IPP) for DOY 062, 063, 065 and 075, year 2014.} 
\begin{center}
\begin{tabular}{@{}c c c c c | c c c c c@{}}
\hline \hline
 Station & \multicolumn{4}{c}{$r_{EW}$ ($\sigma_{r_{EW}}$)} & \multicolumn{4}{c}{$|\Delta \lambda_{IPP}|$ (degrees)}\\ \cline{2-9}%
  & \multicolumn{4}{c}{Year 2014, DOY}& \multicolumn{4}{c}{Year 2014, DOY}\\
   & 062 & 063 & 065 & 075 & 062 & 063 & 065 & 075\\
\hline%
 MRO1 & 0.79(0.05) & 0.73(0.06) & 0.66(0.08) & 0.93(0.02)       & 0.03 & 0.03 & 0.03 & 0.03 \\
 &  &  &  &  &   &   &   &  \\
 MTMA & 0.72(0.07) & 0.65(0.08) & 0.71(0.07) & 0.94(0.02)       & 1.17 & 1.17 & 1.17 & 1.17 \\
 & &  &  &  &  &   &   &   &  \\
 YAR3 & 0.83$^{a}$(0.05) & 0.73(0.06) & 0.77(0.06) & 0.92(0.02) & 1.32 & 1.32 & 1.32 & 1.32 \\
 & &  &  &  &   &   &   &  \\
 WILU & 0.60(0.09) & 0.83(0.04) & 0.54(0.10) & 0.94(0.01)       & 3.54 & 3.54 & 3.54 & 3.54 \\\hline%
  & \multicolumn{4}{c}{$ r_{NS}$ ($\sigma_{r_{NS}}$)} & \multicolumn{4}{c}{$|\Delta \phi_{IPP}|$ (degrees)}\\ 
\hline%
MRO1 & 0.84(0.04) & 0.78(0.05) & 0.87(0.03) & 0.69(0.07)       & 0.10 & 0.65 & 1.51 & 0.65 \\
 &  &  &  &  &   &   &   &  \\
MTMA & 0.77(0.06) & 0.75(0.06) & 0.74(0.06) & 0.46(0.11)       & 1.31 & 2.07 & 2.93 & 0.77 \\
 &  &  &  &  &   &   &   &  \\
YAR3 & 0.21$^{a}$(0.16) & 0.45(0.11) & 0.50(0.10) & 0.35(0.12) & 2.24 & 3.00 & 3.86 & 1.70 \\
 &  &  &  &  &   &   &   &  \\
WILU & 0.89(0.03) & 0.77(0.05) & 0.86(0.04) & 0.64(0.08)       & 0.17 & 0.58 & 1.43 & 0.72 \\
\hline\hline
\end{tabular}
\end{center}
\label{tab:gradcorr}
\tabnote{$^{a}$Partial data available, from 00:00:00 to 18:07:00 UTC (MWA observation window = $\sim$ 11 to 21 UTC)}
\end{table*}

\begin{table*}[hp]
\caption{Inter-station correlation for the EW and NS gradients ($r$), its standard error ($\sigma_{r}$), and IPP separations in longitude ($|\Delta \lambda_{IPP}|$) and latitude ($|\Delta \phi_{IPP}|$) between GPS stations} 
\begin{center}
\begin{tabular}{@{}c c c c c | c c@{}}
\hline \hline
 Station & \multicolumn{4}{c}{$r_{EW}$ ($\sigma_{r_{EW}}$)} & $|\Delta \lambda_{IPP}|$ (degrees)\\ \cline{2-6}%
 & \multicolumn{4}{c}{Year 2014, DOY}&  \\
   & 062 & 063 & 065 & 075  &  \\
\hline%
MRO1-MTMA &  0.88(0.03) & 0.97(0.01) & 0.94(0.02) & 0.99(0.002) & 1.21 \\
     &  &  &   &  &\\
YAR3-MTMA &       -     & 0.83(0.04) & 0.94(0.01) & 0.97(0.01) & 2.50 \\
     &  &  &  &  &\\
MTMA-WILU  & 0.90(0.03) & 0.91(0.02) & 0.92(0.02) & 0.98(0.01) & 2.37 \\
     &   &  &  &  &\\ 
MRO1-YAR3 & -           & 0.89(0.03) & 0.91(0.02) & 0.97(0.01) & 1.29 \\
     &  &  &  &  &\\
MRO1-WILU &  0.88(0.03) & 0.96(0.01) & 0.93(0.02) & 0.97(0.01) & 3.58 \\
     &  &  &  &  &\\
WILU-YAR3 &  -          & 0.86(0.04) & 0.84(0.04) & 0.94(0.02) & 4.87 \\\hline
 	 & \multicolumn{4}{c}{$ r_{NS}$ ($\sigma_{r_{NS}}$)} & $|\Delta \phi_{IPP}|$ (degrees)\\ \hline%
MRO1-MTMA & 0.94(0.02) & 0.92(0.02) & 0.95(0.01) & 0.89(0.03) & 1.42 \\
     &  &  &   &  &\\
YAR3-MTMA &       -    & 0.79(0.05) & 0.88(0.03) & 0.93(0.02) & 0.93 \\
     &  &  &  &  &\\
MTMA-WILU & 0.81(0.05) & 0.87(0.03) & 0.95(0.01) & 0.90(0.03) & 1.49 \\
     &   &  &  &  &\\ 
MRO1-YAR3 &       -    & 0.71(0.07) & 0.83(0.04) & 0.84(0.04) & 2.35 \\
     &  &  &  &  &\\
MRO1-WILU & 0.84(0.04) & 0.91(0.02) & 0.96(0.01) & 0.96(0.01) & 0.07 \\
     &  &  &  &  &\\
WILU-YAR3 &   -        & 0.53(0.10) & 0.74(0.06) & 0.79(0.05) & 2.42 \\
\hline\hline
\end{tabular}
\end{center}
\label{tab:gradcorr2}
\end{table*}

\section{CONCLUSIONS}
\label{sec:conc}
This work presents a single-station approach to estimate $VTEC$ ionosphere gradients. The $VTEC$ and ionosphere gradients were estimated at intervals of 10 minutes. The ionosphere gradients in the EW and NS direction at the GPS station locations are in good agreement with the MWA observed gradients.\\

The ionosphere gradient analysis presented in this research brings forth various questions, namely, the variation of EW and NS gradients with respect to IPP separation, and the elevation dependency reflected by the correlation of gradients between the GPS station, MRO1, and MWA. With the limited data set, these questions cannot not be resolved comprehensively. The future work will focus on including more MWA observation and will attempt to closely probe the above questions in a statistical sense.\\

With our single-station approach, the GPS receiver DCBs can be accurately estimated for any available GPS receiver. Thus our method can be applied to local GPS receiver data for which the DCBs are not publicly available.\\

To develop a regional model for the ionosphere, a multi-station approach needs to be adopted. Furthermore, the spatial resolution of ionosphere gradients is closely related to the scale of the GPS/GNSS receiver network on the ground. The scales at which MWA sees the ionosphere lie between 10 - 100 km. In order to estimate the ionosphere gradients on such scales, dense GPS networks of the order of the ionosphere scales seen by the MWA need to be present around MWA site. The existing GPS network near the MWA site is of scale 150 - 250 km, which limits the GPS based ionosphere research possible for the MWA.\\

In addition to GPS satellite data, other satellite systems can be used to densify the IPPs over the study area. Other global navigation systems like GLONASS (GLObal NAvigation Satellite System), launched by Russia, BeiDou from China, and the European Union's Galileo could be adopted in our model. Of these three global navigation systems, only the GLONASS system is close to completely operational, it currently has 23 operational satellites in orbit. GA receivers are able to capture data from the GLONASS system, and using GLONASS data along with the GPS data has been shown to improve ionospheric modelling \citep{Cos99}.\\

Future work will focus on developing a regional ionosphere model using data from both GPS and GLONASS satellite systems.\\

We find that the Australian SKA site (where the MWA is located) is well suited for low-frequency astronomy. Appendix \ref{app} shows that the conclusions drawn in \citet{Sot13} regarding the suitability of MRO for low-frequency radio astronomy are a result of an incorrect interpretation of CODE maps.

\begin{acknowledgements}
The authors wish to thank Anthony Willis from National Research Council of Canada for the valuable discussions. This scientific work makes use of the Murchison Radio-astronomy Observatory, operated by CSIRO. We acknowledge the Wajarri Yamatji people as the traditional owners of the Observatory site. Support for the MWA comes from the U.S. National Science Foundation (grants AST-0457585, PHY-0835713, CAREER-0847753, and AST-0908884), the Australian Research Council (LIEF grants LE0775621 and LE0882938), the U.S. Air Force Office of Scientific Research (grant FA9550-0510247), and the Centre for All-sky Astrophysics (an Australian Research Council Centre of Excellence funded by grant CE110001020). Support is also provided by the Smithsonian Astrophysical Observatory, the MIT School of Science, the Raman Research Institute, the Australian National University, and the Victoria University of Wellington (via grant MED-E1799 from the New Zealand Ministry of Economic Development and an IBM Shared University Research Grant). The Australian Federal government provides additional support via the Commonwealth Scientific and Industrial Research Organisation (CSIRO), National Collaborative Research Infrastructure Strategy, Education Investment Fund, and the Australia India Strategic Research Fund, and Astronomy Australia Limited, under contract to Curtin University. We acknowledge the iVEC Petabyte Data Store, the Initiative in Innovative Computing and the CUDA Center for Excellence sponsored by NVIDIA at Harvard University, and the International Centre for Radio Astronomy Research (ICRAR), a Joint Venture of Curtin University and The University of Western Australia, funded by the Western Australian State government. 
\end{acknowledgements} 

\begin{appendix}

\section{CODE IONOSPHERE MAP INTERPRETATION DISCREPANCY}
\label{app}

\begin{figure}[hp]
 \centering
\subcaptionbox{CODE IONEX plot in \citet{Sot13}  \label{fig:ionexsot}}{\includegraphics[height=5.7cm,width=6.5cm]{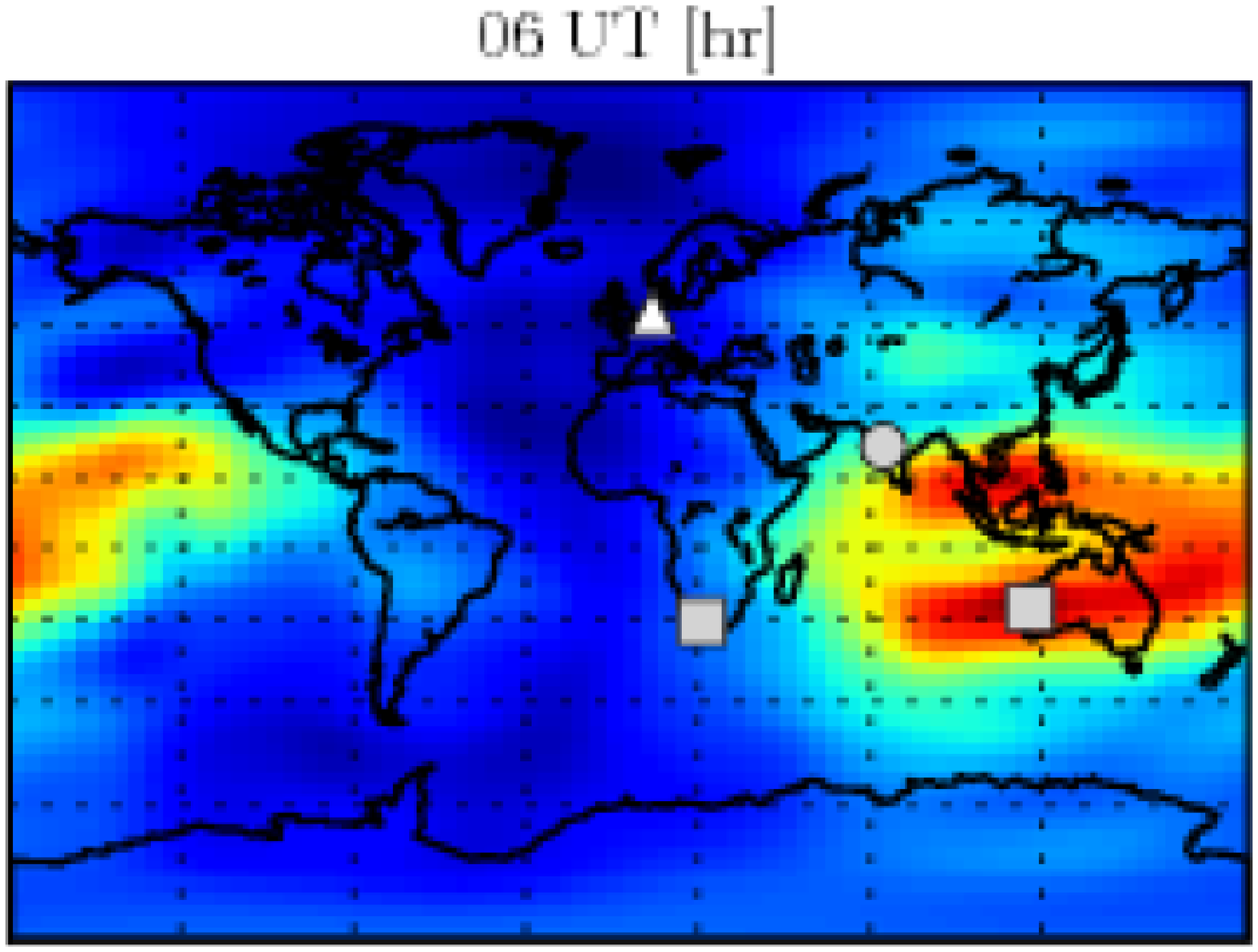}}
\subcaptionbox{CODE IONEX plot by our software\label{fig:ionexref}}{\includegraphics[scale=0.41]{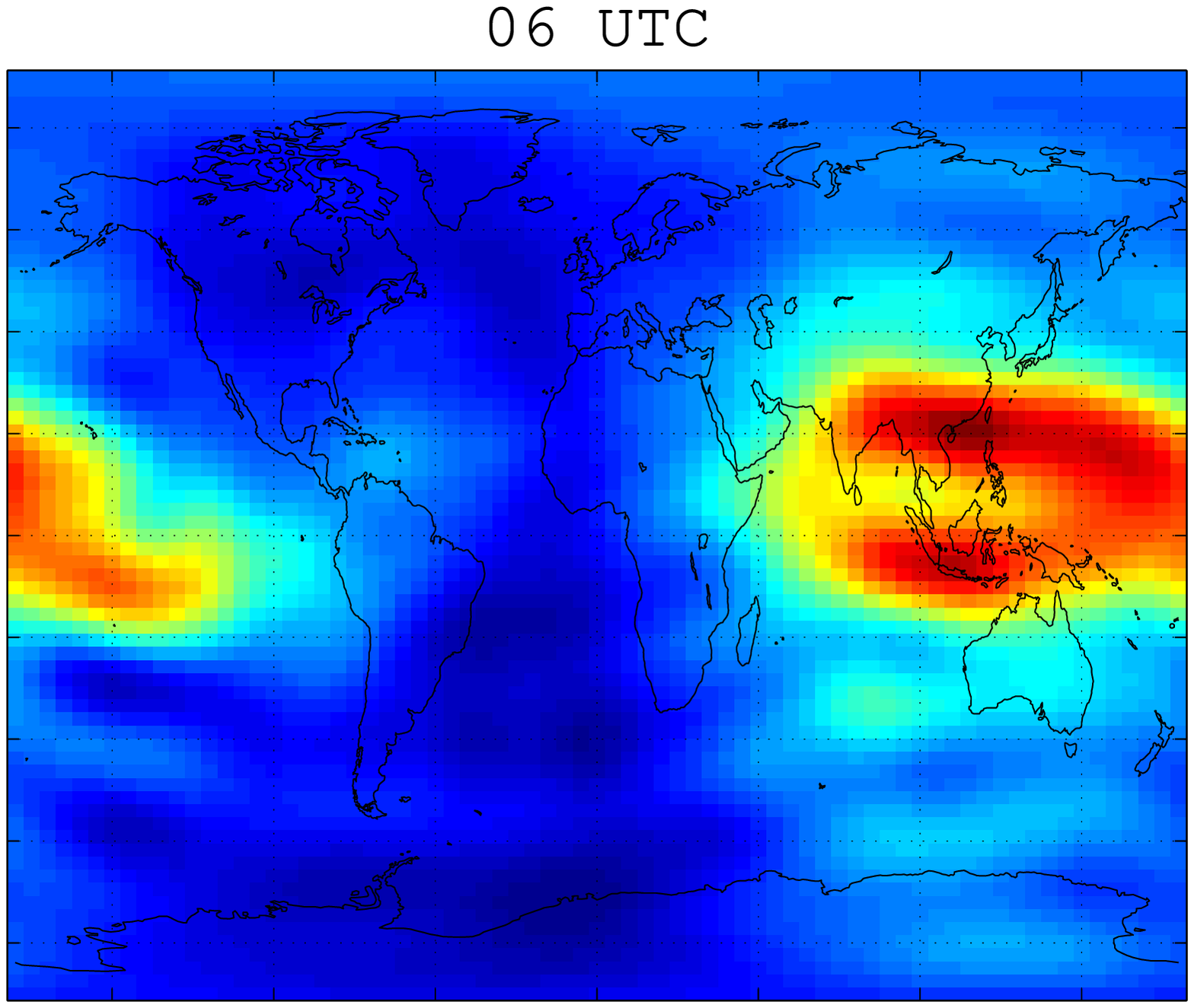}}
     \caption{CODE IONEX plot for 06 UTC by \subref{fig:ionexsot} \citet{Sot13} and \subref{fig:ionexref} by our software for 11 April 2011
        \label{fig:code-anomaly}}
\end{figure}

The CODE maps presented in \citeauthor{Sot13} show a behaviour of the equatorial anomaly different to what is seen in this work. To underline this inconsistency, the CODE maps for the day 11 April 2011 presented by \citeauthor{Sot13} (Figure \ref{fig:ionexsot}) were plotted with our software, Figure \ref{fig:ionexref}. The possible source of inconsistency lies in the way CODE maps are plotted by \citeauthor{Sot13}. On close inspection it is clear that the latitudes provided in CODE IONEX files are inverted in the plot presented by \citeauthor{Sot13}. This results in the equatorial anomaly appearing to pass directly over the MRO, which is not the case. This conclusion is supported in the literature, \citet[][Figure 7.52]{See03}, \citet{Ausion05}.\\

\onecolumn
\section{Figures}

\renewcommand{\thefigure}{\Alph{section}.\arabic{figure}}
\setcounter{figure}{0}  

\begin{figure}[hp]
 \centering
 \subcaptionbox{$VTEC$ at station MRO1\label{fig:vtecmro1063}}{\includegraphics[scale=0.38]{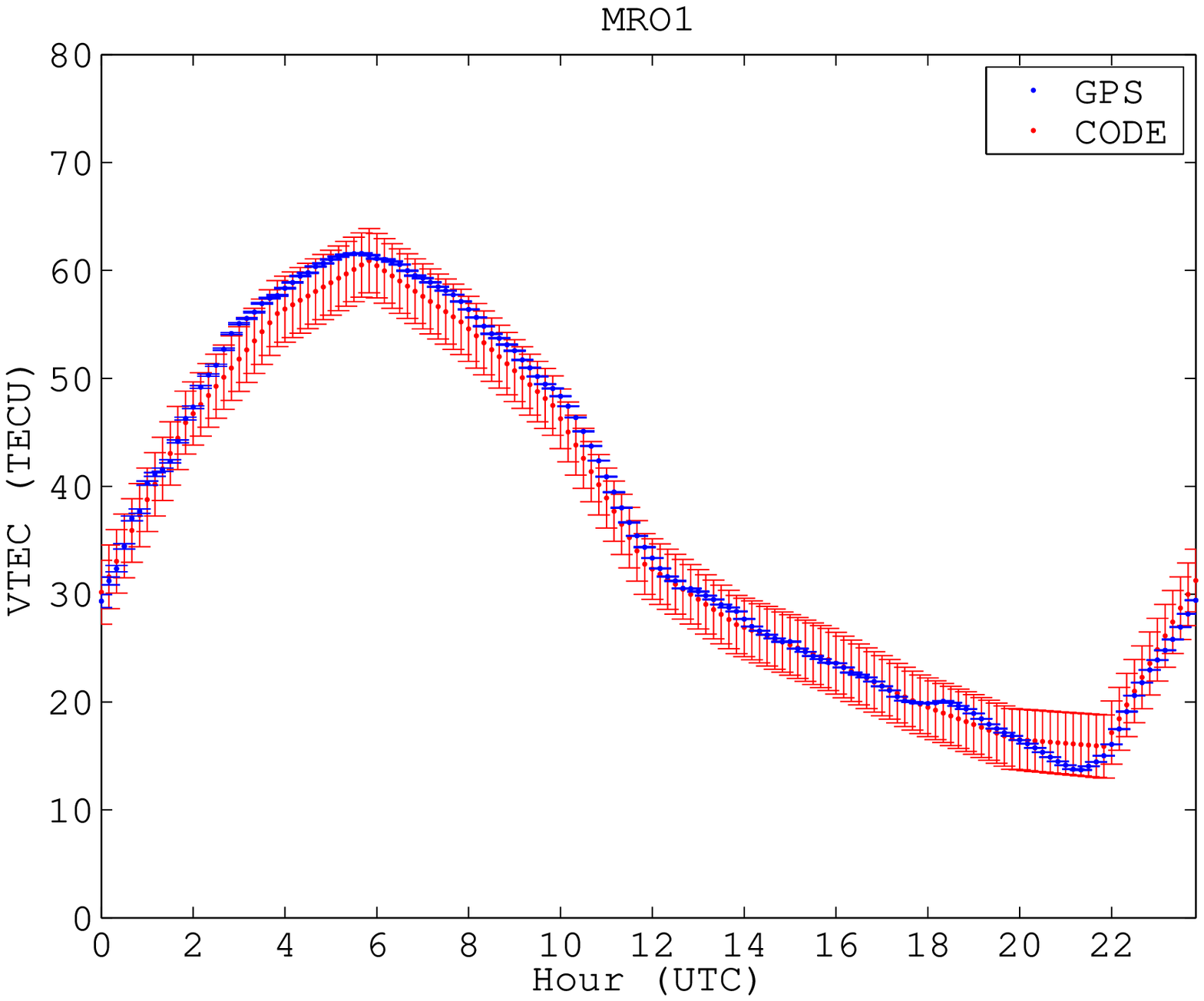}}
  \subcaptionbox{$VTEC$ at station MTMA\label{fig:vtecmtma063}}{\includegraphics[scale=0.38]{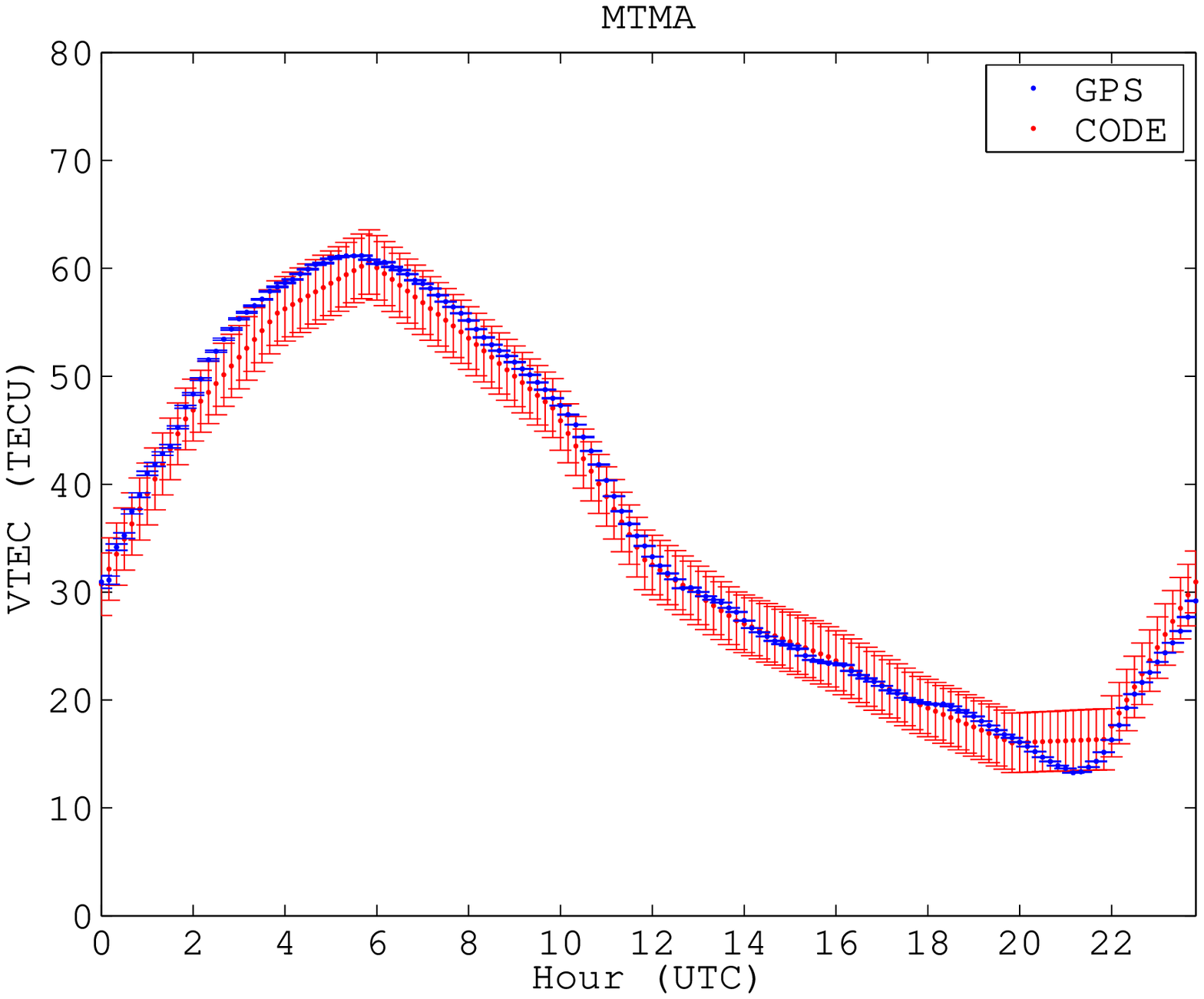}}
  \subcaptionbox{$VTEC$ at station YAR3\label{fig:vtecYAR3063}}{\includegraphics[scale=0.38]{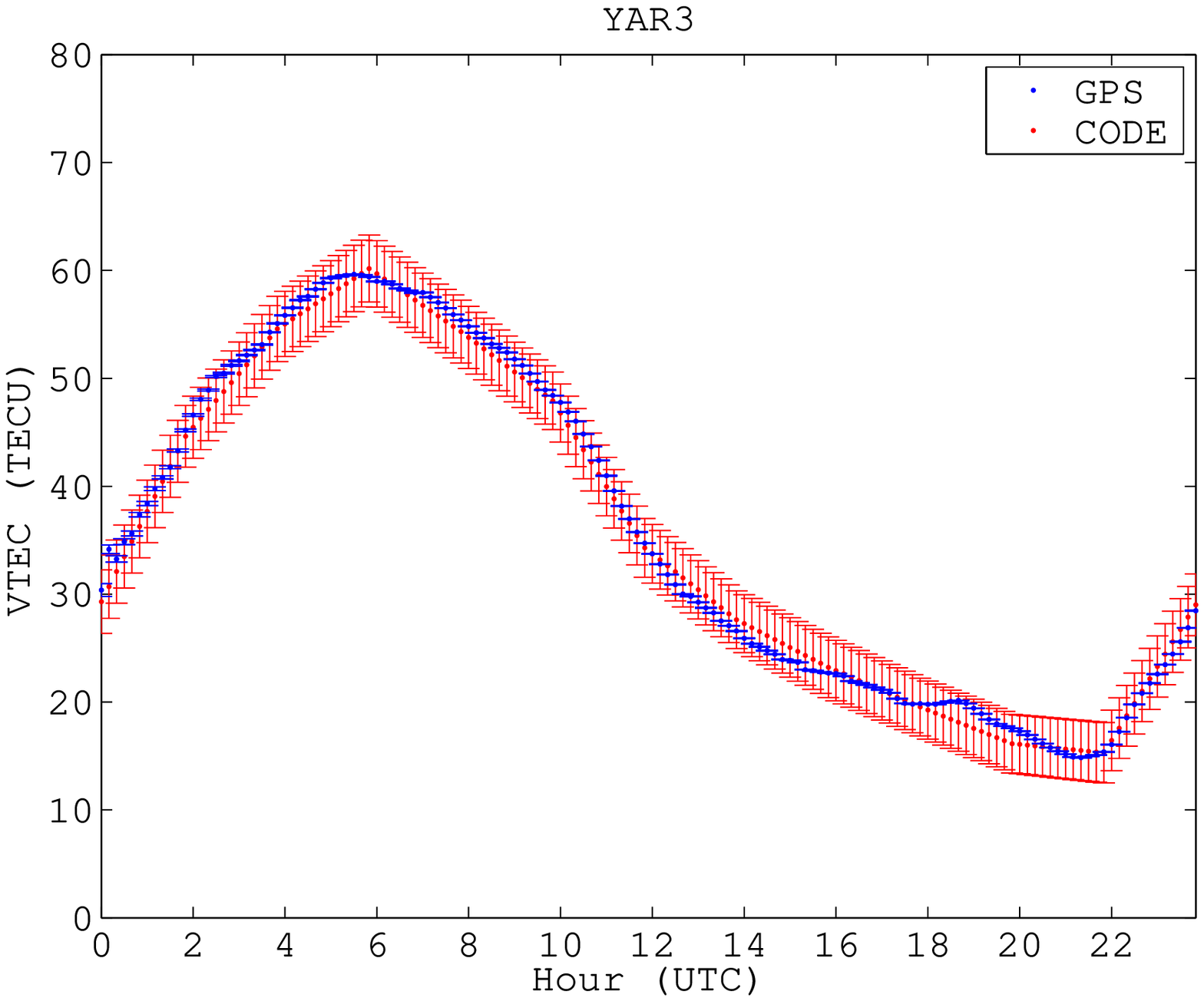}}
  \subcaptionbox{$VTEC$ at station WILU\label{fig:vtecwilu063}}{\includegraphics[scale=0.38]{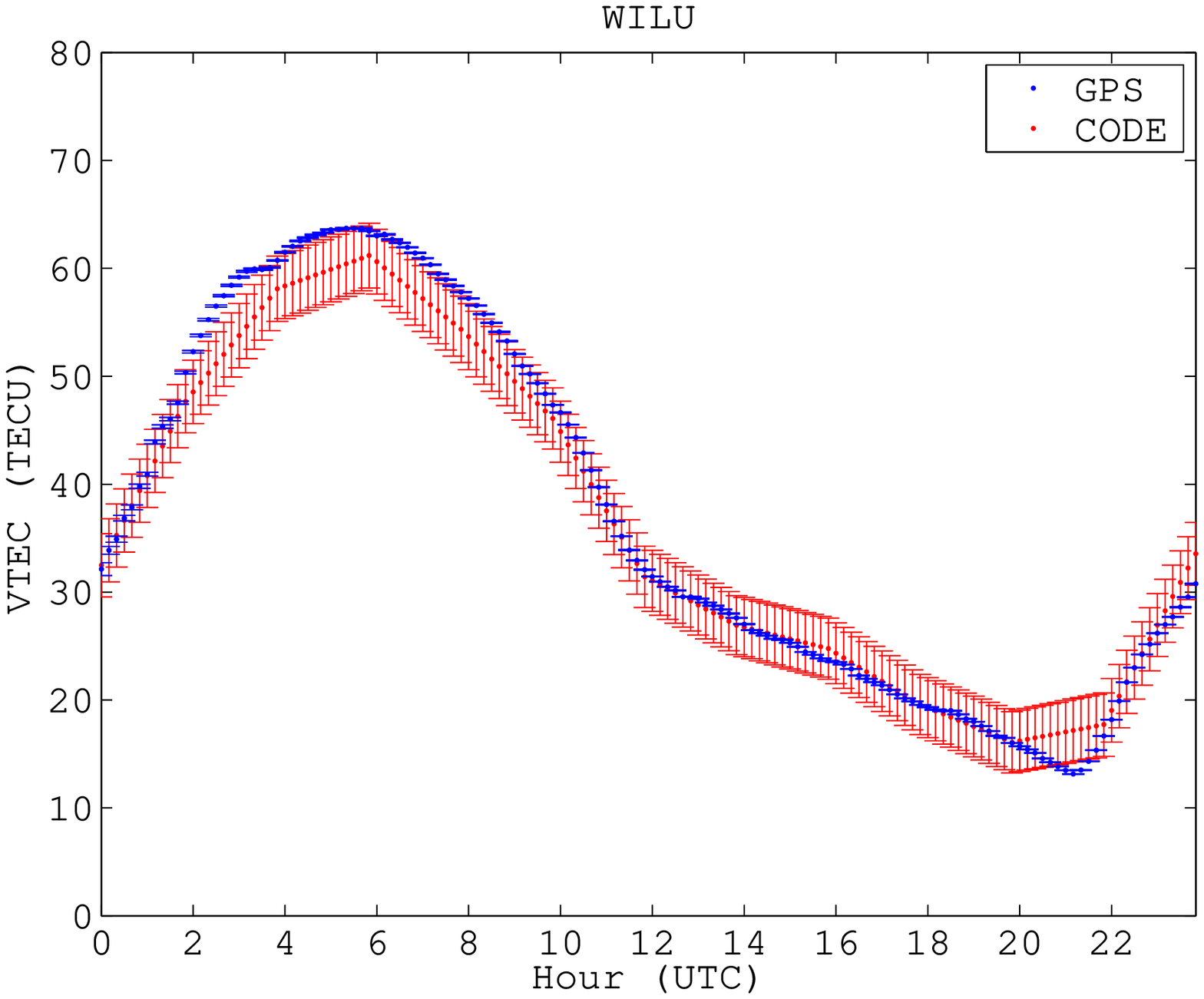}}
     \caption{$VTEC$ at stations MRO1, MTMA, YAR3, and WILU estimated using the method described in the text (blue curve) and CODE IONEX (red curve) on DOY 063, year 2014.
        \label{fig:vtec065}}
\end{figure}

\begin{figure}
 \centering
 \subcaptionbox{$VTEC$ at station MRO1\label{fig:vtecmro1065}}{\includegraphics[scale=0.38]{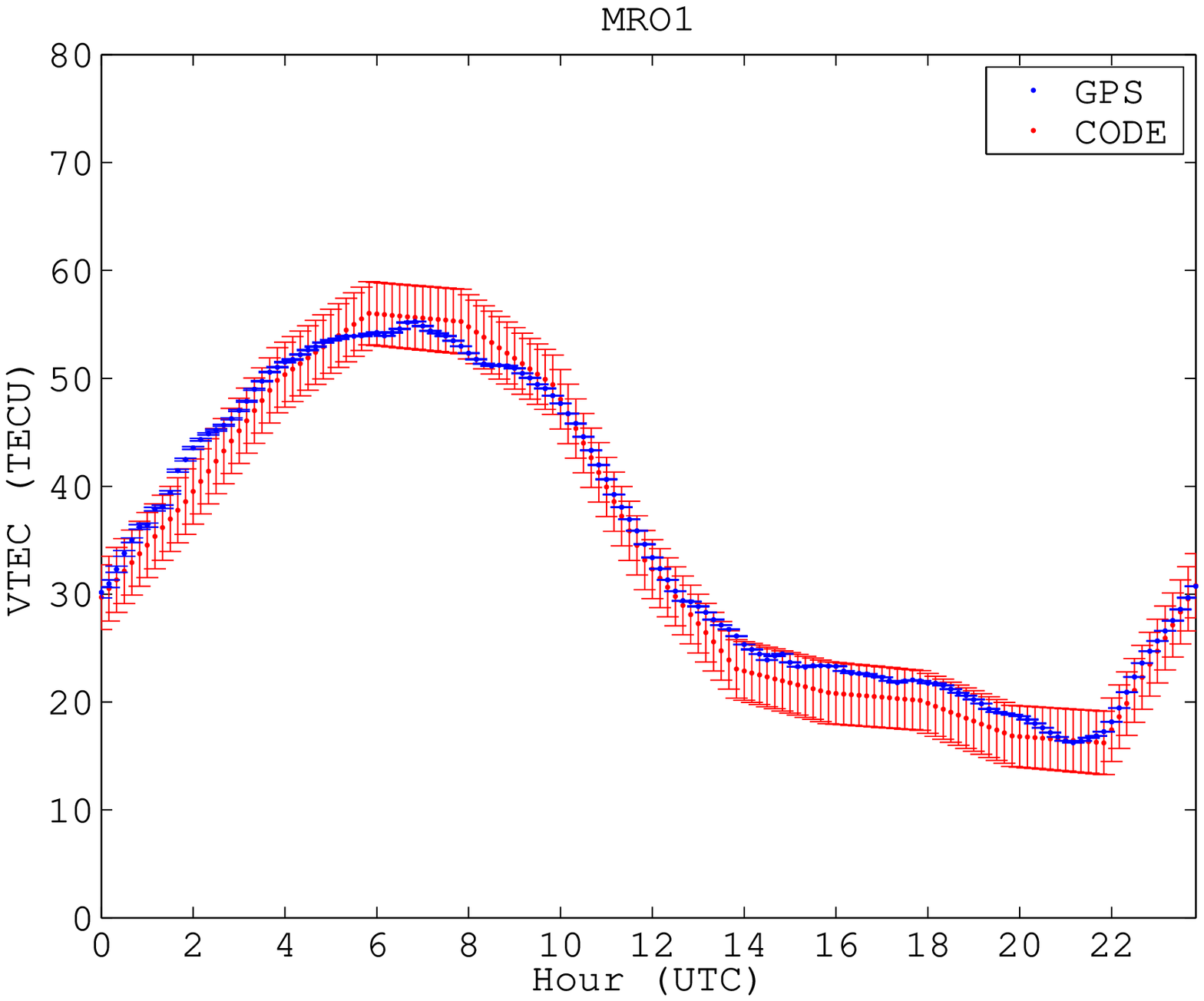}}
 \subcaptionbox{$VTEC$ at station MTMA\label{fig:vtecmtma065}}{\includegraphics[scale=0.38]{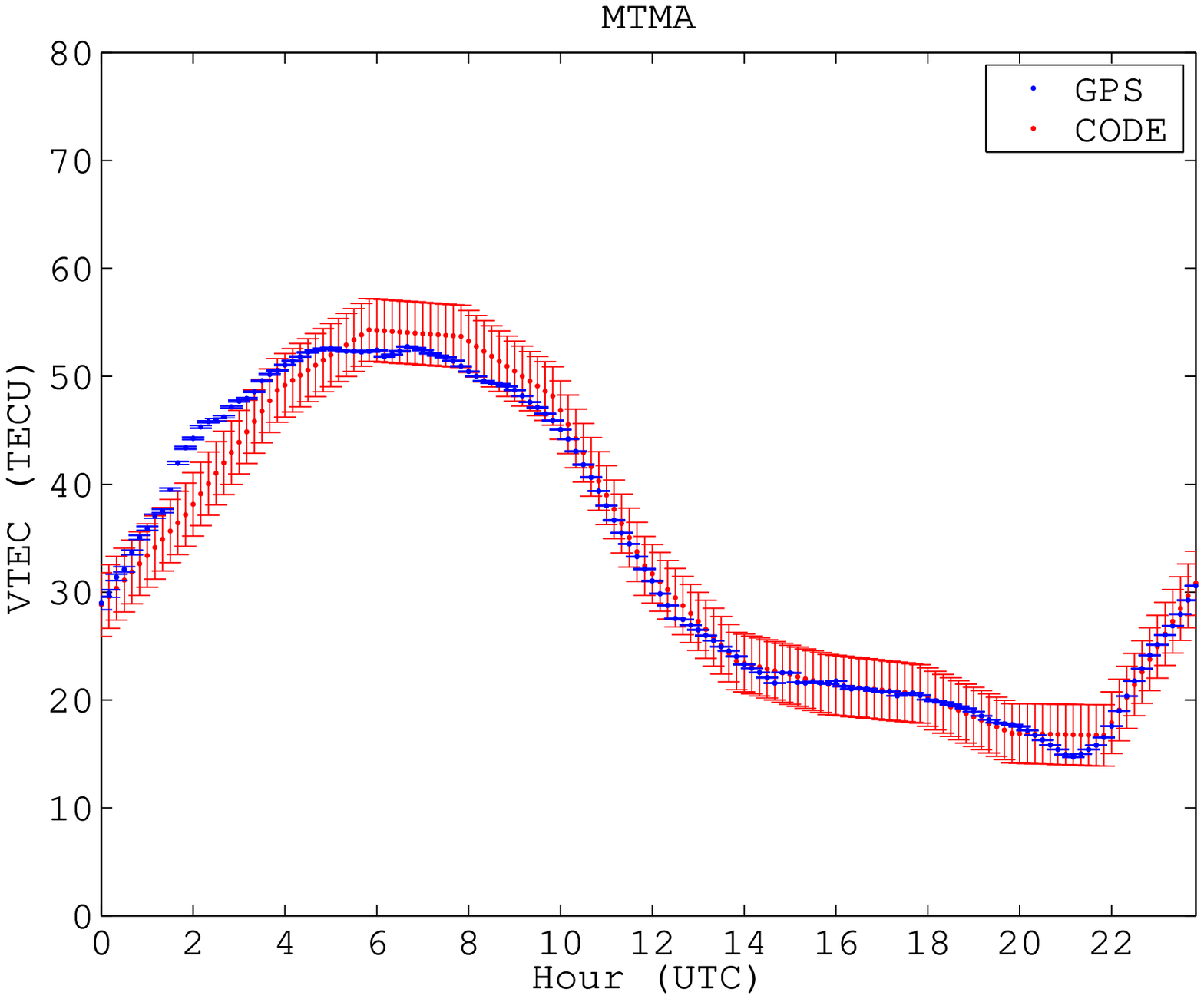}}
 \subcaptionbox{$VTEC$ at station YAR3\label{fig:vtecYAR3065}}{\includegraphics[scale=0.38]{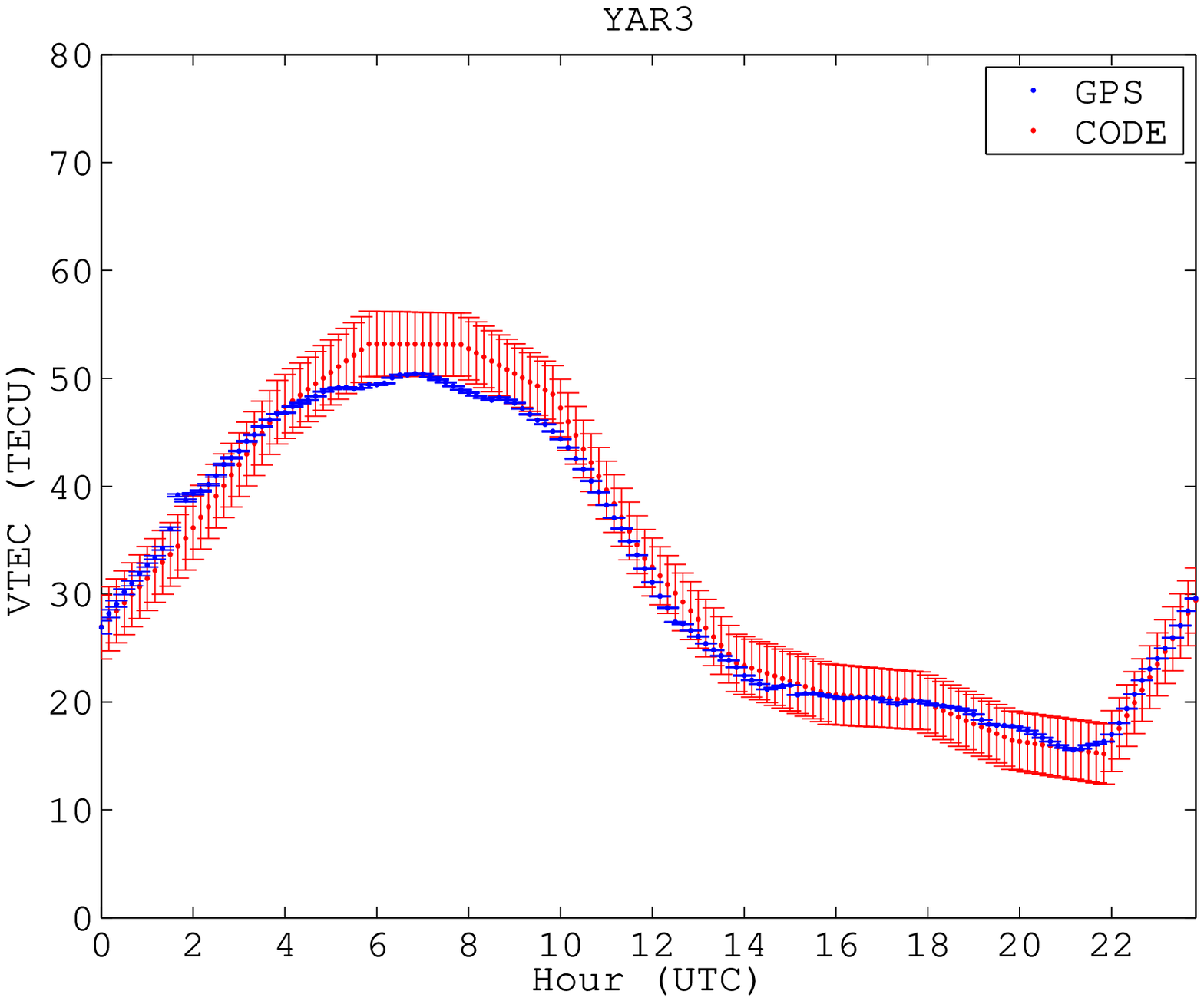}}
  \subcaptionbox{$VTEC$ at station WILU\label{fig:vtecwilu065}}{\includegraphics[scale=0.38]{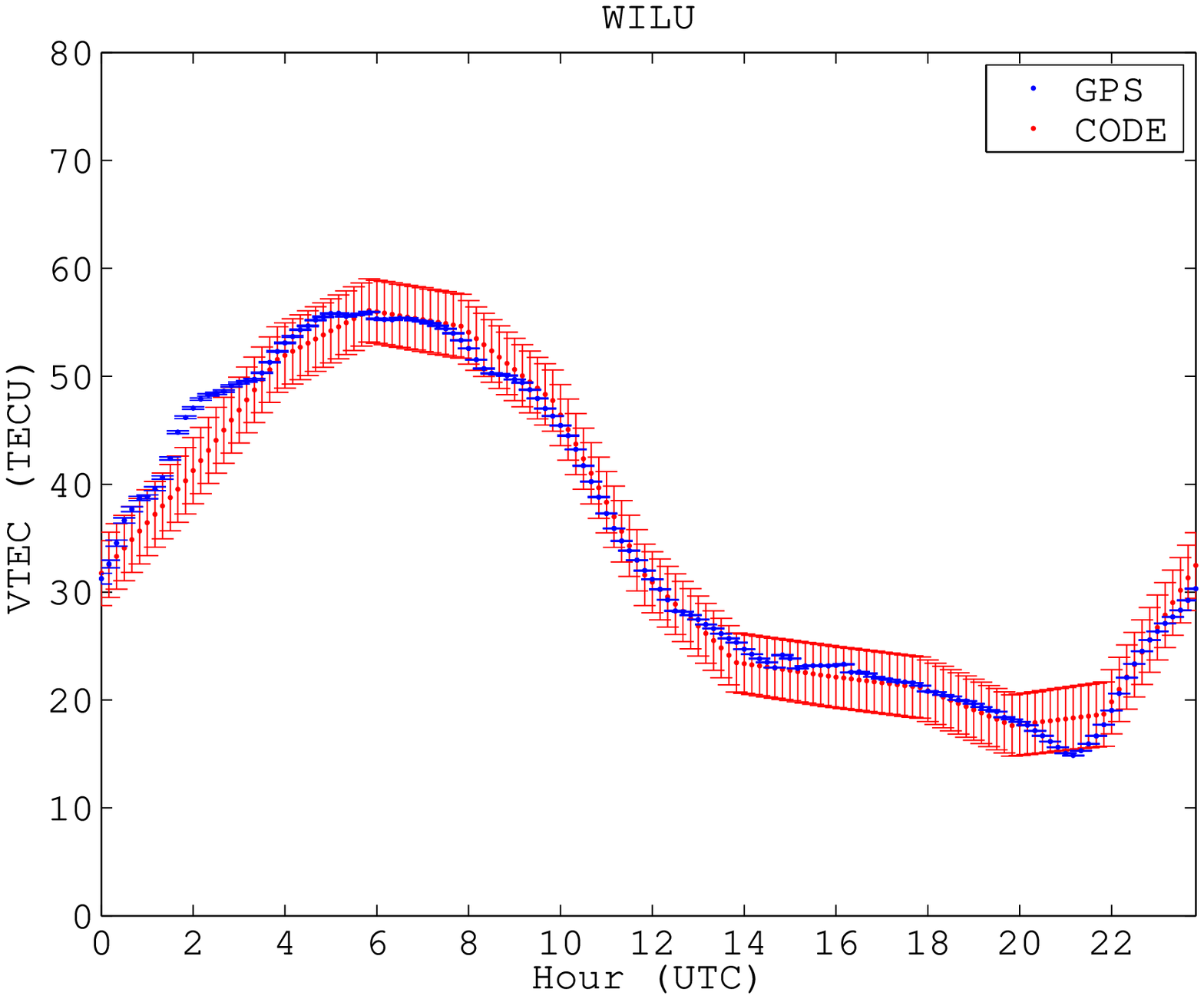}}
     \caption{$VTEC$ at stations MRO1, MTMA, YAR3, and WILU estimated using the method described in the text (blue curve) and CODE IONEX (red curve) on DOY 065, year 2014.
        \label{fig:vtec065}}
\end{figure}

\begin{figure}
 \centering
 \subcaptionbox{$VTEC$ at station MRO1\label{fig:vtecmro1075}}{\includegraphics[scale=0.38]{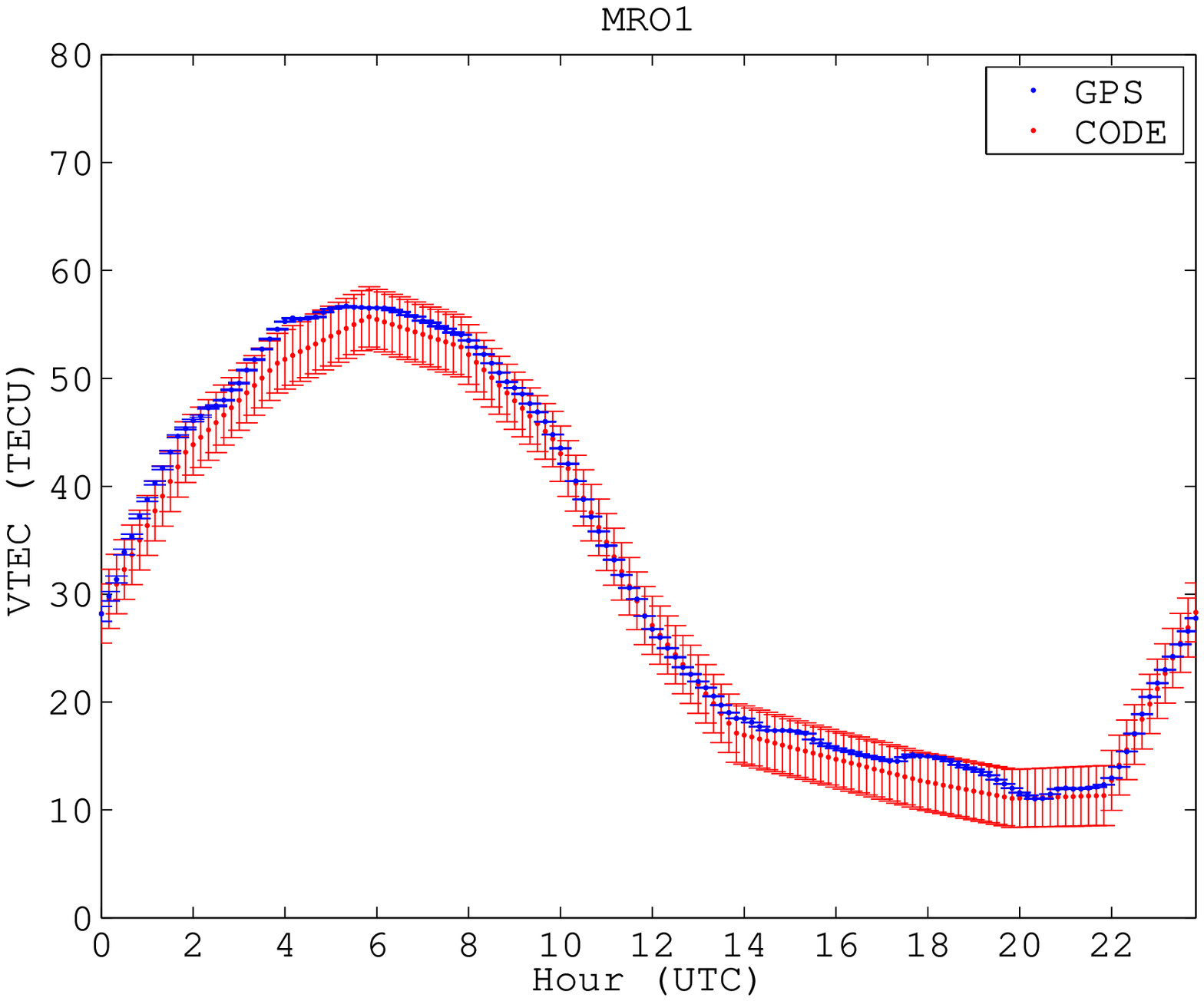}}
 \subcaptionbox{$VTEC$ at station MTMA\label{fig:vtecmtma075}}{\includegraphics[scale=0.38]{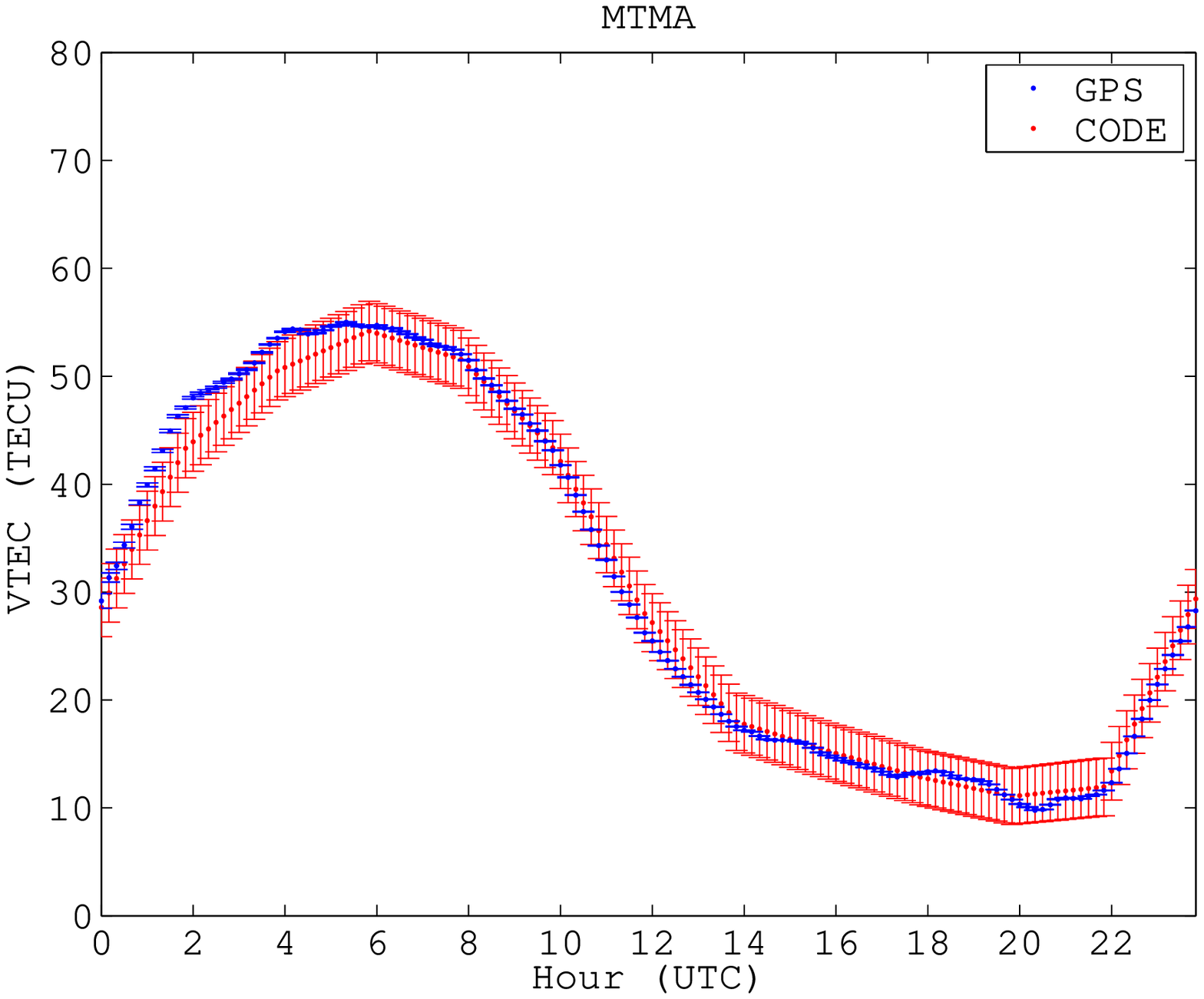}}
 \subcaptionbox{$VTEC$ at station YAR3\label{fig:vtecYAR3075}}{\includegraphics[scale=0.38]{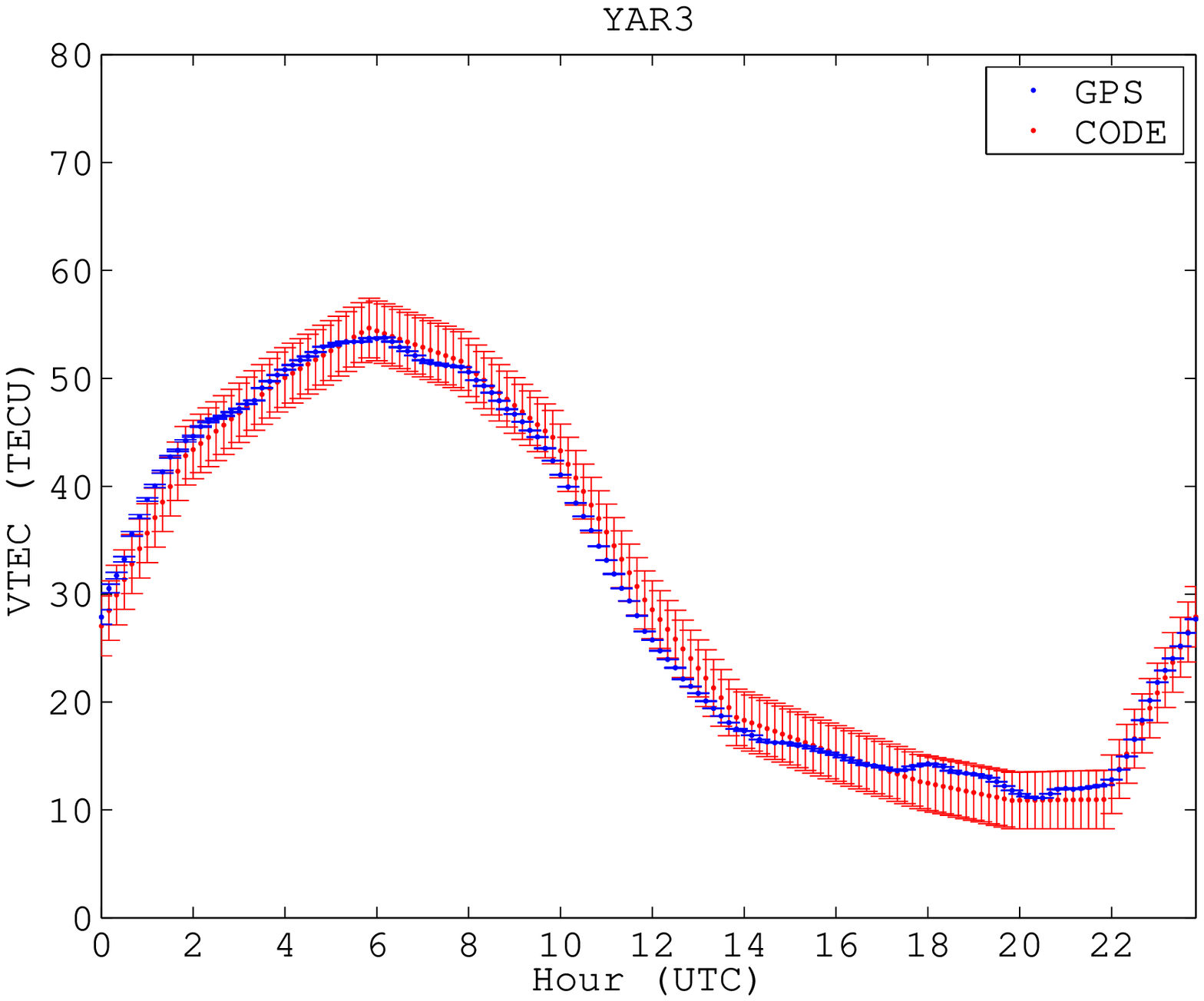}}
 \subcaptionbox{$VTEC$ at station WILU\label{fig:vtecwilu075}}{\includegraphics[scale=0.38]{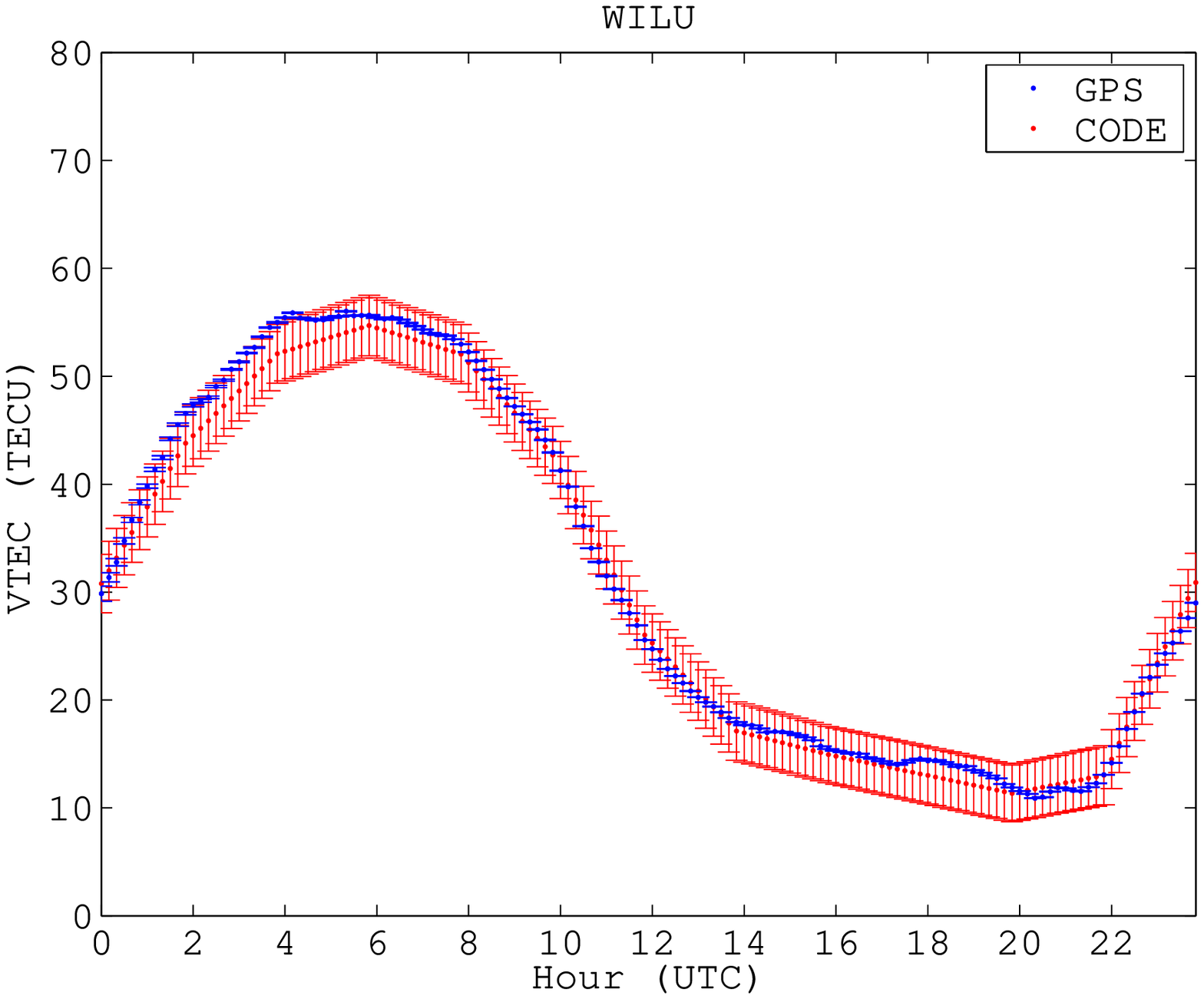}}
     \caption{$VTEC$ at stations MRO1, MTMA, YAR3, and WILU estimated using the method described in the text (blue curve) and CODE IONEX (red curve) on DOY 075, year 2014.
        \label{fig:vtec075}}
\end{figure}

\end{appendix}

\pagebreak
\twocolumn
\newcommand{\pasa}{PASA}
\newcommand{\aj}{AJ}
\newcommand{\apj}{ApJ}
\newcommand{\apjs}{ApJS}
\newcommand{\apjl}{ApJL}
\newcommand{\aap}{A{\&}A}
\newcommand{\aaps}{A{\&}AS}
\newcommand{\mnras}{MNRAS}
\newcommand{\araa}{ARAA}
\newcommand{\pasp}{PASP}
\bibliographystyle{apj}
\bibliography{./Arora_et_al} 

\end{document}